\documentclass[10pt]{article}
\usepackage[utf8]{inputenc}
\usepackage{cite}
\usepackage{hyperref}
\hypersetup{colorlinks=true,linkcolor=bluecyan,citecolor=orange3,urlcolor=green3,pdfencoding=auto,linktocpage}
\usepackage{cancel}
\usepackage{amsmath,amssymb,amsbsy,amstext,amsthm,simplewick,amsfonts,braket}
\usepackage{braket}
\usepackage{mathrsfs}
\usepackage{amsmath}
\usepackage{graphicx}
\usepackage{wrapfig}
\usepackage{upgreek}
\usepackage{bm} %-> bold math
\usepackage{framed}
\usepackage{bbm}
\usepackage{textcomp}
\usepackage{adjustbox}
\usepackage{makecell}
\usepackage{tcolorbox}
\usepackage{empheq}
\usepackage[normalem]{ulem}
\usepackage{enumitem}
\usepackage{array}
\usepackage{dsfont}
\usepackage{ulem}
\usepackage{xcolor}
\definecolor{brown3}{rgb}{0.65,0.16,0.16}
\usepackage{caption}
\usepackage{subcaption}
\usepackage{simpler-wick}
\usepackage{tocloft}
\usepackage{tikz}
\usepackage[compat=1.1.0]{tikz-feynman}
\usepackage{relsize}
\usepackage{float}
\usepackage{xcolor}
\usepackage{tikz}
\usepackage{youngtab}
\usepackage{ytableau}
\usepackage{graphicx}
\usepackage{float}   % for [H]
\usepackage{tikz}
\usetikzlibrary{arrows.meta,backgrounds}

\definecolor{ihBlue}{HTML}{2A6FC4}
\def\scS{0.6}    % box size (cm)
\def\scDx{3.2}   % column spacing (cm)
\def\scDy{1.4}   % row spacing (cm)
\newcommand{\scRow}[3]{%
  \ifnum#2=0
    \node[circle, fill, inner sep=3.8pt] (#1) at ({#2*\scDx},{#3*\scDy}) {};
  \else
    \node[inner sep=0pt, outer sep=0pt, minimum width={#2*\scS cm}, minimum height={\scS cm}]
      (#1) at ({#2*\scDx},{#3*\scDy}) {};
    \draw[scdiagram] (#1.north west) rectangle (#1.south east);
    \ifnum#2>1
      \foreach \i in {2,...,#2}
        \draw[scdiagram] ([xshift={(\i-1)*\scS cm}]#1.north west) -- ++(0,-\scS);
    \fi
  \fi
}

\usepackage{amssymb}
\usetikzlibrary{arrows.meta,backgrounds,calc,positioning}

\definecolor{ihBlue}{HTML}{2A6FC4}
\definecolor{ihRed}{HTML}{C0392B}
\definecolor{ihOrange}{HTML}{CC7A00}   % between layers (colour-blind safe vs red)

\def\s{0.6}                          % box size (cm)
\def\Ux{4.2525} \def\Uy{1.4742}      % "add a box" step
\def\Dx{3.6855} \def\Dy{-1.8711}     % "remove a box" step
\def\Lx{1.764}  \def\Ly{10.332}      % offset of the upper layer
\newcommand{\Pos}[3]{{#1*\Ux+#2*\Dx+#3*\Lx},{#1*\Uy+#2*\Dy+#3*\Ly}}

\tikzset{
  diagram/.style={line width=1.5pt, line join=miter, fill=white},
  arr/.style={-{Stealth[length=10pt,width=7.5pt]}, line width=1.4pt,
              shorten <=4pt, shorten >=5pt},
  lower/.style={arr, ihBlue},
  upper/.style={arr, ihRed},
  between/.style={arr, ihOrange, line width=1.15pt, opacity=0.85},  % add  dash pattern=on 7pt off 4pt  for greyscale print
  more/.style={opacity=0.45},      % stubs: the tree continues
  plane/.style={draw=black!15, line width=0.8pt, rounded corners=14pt},
  keyarr/.style={-{Stealth[length=10pt,width=7.5pt]}, line width=1.4pt, black!75},
  keylabel/.style={font=\LARGE, text=black!80, inner sep=3pt},
  steplabel/.style={font=\LARGE, text=black!60},
  rootlabel/.style={font=\LARGE, inner sep=0pt},
}

\newcommand{\Row}[3]{%
  \pgfmathtruncatemacro{\k}{1+#2-#3}%
  \ifnum\k=0
    \node[circle, fill, inner sep=3.8pt] (#1) at (\Pos{#2}{#3}{0}) {};
  \else
    \node[inner sep=0pt, outer sep=0pt, minimum width={\k*\s cm}, minimum height={\s cm}]
      (#1) at (\Pos{#2}{#3}{0}) {};
    \draw[diagram] (#1.north west) rectangle (#1.south east);
    \ifnum\k>1
      \foreach \i in {2,...,\k}
        \draw[diagram] ([xshift={(\i-1)*\s cm}]#1.north west) -- ++(0,-\s);
    \fi
  \fi
}
\newcommand{\Hook}[3]{%
  \pgfmathtruncatemacro{\k}{2+#2-#3}%
  \node[inner sep=0pt, outer sep=0pt, minimum width={\k*\s cm}, minimum height={2*\s cm}]
    (#1) at (\Pos{#2}{#3}{1}) {};
  \draw[diagram] (#1.north west) -- (#1.north east) -- ++(0,-\s)
    -- ([xshift=\s cm, yshift=\s cm]#1.south west) -- ++(0,-\s)
    -- (#1.south west) -- cycle;
  \draw[diagram] ([yshift=\s cm]#1.south west) -- ++(\s,0);
  \ifnum\k>1
    \foreach \i in {2,...,\k}
      \draw[diagram] ([xshift={(\i-1)*\s cm}]#1.north west) -- ++(0,-\s);
  \fi
}
\ytableausetup{boxsize=0.45em}

\tikzstyle{arrow} = [thick,->,>=stealth]

\usepackage[framemethod=default]{mdframed}
\usetikzlibrary{shapes.geometric, arrows}

\tikzstyle{startstop} = [rectangle, rounded corners, 
minimum width=3cm, 
minimum height=1cm,
text centered, 
draw=black, 
fill=white!30]

\newmdenv[backgroundcolor=gray!15,%
skipabove=5pt,%
skipbelow=5pt,%
leftmargin=2pt,%
rightmargin=2pt,%
innertopmargin=-6pt,%
innerbottommargin=5pt,%
innerleftmargin=5pt,%
innerrightmargin=5pt,%
splittopskip=0pt,%
splitbottomskip=0pt,%
linewidth=0pt,%
nobreak=true]%
{keyeqn}

\newmdenv[backgroundcolor=gray!15,%
skipabove=5pt,%
skipbelow=5pt,%
leftmargin=2pt,%
rightmargin=2pt,%
innertopmargin=-2pt,%
innerbottommargin=5pt,%
innerleftmargin=5pt,%
innerrightmargin=5pt,%
splittopskip=0pt,%
splitbottomskip=0pt,%
linewidth=0pt,%
nobreak=true]%
{keythrm}

\graphicspath{{Fig/}}

\definecolor{lightgreen}{cmyk}{0.2, 0, 0.2, 0.2}
\definecolor{lightgray}{cmyk}{0.1,0.2,0,0.1}
\definecolor{lightgray2}{cmyk}{0.1,0.1,0,0.1}
\definecolor{bluecyan}{RGB}{0, 100, 200}
\definecolor{blue3}{RGB}{31,119,180}
\definecolor{red3}{RGB}{214,39,40}
\definecolor{orange3}{RGB}{255,127,14}
\definecolor{green3}{RGB}{44,160,44}

\definecolor{red2}{RGB}{255,0,0}
\definecolor{green2}{RGB}{0,170,0}
\definecolor{blue2}{RGB}{0,128,255}
\definecolor{magenta2}{RGB}{191,64,191}
\definecolor{purple2}{RGB}{112,48,160}
\definecolor{orange2}{RGB}{255,192,0}

\newcommand\Ccancel[2][red]{
    \let\OldcancelColor\CancelColor
    \renewcommand\CancelColor{\color{#1}}
    \cancel{#2}
    \renewcommand\CancelColor{\OldcancelColor}
}

\newcommand{\red}[1]{\textcolor{red3}{#1}}

\def\Tr{{\rm Tr}}

\newcommand{\be}{\begin{equation}}
\newcommand{\ee}{\end{equation}}
\newcommand{\bea}{\begin{eqnarray}}
\newcommand{\eea}{\end{eqnarray}}
\newcommand{\nn}{\nonumber}

\numberwithin{equation}{section}

\allowdisplaybreaks[1]
\begin{document}

\begin{titlepage}
	\setcounter{page}{1} \baselineskip=15.5 pt 
	\thispagestyle{empty}

 \begin{center}
		{\fontsize{18}{18}\centering {\bf{Symmetries and Soft Theorems of \\ \vspace{0.2cm} Scalar-Vector Theories}}\;}\\
	\end{center}
 
	\vskip 18pt
	\begin{center}
		\noindent
        Zong-Zhe Du,\footnote{\tt z.du1@imperial.ac.uk}$^{,a}$ Kurt Hinterbichler,\footnote{\tt kurt.hinterbichler@case.edu}$^{,b}$ Samanta Saha\footnote{\tt sxs2638@case.edu}$^{,b}$ and
		David Stefanyszyn\footnote{\tt david.stefanyszyn@nottingham.ac.uk}$^{,c}$
	\end{center}
	
	\begin{center}
		\vskip 8pt
        ${}^a$ \textit{Abdus Salam Centre for Theoretical Physics, \\Imperial College, London SW7 2AZ, United Kingdom} \\
        ${}^b$ \textit{CERCA, Department of Physics, \\ Case Western Reserve University, 10900 Euclid Ave, Cleveland, OH, 44106 } \\
		${}^c$ \textit{School of Physics and Astronomy,
			University of Nottingham, University Park, \\ Nottingham, NG7 2RD, United Kingdom} 
	\end{center}

	%=========================================

	\noindent  

We study scalar-vector extensions of the Galileon, in which a coordinate-dependent shift symmetry acts on the vector field.  These theories arise in the scalar-vector sector of the decoupling limit of ghost-free massive gravity: an abelian version arises from pseudo-linear interactions and a non-abelian version arises from dRGT interactions. Using an algebraic classification of possible symmetry algebras, which incorporates inverse Higgs constraints, we show that these massive gravity theories are the unique ones with the scalar transforming under a scalar symmetry and a vector symmetry, and the vector transforming under a two-form symmetry. We construct the theories using the coset construction, which leads to a simpler presentation of their interactions.  We give a complete derivation of their soft theorems, incorporating the full charge conservation equation, and verify that they are satisfied by the amplitudes up to five points.  We investigate the possible extension to higher shift symmetries, confirming the existence of a scalar-vector special Galileon in the abelian case, and finding no such extension in the non-abelian case.

	%=========================================
	
\end{titlepage} 

%\restoregeometry

\newpage
\setcounter{page}{2}
{
	\tableofcontents
}

%=======================================
%=======================================
\section{Introduction}
There are two methods that have been successfully used to classify Minkowski space effective field theories (EFTs): non-linear symmetries and soft limits of scattering amplitudes. Within the context of scalar EFTs, both methods have demonstrated their power as organizing principles. The symmetry approach works by asking how we can extend the Poincar\'{e} algebra with additional generators corresponding to broken symmetries, which are non-linearly realized on the fields \cite{Bogers:2018zeg,Roest:2019oiw,Roest:2019dxy,Hinterbichler:2014cwa}.\footnote{Non-relativistic theories have also been considered \cite{Zoology,Brauner:2020ezm,Grall:2020ibl,Pajer:2018egx}.} One uses an ansatz for the symmetry algebra, guided by \textit{inverse Higgs trees}, that ensures that the resulting symmetries can be realized on the chosen set of Goldstone bosons, i.e. by ensuring that the algebra leads to the appropriate number of inverse Higgs constraints \cite{Ivanov:1975zq}, and are consistent with canonical kinetic terms \cite{Roest:2019oiw}. This approach is based on the coset construction for non-linear realizations \cite{Coleman:1969sm,Callan:1969sn,Volkov:1973vd} (see \cite{Goon:2012dy,Penco:2020kvy} for reviews).\footnote{An alternative symmetry-based perspective comes from brane constructions \cite{deRham:2010eu,Hinterbichler:2010xn,Goon:2011uw,Goon:2011qf,Novotny:2016jkh,Roest:2020vny}.} The soft limit approach instead classifies theories using the soft limits of scattering amplitudes, with different theories distinguished from one another by the rate at which the amplitude vanishes in the soft limit \cite{Adler:1965ga,Cheung:2014dqa,Cheung:2016drk}. Soft limits of scalar EFTs have now been extensively studied \cite{Adler:1965ga,Cheung:2014dqa,Cheung:2021yog,DiVecchia:2015jaq,Cheung:2023qwn,Du:2024hol,Hui:2022dnm,Bittermann:2022nfh,Padilla:2016mno, Cheung:2016drk,CarrilloGonzalez:2019fzc}. These two approaches are related: additional non-linearly realized symmetries typically lead to an enhanced rate at which the soft amplitudes vanish. The algebraic approach has the benefit of being conclusive while the soft limits approach can only provide (very convincing) evidence for the existence of theories. On the other hand, the soft limits approach is on-shell and therefore avoids the complications of field redefinitions and basis changes that are unavoidable in the algebraic approach \cite{Klein:2017npd,deRham:2013hsa}. For these reasons, a hybrid approach is often very useful. 

For scalars, the situation is very well understood.  For theories involving spinning fields, a full classification is not known, but progress has been made \cite{Klein:2018ylk,Cheung:2018oki,Kampf:2021bet,Bonifacio:2019hrj,Elvang:2018dco}. In this work, we study theories involving a scalar $\phi$ and gauge vector $A_{\mu}$ that non-linearly realize symmetry algebras forcing interesting interactions between the two fields, and exhibit interesting soft behavior in the corresponding scattering amplitudes. In \cite{Bonifacio:2019hrj}, an example of such a theory arising in the decoupling limit of massive gravity \cite{deRham:2010ik,deRham:2010kj} (see \cite{Hinterbichler:2011tt,deRham:2014zqa} for reviews) was discussed. The scalar $\phi$ in this theory corresponds to the helicity-zero mode of the massive graviton, and is naturally a Galileon \cite{Nicolis:2008in} with symmetries \cite{Nicolis:2008in,Goon:2012dy}
\begin{align} \label{GalSymm}
    \delta \phi = c + c_{\mu}x^{\mu}\,,
\end{align}
for constant $c_\mu,c$, while the St\"uckelberg transformation on the helicity-1 mode of the massive graviton, captured by the vector field $A_\mu$, is realized as a global non-abelian transformation of the form
\begin{align} \label{TwoFormSymm}
    \delta A_{\mu} = m_{\mu\nu} (x^\nu + \alpha\, \partial^\nu \phi) \,,
\end{align}
where $m_{\mu\nu} = - m_{\nu \mu}$ is a constant two-form symmetry parameter and $\alpha$ is a dimensionful scale that in the massive gravity completion is related to the $\Lambda_3$ strong coupling scale \cite{Bonifacio:2019hrj}. When $\alpha=0$, the algebra becomes ``abelian,'' and this case arises from the decoupling limit of pseudo-linear massive gravity \cite{Hinterbichler:2013eza}. 

We will refer to this class of scalar-vector theories as \textit{scalar-vector Galileons}. The four-dimensional scattering amplitudes and other properties of these theories have been studied in \cite{Kampf:2021bet} using spinor helicity soft theorems.  The leading interactions were also derived directly from the decoupling limit of massive gravity in \cite{Ondo:2013wka}. 

We begin in Section \ref{sec:IHTrees} by setting up the method by which one should look for symmetries within a theory of a scalar and gauge vector.  We introduce two inverse Higgs trees, which fix the commutators between spacetime translations and any other generator, ensuring that we have the necessary number of inverse Higgs constraints. Commutators between two non-Poincar\'{e} generators are in principle only constrained by Lorentz symmetry, but we will also impose the condition that the infinite tower of generators coming from the $U(1)$ gauge symmetry form an ideal of the full symmetry algebra. In Sections \ref{sec:LevelOne} and \ref{sec:LevelTwo}, we then impose the Jacobi identities to constrain the commutators to the allowed subsets. In Section \ref{sec:LevelOne}, we show that the scalar-vector Galileon theory is the only non-trivial theory with interactions between these two fields when the scalar is invariant under a scalar and vector shift, and the vector is invariant under a two-form shift, e.g. we rule out the possibility of a DBI scalar or dilaton coupled to such a gauge field, and we rule out other two-form symmetry transformations for the vector. We then study this theory in detail using the coset construction to find the leading-order interactions, which are Wess-Zumino terms. Our bottom-up symmetry-based approach yields the same interactions found in the decoupling limit of massive gravity \cite{Ondo:2013wka}, but in a more compact form, and shows that they are Wess-Zumino terms. 

In Section \ref{sec:SoftTheorems}, we then perform a detailed derivation of the soft theorems stemming from the non-linearly realized symmetries. In doing so, we utilise the \textit{amplitude combination} proposed in \cite{Du:2024hol, Du:2024soq}, where special combinations of amplitudes appear in the soft theorems. Due to the time-ordered Ward-Takahashi identity, the conserved current hits both asymptotic past and future vacua when Fourier transforming in time, leading to a collaboration between a soft \textit{insertion} and \textit{emission} of the Goldstone mode. The resulting amplitude combinations for the soft scalar are $\mathcal{A}_{E_p} + \mathcal{A}_{-E_p}$ and $\frac{\mathcal{A}_{E_p} - \mathcal{A}_{-E_p}}{E_p}$, and spatial momentum derivatives thereof, where the subscript $E_{p}$ denotes the effective soft energy of the scalar emission. For soft vector particles, the conserved current necessarily carries polarization vectors, rendering the derivation more involved than in the scalar case. In this work, we employ an S-matrix Ward-Takahashi identity
\begin{align}
    \braket{\text{out}|Q S - S Q|\text{in}} = 0\,,
\end{align}
to tackle the problem, allowing us to bypass the off-shell correlation function derivation of soft theorems and directly derive the on-shell identities associated with the non-linear symmetries in $D$ dimensions. Using an example in $D=4$, we show that the amplitude combination is crucial for the soft theorems to hold. 

In Section \ref{sec:SoftBootstrap}, we construct the general four- and five-point tree-level amplitudes for any scalar-vector theory by imposing Lorentz symmetry, gauge invariance and parity invariance up to $2n-2$ derivatives, where $n$ is the number of external legs. We then bootstrap the theory by imposing the soft theorems, showing all five-point Wilson coefficients are fixed in terms of four-point Wilson coefficients for four-vector one-scalar scattering, as well as an additional constraint on the four-point coefficients. We verify that the resulting amplitudes are those of the scalar-vector Galileon theory. 

We extend our symmetry-based analysis in Section \ref{sec:LevelTwo} by looking for special subsets of this theory with larger symmetries. To do so, we include a traceless tensor generator in the algebra such that the scalar becomes a special Galileon \cite{Hinterbichler:2015pqa} and we ask if such a scalar can couple to a gauge field with a two-form symmetry. We find that the two-form symmetry \eqref{TwoFormSymm} is compatible with the special Galileon symmetry only when $\alpha=0$, i.e. in the abelian limit. While an algebra exists in the $\alpha \neq 0$ case, it is incompatible with the tower of gauge symmetries. Our symmetry-based result agrees with the amplitude-based result of \cite{Kampf:2021bet}, which also noted the existence of this special scalar-vector Galileon. Both the scalar and vector transform under the new traceless tensor generator (the vector transforms like a matter field) and we use the symmetries to find the special subset of the scalar-vector Galileon Lagrangian in $D$ dimensions. We also derive the additional soft theorems associated with the traceless tensor symmetry in Section \ref{sec:SoftTheorems}. We conclude and discuss avenues for future work in Section \ref{sec:Conclusions}.

\paragraph{Comparison with previous work} Some of the results we present in this work are already known in the literature, although we arrive at them in new ways, while others are new. We therefore find it useful to summarize how our work fits within the existing literature:
\begin{itemize}
    \item We show that the scalar-vector Galileon theory is the unique theory for which the scalar transforms under a scalar and a vector symmetry while the vector transforms under a two-form symmetry. While this theory was known from \cite{Ondo:2013wka,Bonifacio:2019hrj,Kampf:2021bet}, it was not previously known that the two-form symmetry is incompatible with the DBI and dilaton scalar theories.
    \item We prove that the unique subset of this theory with additional symmetries is the special scalar-vector Galileon and we find the interactions in $D$ dimensions.
    \item The soft theorems we derive are valid in $D$ dimensions, thereby generalizing the $D=4$ results of \cite{Kampf:2021bet}. We also demonstrate the universality of the amplitude combination between soft mode insertion and emission first seen in \cite{Du:2024hol} in the presence of spin.
    \item We present the complete five-point amplitudes for a scalar-vector theory in $D=4$ using the spinor helicity formalism.
\end{itemize}

\section{Algebra Structures and Inverse Higgs Trees}\label{sec:IHTrees}

We begin with the algebraic classification of symmetries of EFTs containing a scalar and a gauge vector.  We consider an algebra that contains the $D$-dimensional Poincaré algebra, consisting of translation generators $P_\mu$ and Lorentz generators $J_{\mu\nu}$, which is linearly realized in the standard way on the scalar and vector Goldstones. The commutators are
\begin{align}
[P_\mu, P_\nu] &= 0 \,,
\\
[J_{\mu\nu}, J_{\rho\sigma}]
&=
\eta_{\mu\rho}J_{\nu\sigma}
-
\eta_{\nu\rho}J_{\mu\sigma}
-
\eta_{\mu\sigma}J_{\nu\rho}
+
\eta_{\nu\sigma}J_{\mu\rho} \,,
\\
[J_{\mu\nu}, P_\rho]
&=
\eta_{\mu\rho}P_\nu
-
\eta_{\nu\rho}P_\mu \,.
\end{align}
We then seek to enlarge the algebra by adding additional generators that are non-linearly realized on the fields. For now, collectively denote these generators by $G_n$. The commutators between $G_n$ and Lorentz are fixed by the tensor index structure of the new generators.  The commutator with translations will be largely fixed by our requirement that the resulting theory contains a single scalar and a single gauge vector: for the spontaneous breaking of spacetime symmetries, Goldstone's theorem does not apply, and some of the would-be Goldstones associated with the broken generators (henceforth referred to as ``inessential Goldstones'') can be eliminated in favor of ``essential Goldstones.'' Only the latter are necessary to realize the full algebra. If the algebra contains a commutator of the form $[P_{\mu}, G_n] \supset  G_{m}$, i.e. the commutator between a non-Poincar\'{e} generator and translations gives back a different non-Poincar\'{e} generator, the Goldstone associated with $G_n$ can be eliminated in favor of the Goldstone associated with $G_m$. Such conditions are referred to as \textit{inverse Higgs constraints} (IHCs). The coset construction efficiently captures the presence of such IHCs and ensures they are consistent with the symmetry algebra: we introduce a field for each broken generator and the coset produces one-forms of which some can be set to zero to yield the IHCs \cite{Ivanov:1975zq,Low:2001bw, Goon:2012dy}. The structure of the $[P_{\mu}, G_{n}]$ commutators can then be fixed by \textit{inverse Higgs trees} which for a given set of essential Goldstones, tell us which generators can be included in the algebra \cite{Roest:2019oiw,Roest:2019dxy}.

\begin{figure}[H]
  \centering
  \resizebox{0.65\linewidth}{!}{%
  \begin{tikzpicture}[
      scdiagram/.style={line width=1.5pt, line join=miter, fill=white},
      sclink/.style={-{Stealth[length=10pt,width=7.5pt]}, line width=1.4pt, ihBlue,
                     shorten <=4pt, shorten >=5pt},
      scplane/.style={draw=black!15, line width=0.8pt, rounded corners=14pt},
      scstep/.style={font=\LARGE, text=black!60},
    ]
    % diagrams (name = N<column><row>; column = number of boxes)
    \foreach \c/\r in {0/4,0/2,0/0, 1/3,1/1, 2/2,2/0, 3/1, 4/0}
      {\scRow{N\c\r}{\c}{\r}}

    % faint shaded plane behind the tree
    \begin{scope}[on background layer]
      \path[scplane, fill=ihBlue!5]
        (-0.55,7.10) -- (-0.55,-0.70) -- (14.40,-0.70) -- (14.40,0.56) -- cycle;
    \end{scope}

\foreach \u/\v in {N13/N04,
                   N22/N13, N02/N13,
                   N31/N22, N11/N22, N11/N02,
                   N40/N31, N20/N31, N20/N11, N00/N11}
  \draw[sclink] (\u) -- (\v);

    % step numbers beside the rows, and the starting field
    \foreach \r in {0,...,4}
      {\pgfmathtruncatemacro{\n}{4-\r}\node[scstep] at (-2.0,{\r*\scDy}) {$\n$};}
    \node[scstep] at (-2.0,{4*\scDy+1.2}) {Level $n$};
    \node[font=\LARGE, inner sep=0pt, anchor=south] at ([yshift=8pt]N04.north) {$\phi$};
  \end{tikzpicture}%
  }
  \caption{Scalar inverse Higgs tree written in terms of Lorentz irreducible representations \cite{Roest:2019oiw}. }
  \label{fig:scalarIH}
\end{figure}
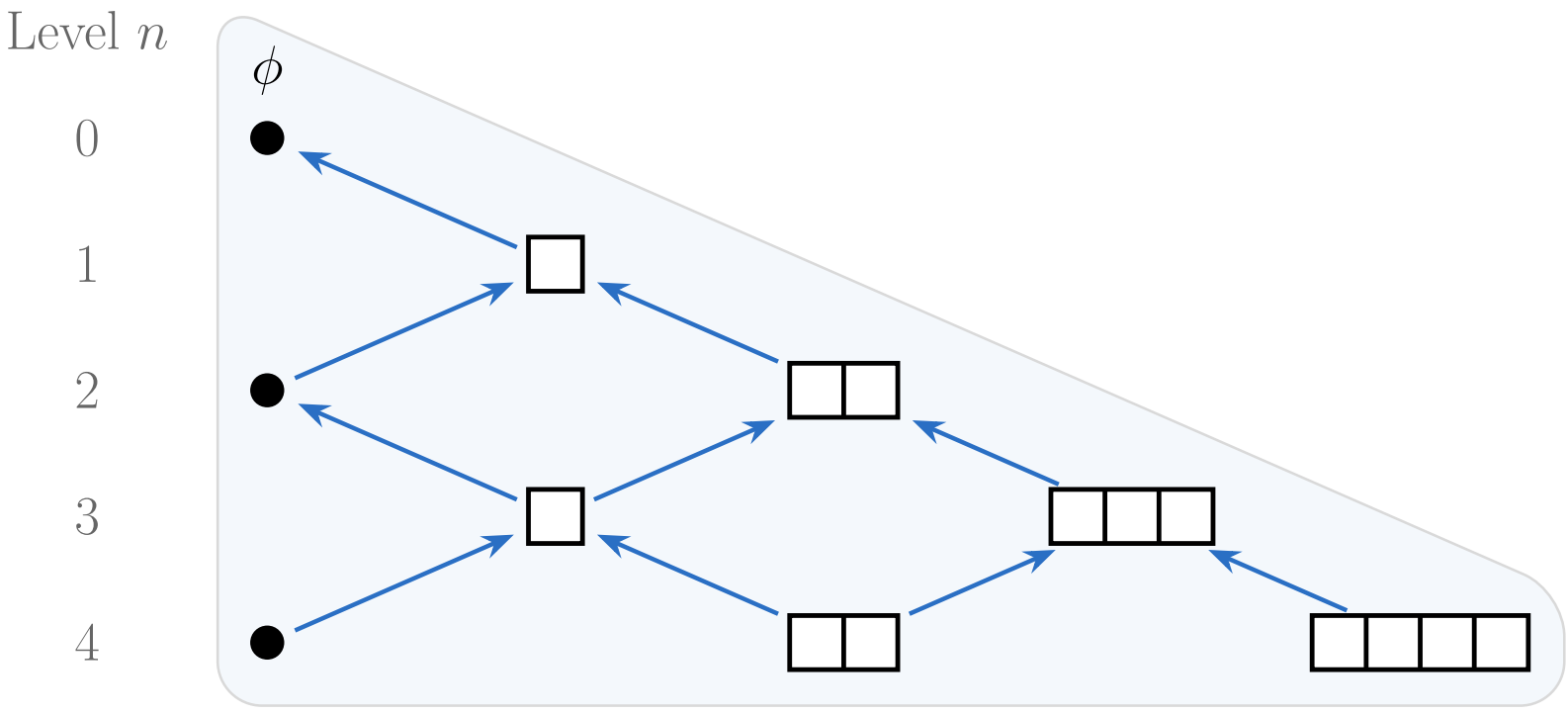

\paragraph{Scalar Inverse Higgs Tree:} Our essential Goldstones are a scalar and a gauge vector. For the scalar, any non-linearly realized symmetry acts on the field as \cite{Hinterbichler:2014cwa}
\begin{equation} \label{ScalarTransformation}
\delta_n \phi
=
c_{\mu_1\cdots\mu_n}\,
x^{\mu_1}\cdots x^{\mu_n}
+\cdots,
\end{equation}
where the $c_{\mu_1\cdots\mu_n}$ are constant, symmetric symmetry parameters, and the $+ \ldots$ allows for field-dependent terms coming from interactions. This general symmetry expression informs us of the Lorentz representations of the generators that can appear in the inverse Higgs tree for the scalar with each generator being fully symmetric. So, at the zeroth level in the tree (where the level refers to the power of $x^{\mu}$ in \eqref{ScalarTransformation}) only a scalar generator can be included. This is the generator associated with $\phi$ in the coset. At the first level only a vector can be included, while at the second level a scalar and a traceless tensor can be included. This structure is shown in Figure \ref{fig:scalarIH} with each node in the inverse Higgs tree representing a Lorentz irreducible representation and an arrow illustrating a connection between these generators using translations. Acting on \eqref{ScalarTransformation} with a translation simply reduces the order of the symmetry by one, and so takes us down one level in the tree. 

\usetikzlibrary{positioning}

Although each generator in Figure \ref{fig:scalarIH} is allowed by Lorentz and the IHCs, as argued in \cite{Roest:2019oiw}, only the traceless coefficient tensors in \eqref{ScalarTransformation} are symmetries of the canonical scalar kinetic term, so only the traceless generators, those on the far-right diagonal of Figure \ref{fig:scalarIH}, are consistent with a canonical kinetic term of the scalar. In this work we will restrict to those generators. Furthermore, although our methods are general, we will truncate the tree at the second level, so we allow for a scalar $C$, a vector $C_{\mu}$, and a traceless tensor $C_{\mu\nu}$. It is known that for a lone scalar there are no non-trivial algebras beyond this truncation \cite{Bogers:2018zeg,Roest:2019oiw,Cheung:2014dqa}. The commutators with Lorentz are therefore \begin{align}
[J_{\mu\nu}, C]
&=
0 \,,
\\
[J_{\mu\nu}, C_\rho]
&=
\eta_{\mu\rho}C_\nu
-
\eta_{\nu\rho}C_\mu \,,
\\
[J_{\mu\nu}, C_{\rho\sigma}]
&=
\eta_{\mu\rho}C_{\nu\sigma}
-
\eta_{\nu\rho}C_{\mu\sigma}
+
\eta_{\mu\sigma}C_{\nu\rho}
-
\eta_{\nu\sigma}C_{\mu\rho} \,, \label{TensorAndLorentz}
\end{align}
and the inverse Higgs tree tells us that the commutators with translations are 
\begin{align}
[P_\mu, C]
&=
a_1 P_\mu \,, \\
[P_\mu, C_\nu]
&=
\eta_{\mu\nu}C
+
a_2 J_{\mu\nu} \,,
\\
[P_{\mu}, C_{\nu \rho}] &= \eta_{\mu\nu} C_{\rho} + \eta_{\mu \rho}C_{\nu} - \frac{2}{D} \eta_{\nu \rho} C_{\mu} + a_3\left( \eta_{\mu\nu} P_{\rho} + \eta_{\mu \rho}P_{\nu} - \frac{2}{D} \eta_{\nu \rho} P_{\mu} \right) \,. \label{TensorAndTranslations}
\end{align}
Note that although the inverse Higgs tree dictates which non-linear generators appear on the right-hand side of these commutators, it does not dictate which linearly-realized generators can appear so we allow for all possibilities, hence the arbitrary constants $a_1,a_2,a_3$. If the generator $C_{\mu}$ appears in the tree, we must have $[P_{\mu}, C_{\nu}] \supset \eta_{\mu\nu}C$ so that the associated Goldstone can be eliminated in favor of $\phi$. We therefore don't include a free coefficient in that part of the commutator since it can always be rescaled to unity. The same applies for the tensor generator at level two.   

\begin{figure}[H]
  \centering
  \providecolor{ihBlue}{HTML}{3B6EC6}
  \providecolor{ihRed}{HTML}{C0392B}
  % geometry (cm): box size, one level horizontally, one step within a layer,
  % height of the hook layer
  \newcommand{\vtBx}{0.7}
  \newcommand{\vtLx}{5.4}
  \newcommand{\vtLy}{2.0}
  \newcommand{\vtOy}{13}
  % internal box walls of a row of \vtk boxes in node #1
  \newcommand{\vtWalls}[1]{%
    \ifnum\vtk>1
      \pgfmathtruncatemacro{\vtkm}{\vtk-1}%
      \foreach \i in {1,...,\vtkm}
        \draw[line width=1.5pt]
          ([xshift={\i*\vtBx*1cm}]#1.south west) -- ([xshift={\i*\vtBx*1cm}]#1.north west);%
    \fi}
  % one-row diagram [k], k = 1+a-b boxes, at level a+b  (k = 0: a dot)
  \newcommand{\vtRow}[3]{%
    \pgfmathtruncatemacro{\vtk}{1+#2-#3}%
    \ifnum\vtk=0
      \node[vtdot] (#1) at ({(#2+#3)*\vtLx},{(#2-#3)*\vtLy}) {};%
    \else
      \node[vtbox, minimum width={\vtk*\vtBx*1cm}] (#1)
        at ({(#2+#3)*\vtLx},{(#2-#3)*\vtLy}) {};%
      \vtWalls{#1}%
    \fi}
  % hook [k,1], k = 2+a-b, at level a+b;  node #1 = first row, #1-h = second-row box
  \newcommand{\vtHook}[3]{%
    \pgfmathtruncatemacro{\vtk}{2+#2-#3}%
    \node[vtbox, minimum width={\vtk*\vtBx*1cm}] (#1)
      at ({(#2+#3)*\vtLx},{(#2-#3)*\vtLy+\vtOy}) {};%
    \vtWalls{#1}%
    \node[vtbox, minimum width={\vtBx*1cm}, anchor=north west] (#1-h) at (#1.south west) {};}
  \resizebox{0.85\linewidth}{!}{%
  \begin{tikzpicture}[
      vtbox/.style={draw, rectangle, line width=1.5pt, fill=white, inner sep=0pt,
                    outer sep=0pt, minimum height={\vtBx*1cm}},
      vtdot/.style={circle, fill=black, inner sep=0pt, outer sep=0pt, minimum size=0.34cm},
      vtarrow/.style={-{Stealth[length=10pt,width=7.5pt]}, line width=1.4pt,
                      shorten <=4pt, shorten >=5pt},
      vtlower/.style={vtarrow, ihBlue},
      vtupper/.style={vtarrow, ihRed},
      vtbetween/.style={vtarrow, black!45},
      vtstub/.style={dash pattern=on 5pt off 3pt, shorten <=0pt},
      vtplane/.style={draw=black!15, line width=0.8pt, rounded corners=14pt},
      vtstep/.style={font=\Huge, text=black!60, scale=1.3, transform shape},
    ]
    % ---- diagrams: B<a><b> one-row (lower layer), T<a><b> hooks (upper layer) ----
    % level = a+b;  rows have 1+a-b boxes, hooks have 2+a-b boxes in the first row
    \foreach \a/\b in {0/0,1/0,2/0,3/0, 0/1,1/1,2/1,3/1, 1/2,2/2,3/2, 2/3,3/3}
      {\vtRow{B\a\b}{\a}{\b}}
    \foreach \a/\b in {0/1,1/1,2/1,3/1, 1/2,2/2,3/2, 2/3,3/3}
      {\vtHook{T\a\b}{\a}{\b}}

    % ---- shaded planes behind each layer ----
    \begin{scope}[on background layer]
      \path[vtplane, fill=ihBlue!5]
        (-1.10,-1.10) -- (4.48,-2.92) -- (27.92,-2.92) -- (33.50,-1.10) -- (33.50,1.10)
        -- (18.35,7.10) -- (14.05,7.10) -- (3.95,3.10) -- (-1.10,1.10) -- cycle;
      \path[vtplane, fill=ihRed!5]
        (4.30,9.20) -- (28.10,9.20) -- (33.15,11.20) -- (33.85,11.90) -- (33.85,14.10)
        -- (23.75,18.10) -- (19.45,18.10) -- (4.30,12.10) -- cycle;
    \end{scope}

    % ---- arrows: X -> Y means [P_mu, X] contains Y (level n+1 -> level n) ----
    % lower layer: P removes a box ...
    \foreach \u/\v in {B10/B00, B20/B10, B30/B20,
                       B11/B01, B21/B11, B31/B21,
                       B22/B12, B32/B22,
                       B33/B23}
      \draw[vtlower] (\u) -- (\v);
    % ... or adds a box
    \foreach \u/\v in {B01/B00, B11/B10, B21/B20, B31/B30,
                       B12/B11, B22/B21, B32/B31,
                       B23/B22, B33/B32}
      \draw[vtlower] (\u) -- (\v);
    % upper layer: P removes a box from the first row ...
    \foreach \u/\v in {T11/T01, T21/T11, T31/T21,
                       T22/T12, T32/T22,
                       T33/T23}
      \draw[vtupper] (\u-h) -- (\v);
    % ... or adds one
    \foreach \u/\v in {T12/T11, T22/T21, T32/T31,
                       T23/T22, T33/T32}
      \draw[vtupper] (\u) -- (\v);
    % between the layers: hook [k,1] -> row [k] (P removes the second-row box) ...
    \foreach \u/\v in {T01/B00, T11/B10, T21/B20, T31/B30,
                       T12/B11, T22/B21, T32/B31,
                       T23/B22, T33/B32}
      \draw[vtbetween] (\u-h) -- (\v);
    % ... and row [k] -> hook [k,1] (P adds a second-row box)
    \foreach \u/\v in {B11/T01, B21/T11, B31/T21,
                       B22/T12, B32/T22,
                       B33/T23}
      \draw[vtbetween] (\u) -- (\v-h);

    % ---- the tree continues: arrows coming in from generators beyond the window ----
    \foreach \a/\b in {4/0,4/1,4/2,4/3,3/4}
      \coordinate (gB\a\b) at ({(\a+\b)*\vtLx},{(\a-\b)*\vtLy});
    \foreach \a/\b in {4/1,4/2,4/3,3/4}
      \coordinate (gT\a\b) at ({(\a+\b)*\vtLx},{(\a-\b)*\vtLy+\vtOy});
    \foreach \b in {0,...,3}
      \draw[vtlower, vtstub] ($(B3\b)!0.5!(gB4\b)$) -- (B3\b);
    \draw[vtlower, vtstub] ($(B33)!0.5!(gB34)$) -- (B33);
    \draw[vtbetween, vtstub] ($(B33)!0.16!(gT34)$) -- (B33);
    \foreach \b in {1,2,3} {
      \draw[vtupper, vtstub] ($(T3\b)!0.5!(gT4\b)$) -- (T3\b);
      \draw[vtbetween, vtstub] ($(T3\b-h)!0.16!(gB4\b)$) -- (T3\b-h);
    }
    \draw[vtupper, vtstub] ($(T33)!0.5!(gT34)$) -- (T33);

    % ---- level numbers under the columns, and the starting field ----
    \node[vtstep, anchor=east] at (-1.4,-4.4) {Level $n$};
    \foreach \n in {0,...,6}
      \node[vtstep] at ({\n*\vtLx},-4.4) {$\n$};
    \node[font=\Huge, scale=1.3, transform shape, anchor=east] at ([xshift=-10pt]B00.west) {$A_\mu$};
  \end{tikzpicture}%
  }
  \caption{Vector inverse Higgs tree written in terms of Lorentz irreducible
    representations.}
  \label{fig:ih-vectortree}
\end{figure}
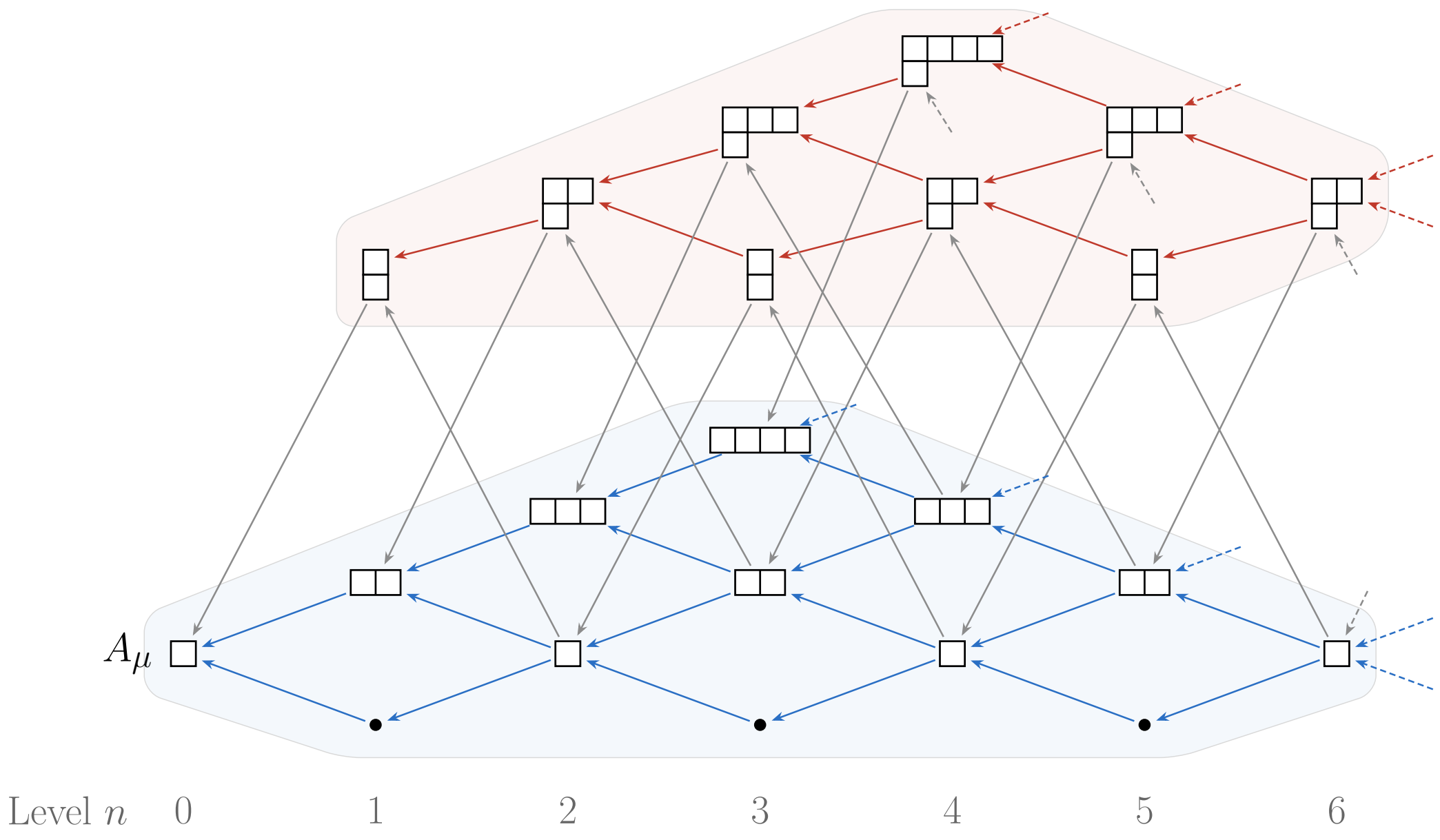

\paragraph{Vector Inverse Higgs Tree:} For a vector $A_{\mu}$, any non-linearly realized symmetry transformation takes the form
\begin{equation}
\delta_n A_\mu
=
d_{\mu,\nu_1\cdots\nu_n}
x^{\nu_1}\cdots x^{\nu_n}
+\cdots \,,
\end{equation}
where $d_{\mu,\nu_1\cdots\nu_n}$ is symmetric in the indices
$\nu_1,\ldots,\nu_n$. Unlike for the scalar case, here there are two distinct branches at each order: either $d_{\mu,\nu_1\cdots\nu_n}$ is completely symmetric, or it is of mixed symmetry. Since we want this vector Goldstone to be a gauge field, we need to identify the fully-symmetric branch with the tower of polynomial $U(1)$ gauge transformations
\begin{align}
\delta A_\mu=\partial_\mu\lambda(x) \,,
\end{align}
where the gauge parameter admits the Taylor expansion
\begin{align}
\lambda(x)
=
\sum_{l=0}^{\infty}
\lambda_{\mu_1\cdots\mu_l}
x^{\mu_1}\cdots x^{\mu_l} \,,
\end{align}
with constant, symmetric coefficients $\lambda_{\mu_1\cdots\mu_l}$. Each polynomial appearing in the expansion of $\lambda(x)$ defines a corresponding symmetric gauge generator $B_{\mu_1\cdots\mu_l}$ which must be included in the vector inverse Higgs tree. The actions of these generators are
\begin{align}
\delta_{(B_\mu)}A_\nu
&=
\lambda_\mu\delta^\mu_{\nu} \,,
\\
\delta_{(B_{\mu_1\mu_2})}A_\nu
&=
2\lambda_{\mu_1\mu_2}\,
\delta_\nu^{(\mu_1}x^{\mu_2)} \,,
\\
&\vdots \\
\delta_{(B_{\mu_1\cdots\mu_l})}A_\nu
&=
l\,\lambda_{\mu_1\cdots\mu_l}
\delta_\nu^{(\mu_1}
x^{\mu_2}\cdots x^{\mu_l)} \,, \\
&\vdots\,
\end{align}
The commutators of these gauge generators with Lorentz are fixed by their index structure, while the commutators with translations take the schematic form $[P,B_l] = B_{l-1}$ thereby ensuring we have the appropriate IHCs to reduce us to a gauge vector only \cite{Goon:2014ika}. The commutators with Poincar\'{e} are therefore completely fixed by the structure of the $U(1)$ gauge symmetry. This branch of the inverse Higgs tree is shown in the lower part of Figure \ref{fig:ih-vectortree}.

The second branch of the inverse Higgs tree corresponds to new non-linearly realized symmetries that are not gauge transformations and consists of generators with mixed symmetry,
$M_{[\mu,\nu_1]\nu_2\cdots\nu_n}$. At level one of this branch the only possible generator is a two-form acting on the gauge vector as
\begin{align}
\delta A_\mu
=
m_{\mu\nu}x^\nu+\cdots \,,
\end{align}
where $m_{\mu\nu}$ is antisymmetric and the dots denote possible field-dependent terms. At higher levels, there are similar structures involving mixed symmetry tensors as shown in the upper part of Figure \ref{fig:ih-vectortree}. Only the traceless ones are consistent with the Maxwell kinetic term \cite{Klein:2018ylk}. In this work, we restrict ourselves to this first level two-form generator, $M_{\mu\nu}$. Its commutator with Lorentz is
\begin{align}
[J_{\mu\nu},M_{\rho\sigma}]
&=
\eta_{\mu\rho}M_{\nu\sigma}
-
\eta_{\nu\rho}M_{\mu\sigma}
-
\eta_{\mu\sigma}M_{\nu\rho}
+
\eta_{\nu\sigma}M_{\mu\rho} \,,
\end{align}
while the inverse Higgs tree dictates its commutator with translations to be 
\begin{align} \label{MInverseHiggs}
[M_{\mu\nu},P_\rho]
=
\left(
\eta_{\mu\rho}B_\nu
-
\eta_{\nu\rho}B_\mu
\right) + b_1 \left(
\eta_{\mu\rho}P_\nu
-
\eta_{\nu\rho}P_\mu
\right)   \,.
\end{align}
Here we already see a connection between the two branches: since $B_{\mu}$ is the generator corresponding to the essential Goldstone $A_{\mu}$, all other generators on each branch of the tree must eventually reach $B_{\mu}$ via translations, as shown in Figure \ref{fig:ih-vectortree}. 

\paragraph{Summary of the set-up:} To summarize, in the scalar tree we include $C, C_{\mu}$ and $C_{\mu\nu}$, while in the vector tree we include the full tower of generators of the $U(1)$ gauge symmetry denoted by $B_{\mu_1 \ldots \mu_l}$ and an additional two-form generator $M_{\mu\nu}$ that generates a non-gauge, non-linearly realized symmetry. The commutators with Lorentz are fixed by the index structure of each generator while commutators with translations are fixed by the inverse Higgs trees. For the remaining commutators between the new generators, we allow for all possible structures and then fix them using Jacobi identities.  In the following section we implement this procedure and find the possible algebras in the various cases of interest.

\section{Level 1: Scalar-Vector Galileon Theory} \label{sec:LevelOne}
We begin by truncating the scalar tree at level one i.e. we don't include the generator $C_{\mu\nu}$.  
\subsection{Commutators and Jacobi Identities}
Having fixed the commutators required by the inverse Higgs trees together with Lorentz invariance, the remaining structure of the algebra is determined by the commutators among the non-linear generators. We assume that the gauge generators form an ideal of the full symmetry algebra, meaning that a commutator between a gauge generator and any other generator produces a gauge generator. In practice, including the full tower of gauge symmetries from the outset is difficult, so our approach will be to include only the leading generator $B_{\mu}$, which generates a constant shift on $A_\mu$ and is associated with it in the coset, and to check that our resulting algebras can be enlarged by the addition of all other gauge generators. The most general ansatz for the commutators between $B_{\mu}$ and non-Poincar\'{e} generators is therefore
\begin{align}
[B_\mu,C]
&=
g_1B_\mu \,,
\\
[C_\mu,B_\nu]
&=
0 \,,
\\
[M_{\mu\nu},B_\rho]
&=
g_2
\left(
\eta_{\mu\rho}B_\nu
-
\eta_{\nu\rho}B_\mu
\right) \,, \\
[B_{\mu},B_{\nu}] &= 0 \,.
\end{align}
The remaining commutators involving the non-linear generators are
\begin{align}
[C_\mu,C]
&= a_4 C_\mu + a_5 P_\mu + a_6 B_{\mu} \,,
\\
[C_\mu,C_\nu]
&= a_7 J_{\mu\nu} + a_8 M_{\mu\nu} \,,
\\
[M_{\mu\nu},C]
&= b_2 M_{\mu\nu} + b_3 J_{\mu\nu} \,,
\\
[M_{\mu\nu},C_\rho]
&= b_4 (\eta_{\mu\rho}C_\nu-\eta_{\nu\rho}C_\mu)
 + b_5 (\eta_{\mu\rho}P_\nu-\eta_{\nu\rho}P_\mu)
 + b_6 (\eta_{\mu\rho}B_\nu-\eta_{\nu\rho}B_\mu) \,,
\\
[M_{\mu\nu},M_{\rho\sigma}]
&= b_7 (\eta_{\mu\rho}M_{\nu\sigma}
 -\eta_{\nu\rho}M_{\mu\sigma}
 -\eta_{\mu\sigma}M_{\nu\rho}
 +\eta_{\nu\sigma}M_{\mu\rho}) \nonumber \\
&~~~~~~~~~~~~ + b_8 (\eta_{\mu\rho}J_{\nu\sigma}
 -\eta_{\nu\rho}J_{\mu\sigma}
 -\eta_{\mu\sigma}J_{\nu\rho}
 +\eta_{\nu\sigma}J_{\mu\rho})\,.
\end{align}
We now impose the Jacobi identities on these commutators starting with the $(PPG)$ identities then moving to $(PGG)$ and finally $(GGG)$. We note that it is possible to set some of these parameters to zero without loss of generality using basis changes. We will do this after imposing the Jacobi identities. 

The only non-trivial case at $(PPG)$ is $(P_{\mu}, P_{\nu},C_{\rho})$, yielding $a_1=a_2$. (This condition is familiar from the conformal algebra that can be realized on the dilaton.) Moving to $(PGG)$, the non-trivial constraints that arise are 

\begin{align}
(P_\mu,C,C_\nu):\qquad
& a_1+a_4=0 \,,
\\
(P_\rho,C,M_{\mu\nu}):\qquad
& a_1+b_2-g_1=0,
\qquad b_1b_2+b_3=0 \,,
\\
(P_\rho,C_\mu,C_\nu):\qquad
& a_2+a_4=0,
\qquad a_5-a_7-b_1a_8=0,
\qquad a_6-a_8=0 \,,
\\
(P_\mu,C_\nu,B_\rho):\qquad
& a_2=0,
\qquad g_1=0 \,,
\\
(P_\rho,C_\sigma,M_{\mu\nu}):\qquad
& a_2=0,
\qquad b_2=b_3=0,
\qquad b_4-b_1=0 \,,
\\
(P_\lambda,M_{\mu\nu},M_{\rho\sigma}):\qquad
& b_1+g_2-b_7=0,
\qquad b_1^2-b_1b_7-b_8=0 \,.
\end{align}
The remaining identities of this class,
\[
(P_\mu,C,C),\qquad
(P_\mu,C,B_\nu),\qquad
(P_\mu,B_\nu,B_\rho),\qquad
(P_\rho,B_\sigma,M_{\mu\nu}) \,,
\]
are satisfied identically. Note that these identities have fixed $a_1=a_2=0$, which rules out the possibility of the scalar being a dilaton. This is ultimately due to the vanishing of the $[C_{\mu},B_{\nu}]$ commutator. However, we have checked that including the next generator in the gauge part of the vector inverse Higgs tree i.e. $B_{\mu\nu}$ does not alter this conclusion. Combining these constraints gives
\begin{align}
a_1=a_2=a_4= b_2 = b_3 = g_1 &= 0 \,, \\ 
b_4 - b_1 &= 0 \,, \\
g_2 + b_1 - b_7 & = 0 \,, \\
b_8- b_1(b_1-b_7) & = 0 \,, \\
a_5- a_7 - b_1a_8 & = 0 \,, \\
a_6- a_8 & = 0 \,.
\end{align}
Moving to the $(GGG)$ identities, we have (without yet imposing any of the previous conditions)
\begin{align}
(C,C_\mu,C_\nu):\qquad
& 2a_2a_5+2a_4a_7-a_8b_3=0,
\qquad a_8(2a_4-b_2)=0 \,,
\\
(C,C_\rho,M_{\mu\nu}):\qquad
& (a_1-a_4-b_2)b_5+(b_4-b_1)a_5=0 
\nonumber\\
& (a_4+b_2-g_1)b_6+a_5+(g_2-b_4)a_6=0 \,,
\nonumber\\
& b_2b_4+b_3=0 \,,
\\
(C,B_\rho,M_{\mu\nu}):\qquad
& b_2g_2+b_3=0 \,,
\\
(C,M_{\mu\nu},M_{\rho\sigma}):\qquad
& b_2b_7+2b_3=0,
\qquad 2b_2b_8-b_3b_7=0 \,,
\\
(C_\mu,C_\nu,B_\rho):\qquad
& a_7+a_8g_2=0 \,,
\\
(C_\rho,C_\sigma,M_{\mu\nu}):\qquad
& a_2b_5+a_7b_4-a_8b_8=0,
\nonumber\\
& a_7+a_8(b_7-b_4)=0,
\qquad b_5=0 \,,
\\
(C_\lambda,M_{\mu\nu},M_{\rho\sigma}):\qquad
& (b_1+b_4-b_7)b_5=0,
\nonumber\\
& b_4^2-b_4b_7-b_8=0,
\nonumber\\
& (b_4-b_7+g_2)b_6+b_5=0 \,,
\\
(B_\lambda,M_{\mu\nu},M_{\rho\sigma}):\qquad
& g_2^2-b_7g_2-b_8=0 \,.
\end{align}
All other $(GGG)$ identities
are satisfied identically.
Upon substituting the constraints from  $PPG$ and $PGG$, the $GGG$ identities supply only two additional independent constraints,
\begin{align}
  a_7+a_8g_2=0,\qquad b_5=0 \,.  
\end{align}
The complete set of constraints can then be expressed as 
\[
\boxed{
\begin{gathered}
a_1=a_2=a_4=0,\qquad
b_2=b_3=b_5=0,\qquad g_1=0,\\
b_4=b_1,\qquad
g_2=b_7-b_1,\qquad
b_8=b_1(b_1-b_7),\\
a_6=a_8,\qquad
a_7=a_8(b_1-b_7),\qquad
a_5=a_8(2b_1-b_7).
\end{gathered}}
\]
The coefficients $a_8$, $b_1$, $b_6$, and $b_7$ remain unconstrained by the Jacobi identities. 

As we alluded to earlier, we haven't yet utilised our freedom to make basis changes to set some parameters to zero. If we go back to our general ansatz for the commutators, we can define 
\begin{align}
\widetilde{M}_{\mu\nu} = M_{\mu\nu} - b_1 J_{\mu \nu} \,,
\end{align}
to eliminate $P_{\mu}$ from the RHS of \eqref{MInverseHiggs}. Since every other commutator involving $M_{\mu\nu}$ is general and therefore unaffected by this basis change, we can fix $b_1 = 0$ without loss of generality from the outset. This reduces the non-zero commutators to 
\begin{align}
[P_{\mu},C_{\nu}] &= \eta_{\mu\nu} C \,, \\
[M_{\mu\nu},P_\rho] &=
\eta_{\mu\rho}B_\nu-\eta_{\nu\rho}B_\mu  \,, \\
[M_{\mu\nu},B_\rho] &= b_7 \left( \eta_{\mu\rho}B_\nu - \eta_{\nu\rho}B_\mu\right)  \,, \\
[C_{\mu},C] &= -a_8(b_7 P_{\mu} - B_{\mu}) \,,\\
[C_{\mu},C_{\nu}] &= -a_8(b_7 J_{\mu\nu} - M_{\mu\nu}) \,, \\
[M_{\mu\nu},C_\rho] &=
b_6(\eta_{\mu\rho}B_\nu-\eta_{\nu\rho}B_\mu)  \,, \\
[M_{\mu\nu},M_{\rho\sigma}]
&= b_7 (\eta_{\mu\rho}M_{\nu\sigma}
 -\eta_{\nu\rho}M_{\mu\sigma}
 -\eta_{\mu\sigma}M_{\nu\rho}
 +\eta_{\nu\sigma}M_{\mu\rho}) \,.
\end{align}
We therefore have a three-parameter family of algebras, but we don't expect them all to be necessarily realized on a single scalar and a gauge vector, since compatibility with the remaining tower of gauge symmetries is non-trivial. To make further progress, we now assume that the generators in the scalar tree i.e. $C$ and $C_{\mu}$, commute with the gauge generators. This is what we have found so far and seems a reasonable assumption given that the scalar does not transform under the gauge symmetry. If we then add $B_{\mu\nu}$ to the algebra, the condition $[B_{\mu\nu},P_{\rho}] \neq 0$ combined with the $(C, C_{\mu},B_{\rho \sigma})$ Jacobi identity requires $a_8 b_7 = 0$. If $a_8 = 0$ and $b_7 \neq 0$, we can then perform the basis change 
\begin{align}
\widetilde C_\mu
&=C_\mu-\frac{b_6}{b_7}B_\mu \,,
\end{align}
to fix 
\begin{align} 
[M_{\mu\nu},\widetilde C_\rho]
=0 \,.
\end{align}
This decouples the scalar tree from the vector tree and the non-abelian nature of the vector algebra is ruled out by the no-go theorem of \cite{Klein:2018ylk} (i.e. adding the remaining gauge symmetries fixes $b_7=0$).\footnote{The absence of special gauge theories is also seen in amplitudes where no enhanced Adler zeros are possible \cite{Elvang:2018dco}.} We therefore require $b_7 = 0$, and the non-zero commutators are 
\begin{align}
[P_{\mu},C_{\nu}] &= \eta_{\mu\nu} C \,, \\
[M_{\mu\nu},P_\rho] &=
\eta_{\mu\rho}B_\nu-\eta_{\nu\rho}B_\mu  \,, \\
[C_{\mu},C] &= a_8 B_{\mu} \,,\\
[C_{\mu},C_{\nu}] &= a_8 M_{\mu\nu} \,, \label{CCcommutator} \\
[M_{\mu\nu},C_\rho] &=
b_6(\eta_{\mu\rho}B_\nu-\eta_{\nu\rho}B_\mu)  \,.
\end{align}

At this stage it is clear that this algebra is consistent with the full tower of gauge symmetries as long as those generators commute with $C,C_{\mu}$ and $M_{\mu\nu}$, and since we have all necessary inverse Higgs constraints, this algebra can be realized on a single scalar and a single gauge vector. However, it is not guaranteed that the realization is compatible with canonical kinetic terms: when we constructed the scalar tree, we only included generators for which the corresponding symmetries are compatible with $(\partial \phi)^2$, and similarly for the vector tree, we only included those compatible with $F_{\mu\nu}^2$, but our approach does not guarantee that a transformation of the scalar/vector under a generator in the vector/scalar tree is compatible with canonical kinetic terms. It is clear from \eqref{CCcommutator} that when $a_8 \neq 0$, the vector will transform non-trivially under the action of $C_{\mu}$ (since by construction it transforms non-trivially under the action of $M_{\mu\nu}$) and we need to check that this transformation is compatible with the standard $F_{\mu\nu}^2$ kinetic term of the vector. To do this, we parametrize the coset element as
\begin{align}
    \gamma = e^{x^{\mu}P_{\mu}}e^{\phi C} e^{\xi^{\mu} C_{\mu}}e^{A^{\mu}B_{\mu}}e^{-{1\over 2}G^{\mu\nu}M_{\mu\nu}}\,.
\end{align}
We don't include the tower of gauge symmetries and their presence will not change the transformation $\delta_{(C_{\mu})}A_{\nu}$. $\phi$ and $A_{\mu}$ are the essential Goldstones that are going to be required to realize the symmetry non-linearly, while $\xi_{\mu}$ and $G_{\mu\nu}$ are inessential Goldstones that are not required to realize the symmetry and will ultimately be removed by inverse Higgs constraints in favor of the essential ones. The inessential Goldstone $G_{\mu\nu}$ is a two-form and therefore satisfies $G_{\mu\nu} = - G_{\nu\mu}$. We now compute $ \tilde{\gamma} = e^{c^{\mu}C_{\mu}}  \gamma$ and write it in the same way as $\gamma$ to extract the various transformation rules. At linear order in $c_{\mu}$ and to leading order in fields we find 
\begin{align}
 \delta_{(C_{\mu})}A_{\nu} = a_8 c_{\nu} \phi + \mathcal{O}(\phi^2) \,.   
\end{align} 
It is then simple to check that this symmetry transformation is incompatible with the Maxwell kinetic term since
\begin{align}
\delta_{(C_{\mu})}(F_{\mu\nu}^2) = 2 a_8 F^{\mu\nu}(c_{\nu} \partial_{\mu} \phi - c_{\mu}\partial_{\nu} \phi) \,,
\end{align}
which is not a total derivative. Given that $\phi$ also transforms under this symmetry, this unwanted term could in principle be cancelled by a three-derivative interaction $A \phi \phi$, however there are no such interactions that are gauge invariant since both $F^{\mu\nu} \phi \partial_{\mu}\partial_{\nu}\phi$ and $F^{\mu\nu}\partial_{\mu}\phi \partial_{\nu} \phi$ vanish by symmetry.\footnote{We note that higher-derivative possibilities exist: the variation of the kinetic term $F^{\mu\nu} \Box F_{\mu\nu}$ could collaborate with $F^{\mu\nu} \partial_{\mu} \Box \phi \partial_{\nu} \phi$. Although this theory is clearly sick when expanded around a Poincar\'{e}-invariant background, ghost-free fluctuations might be possible on symmetry-breaking backgrounds.}

Having canonical and diagonalised kinetic terms therefore requires us to fix $a_8 = 0$, leaving us with 
\begin{align}
[P_{\mu},C_{\nu}] &= \eta_{\mu\nu} C \,, \\
[M_{\mu\nu},P_\rho] &=
\eta_{\mu\rho}B_\nu-\eta_{\nu\rho}B_\mu  \,, \\
[M_{\mu\nu},C_\rho] &=
b_6(\eta_{\mu\rho}B_\nu-\eta_{\nu\rho}B_\mu)  \,,
\end{align}
as the non-zero commutators. The scalar sector algebra is therefore that of the Galileon, with dilaton and DBI possibilities ruled out. 

This algebra arises in the decoupling limit of massive gravity with the $b_6=0$ case arising from the decoupling limit of pseudo-linear massive gravity while the $b_6 \neq 0$ case comes from the decoupling limit of dRGT \cite{Bonifacio:2019hrj}. In the massive gravity decoupling limit, this two-form symmetry arises from the Lorentz part of the St\"uckelberg symmetry \cite{Bonifacio:2019hrj}. It is interesting that the only non-trivial algebra we have found in this case has a massive gravity realization. For the remainder of this section, we study this algebra in more detail by deriving the symmetry transformations and leading interactions using the coset construction.

\subsection{Coset Construction}
For reasons that will become clear soon, we parametrize the commutators by 
\begin{align}\label{MGSVTcommutators1}
[P_{\mu}, C_{\nu}] = \eta_{\mu \nu} C, \qquad [P_{\mu}, M_{\rho \sigma}] = \frac{1}{2}\left(\eta_{\mu \rho} B_{\sigma}  - \eta_{\mu \sigma} B_{\rho} \right), \qquad [C_{\mu}, M_{\rho \sigma}] =- \frac{\alpha}{2} \left( \eta_{\mu \rho} B_{\sigma}  - \eta_{\mu \sigma} B_{\rho} \right) \,,
\end{align}
with a single constant $\alpha$.  When $\alpha \neq 0$, the final commutator ties together the scalar and the vector, so (with an abuse of terminology) we call the $\alpha\not=0$ case ``non-abelian'' and the $\alpha=0$ case ``abelian.'' 
\subsubsection{Maurer-Cartan Form, Inverse Higgs Solutions and Symmetry Transformations} \label{sec:CosetSymmetries}
In going through the coset construction, we work with these generators only and check consistency with the remaining gauge symmetries as we go. There are many choices for how we parametrize the coset element and we choose to work with 
\begin{align} \label{CosetElement}
    \gamma = e^{x^{\mu}P_{\mu}}e^{\phi C} e^{\xi^{\mu} C_{\mu}}e^{A^{\mu}B_{\mu}}e^{-\frac{1}{2} G^{\mu\nu}M_{\mu\nu}}\,,
\end{align}
as we did earlier since, as we will see, it leads to a compact form for the symmetry transformations. Invariant actions follow from the Maurer-Cartan form, which in this case can be computed straightforwardly,
\begin{align}
\gamma^{-1} d \gamma = \omega^{\mu}_{P} P_{\mu}   + \omega_{C} C + \omega^{\mu}_{C} C_{\mu} + \omega^{\mu}_{B} B_{\mu} + \omega^{\mu\nu}_{M} M_{\mu\nu} \,, 
\end{align}
where 
\begin{align}
\omega^{\mu}_{P} &= dx^{\mu} \,, \label{OneForm1} \\
\omega_{C} & = d \phi + \xi_{\mu} dx^{\mu} \,, \label{OneForm2} \\
\omega^{\mu}_{C} &= d \xi^{\mu} \,, \label{OneForm3} \\
\omega^{\mu}_{B} &= d A^{\mu} +{1\over 2} G^{\mu\nu} dx_{\nu} - {1\over 2} \alpha G^{\mu\nu} d \xi_{\nu} \,, \label{OneForm4} \\
\omega^{\mu\nu}_{M} &= -{1\over 2} d G^{\mu\nu} \label{OneForm5} \,.
\end{align}
Note that the expression for $\omega^{\mu}_{B}$ contains linear and quadratic terms and therefore we expect for $\alpha \neq 0$ the symmetry will relate operators with different powers of the fields. Setting $\omega_{C} = 0$ yields the familiar inverse Higgs constraint of the Galileon \cite{Goon:2012dy} which can be solved to give 
\begin{align} \label{GalileonIHC}
    \xi_{\mu} = - \partial_{\mu} \phi \,.
\end{align}
To eliminate $G_{\mu\nu}$, we take $\omega^{\mu}_{B}$, pull back to spacetime, and set the anti-symmetric part of the resulting object to zero. This yields the constraint
\begin{align}
-F^{\mu\nu} +G^{\mu\nu} + \alpha G^{\rho [\mu}\partial^{\nu]}\xi_{\rho} = 0 \,,
\end{align}
where $F_{\mu\nu} = \partial_{\mu}A_{\nu} - \partial_{\nu}A_{\mu}$ is the standard Maxwell field strength and we anti-symmetrize with weight one i.e. $A_{[\mu\nu]} = \frac{1}{2}(A_{\mu\nu} - A_{\nu\mu})$. The symmetric part of the pullback of $\omega^{\mu}_{B}$ would be set to zero if we had included the next generator in the tower of gauge symmetries, see e.g. \cite{Klein:2018ylk,Goon:2014paa}. If we impose \eqref{GalileonIHC}, we have
\begin{align} \label{VectorIHC}
-F^{\mu\nu} +  G^{\mu\nu} - \alpha G^{\rho [\mu}\Pi^{\nu]}{}_{\rho} = 0 \,,
\end{align}
where we have defined $\Pi_{\mu\nu} \equiv \partial_{\mu}\partial_{\nu} \phi$. The solution to this equation yields an expression for $G_{\mu\nu}$ in terms of $F_{\mu\nu}$ and $\Pi_{\mu\nu}$, and will therefore be invariant under the full tower of gauge symmetries even though we only included the leading one (the constant shift on $A_{\mu}$). Working perturbatively in $\alpha$ (or in $\Pi$), the solution is
\begin{align} \label{PertGSolution}
    G^{\mu\nu} = F^{\mu\nu} +\alpha F^{\rho [\mu} \Pi^{\nu]}{}_{\rho} + \mathcal{O}(\alpha^2) \,.
\end{align}
This can be extended to all orders in $\alpha$ and written as follows using matrix notation, following a similar construction in \cite{Ondo:2013wka}:
\begin{align} \label{fullGSolution}
    G= \int_0^\infty du \, e^{-u} e^{-{u\alpha\over 2}  \Pi}F e^{-{u\alpha\over 2}\Pi}=\sum_{n=0}^\infty  \left(-{\alpha\over 2}\right)^n  \sum_{m=0}^n \left(\begin{array}{c}n \\m\end{array}\right)  \Pi^{n-m} F \Pi^{m}\,. 
\end{align}
These inverse Higgs solutions can be used to reduce any invariant action to one using $\phi$ and $A_{\mu}$ only. 

The coset construction also provides us with a method of deriving the symmetry transformations of the Goldstones, as we saw earlier. We act on the coset element \eqref{CosetElement} from the left with $e^{gG}$ where $g$ is the symmetry parameter and $G$ is the generator we are interested in, and write the resulting product in the same form as the original coset element. For the generator $C$ we have 
\begin{align}
    e^{cC} \gamma = e^{x^{\mu}P_{\mu}}e^{(\phi+c) C} e^{\xi^{\mu} C_{\mu}}e^{A^{\mu}B_{\mu}}e^{-{1\over 2} G^{\mu\nu}M_{\mu\nu}} \,,
\end{align}
thereby yielding the symmetry transformations 
\begin{align}
\delta_{C} \phi = c, \qquad \delta_{C}\xi_{\mu} = \delta_{C} A_{\mu} = \delta_{C}G_{\mu\nu} = 0 \,.
\end{align}
This result is particularly simple because the generator $C$ commutes with all other generators. A little more work is required for $C_{\mu}$: 
\begin{align}
    e^{c^{\mu}C_{\mu}} \gamma = e^{x^{\mu}P_{\mu}}e^{(\phi - c_{\mu}x^{\mu}) C} e^{(\xi^{\mu}+c^{\mu}) C_{\mu}}e^{A^{\mu}B_{\mu}}e^{-{1\over 2}G^{\mu\nu}M_{\mu\nu}} \,,
\end{align}
thereby yielding
\begin{align}
\delta_{(C_\mu)} \phi = -c_{\mu}x^{\mu}, \qquad \delta_{(C_\mu)}\xi_{\mu} = c_{\mu}, \qquad \delta_{(C_{\mu})} A_{\mu} = \delta_{(C_\mu)}G_{\mu\nu} = 0 \,.
\end{align}
In arriving at this result we used
\begin{align}
    e^{c^{\mu}C_{\mu}} e^{x^{\mu}P_{\mu}} =  e^{x^{\mu}P_{\mu}}e^{c^{\mu}C_{\mu}} e^{- c^{\mu}x_{\mu} C} \,,
\end{align}
which follows from the BCH formula and $[P_{\mu}, C_{\nu}] = \eta_{\mu\nu} C$. As a consistency check, we see that the transformations of $\phi$ and $\xi_{\mu}$ are consistent with the inverse Higgs solution \eqref{GalileonIHC}. For $B_{\mu}$ the process is again straightforward because this generator commutes with all other generators. The symmetry transformations are 
\begin{align}
\delta_{(B_{\mu})} \phi = \delta_{(B_{\mu})}\xi_{\mu} = 0, \qquad \delta_{(B_{\mu})}A_{\mu} = b_{\mu}, \qquad \delta_{(B_{\mu})}G_{\mu\nu} = 0 \,.
\end{align}
Finally, we have the symmetry generated by $M_{\mu\nu}$. Again using the BCH formula and the commutators \eqref{MGSVTcommutators1} we find
\begin{align} \label{MSymmetries}
\delta_{(M_{\mu\nu})} \phi = \delta_{(M_{\mu\nu})}\xi_{\mu} = 0, \qquad \delta_{(M_{\mu\nu})}A_{\mu} = m_{\mu\nu} (x^{\nu} + \alpha \partial^{\nu}\phi), \qquad \delta_{(M_{\mu\nu})}G_{\mu\nu} = -2m_{\mu\nu} \,,
\end{align}
where in the expression for $\delta_{(M_{\mu\nu})}A_{\mu}$ we used the inverse Higgs solution \eqref{GalileonIHC}. Using 
\begin{align}
\delta F_{\mu\nu} = -2 m_{\mu\nu} + 2 \alpha m_{\rho [\mu}\Pi_{\nu]}{}^{\rho} \,,
\end{align}
it is straightforward to check that these symmetry transformations are consistent with the inverse Higgs constraint \eqref{VectorIHC}. Despite the somewhat complicated inverse Higgs solution for $G_{\mu\nu}$, its transformation rule in \eqref{MSymmetries} is very simple, matching that of $F_{\mu\nu}$ in the abelian case.  

Note that in deriving each of these symmetry transformations, we did not need to assume that the symmetry parameters were infinitesimal i.e. these are the finite forms of the transformations. This is a consequence of the commutators \eqref{MGSVTcommutators1}, since any generator appearing on the RHS of these equations commutes with all other generators. This means any BCH expressions truncate at the first non-trivial order.

\subsubsection{A Comment on Including All Gauge Symmetries}

We have not included the full tower of gauge symmetries in the coset element since, as we shall see, it is more straightforward to impose the gauge symmetries by hand. However, for this approach to work, it is important to check that the one-forms and global symmetry transformations we have just derived are unchanged if we did include the gauge symmetries, and that the gauge symmetries act in the usual way on $A_\mu$ only. The gauge symmetry commutators are schematically $[P, B_{l}] = B_{l-1}$ and if we include them we can write the new coset element as
\begin{align} \label{CosetWithGauge}
    \tilde{\gamma} = \gamma e^{\text{Gauge}}\,,
\end{align}
where $\gamma$ is given in \eqref{CosetElement} and $e^{\text{Gauge}}$ schematically denotes contractions between the generators of the tower of gauge symmetries and the corresponding inessential Goldstones that would be removed in favor of the gauge vector via inverse Higgs constraints as in \cite{Goon:2014paa,Klein:2018ylk}. Whether we parametrize $e^{\text{Gauge}}$ as a product of exponentials each with a single generator, as one exponential containing the full sum of generators, or indeed anything in between, is immaterial since the gauge symmetries all commute amongst themselves. Given that the generators of the gauge symmetries commute with all the non-Lorentz generators other than translations $P_\mu$, in which case the commutator returns another gauge symmetry, we have 
\begin{align}
    \tilde{\gamma}^{-1} d \tilde{\gamma} = \gamma^{-1} d \gamma + \omega_{\text{Gauge}} \,,
\end{align}
meaning that the one-forms \eqref{OneForm1} - \eqref{OneForm5} are unchanged and the only difference is the addition of new one-forms corresponding to the gauge symmetries. Note that the inverse Higgs solutions that remove the inessential Goldstones of the gauge symmetries will depend on $\partial_{\mu} \xi_{\nu}$, and therefore ultimately depend on $\Pi_{\mu\nu}$, and $G_{\mu\nu}$. This is because those fields appear in \eqref{OneForm4}. Furthermore, with the parametrization \eqref{CosetWithGauge}, the symmetry transformations derived in the absence of the gauge symmetries are unchanged and the inessential Goldstones of the gauge symmetries do not transform under those symmetries. This is consistent with the inverse Higgs solutions thanks to the dependence on $\Pi$ and $G$ in those constraints.  If we act on \eqref{CosetElement} to the left with a gauge symmetry generator, we find the expected transformation rules, i.e. those obtained if we did not include $C, C_\mu$ and $M_{\mu\nu}$ in the coset, since the gauge symmetry generators commute with these generators.

In summary, the one-forms and symmetry transformations we derived above are unchanged by the inclusion of the gauge symmetries, and the gauge symmetries act in the usual way on $A_\mu$ only. We can therefore work with the restricted algebra that does not contain the full tower of gauge symmetries and impose the gauge symmetries by hand.

\subsubsection{A Comment on Basis Changes}

Before moving on to derive the resulting theory, we first want to make sure that we cannot set $\alpha=0$ by a basis change, so that it is a genuine parameter in the theory. In the absence of the gauge symmetries, it is clear that we can: from the form of the commutators \eqref{MGSVTcommutators1} the basis change $\tilde{C}_\mu = C_{\mu} + \alpha P_{\mu}$, with all other generators unchanged, brings these commutators to the form 
\begin{align}\label{MGSVTcommutators}
[P_{\mu}, \tilde{C}_{\nu}] = \eta_{\mu \nu} C, \qquad [P_{\mu}, M_{\rho \sigma}] = \frac{1}{2}\left(\eta_{\mu \rho} B_{\sigma}  - \eta_{\mu \sigma} B_{\rho} \right), \qquad [\tilde{C}_{\mu}, M_{\rho \sigma}] =0 \,,
\end{align}
which has achieved $\alpha = 0$.   Now consider the addition of the gauge symmetries. In the original basis, $C_{\mu}$ commutes with all gauge symmetry generators, but this is not the case in the new basis. Schematically we have 
\begin{align}
[C_{\mu}, \text{Gauge}] = 0 \longrightarrow [\tilde{C}_{\mu}, \text{Gauge}] = \alpha\, \text{Gauge} \,.
\end{align}
A consequence of this is that in the new basis, the gauge symmetries do not act on the gauge vector in the usual way. We can see this explicitly by working with an example. Consider the original basis and the gauge symmetry generated by the symmetric tensor $B_{\mu\nu}$ i.e. $\delta_{(B_{\mu\nu})} A_{\mu} = b_{\mu \nu} x^{\nu}$ which comes from $\delta A_{\mu} = \partial_{\mu} \lambda$ with $\lambda = \frac{1}{2}b_{\mu\nu} x^{\mu}x^{\nu}$. This generator commutes with all generators other than translations for which we have 
\begin{align}
    [P_{\mu}, B_{\rho \sigma}] = \frac{1}{2}(\eta_{\mu \rho}B_{\sigma} + \eta_{\mu \sigma}B_{\rho}) \,.
\end{align}
If we parametrize the coset element as $\tilde{\gamma} = \gamma e^{\psi^{\mu\nu} B_{\mu\nu}}$ we get the desired transformation rule on $A_{\mu}$ when we act from the left with $e^{b^{\mu\nu} B_{\mu\nu}}$ and no transformation on the scalar, as expected. After the basis change we get the additional commutator
\begin{align}
    [\tilde{C}_{\mu}, B_{\rho \sigma}] = \frac{\alpha}{2}(\eta_{\mu \rho}B_{\sigma} + \eta_{\mu \sigma}B_{\rho}) \,.
\end{align}
If we still parametrize the coset element as $\tilde{\gamma} = \gamma e^{\psi^{\mu\nu} B_{\mu\nu}}$ (with the $C_{\mu}$ generator in $\gamma$ replaced by $\tilde{C}_{\mu}$), the transformations generated by $B_{\mu\nu}$ are now   
\begin{align}
    \delta_{(B_{\mu \nu})} \phi = 0, \qquad \delta_{(B_{\mu \nu})} A_{\mu} =- b_{\mu\nu}( x^{\nu} + \alpha \xi^{\nu}) \,.
\end{align}
In this case we cannot write the transformation on $A_{\mu}$ as $\delta A_{\mu} = \partial_{\mu} \lambda$ (before or after using $\xi_{\mu} = -\partial_{\mu} \phi$) so the basis change has altered the form of the gauge symmetries. This tells us that we cannot simply fix $\alpha = 0$ when we have the gauge symmetries as part of the full algebra. We might be tempted to try different parametrizations of the coset element so that the gauge symmetries act in the usual way. For example, we can maintain $\delta_{(B_{\mu \nu})} A_{\mu} = b_{\mu\nu}x^{\nu}$ by working with 
\begin{align} \label{CosetElement2}
    \tilde{\gamma} = e^{x^{\mu}P_{\mu}} e^{\psi^{\mu\nu}B_{\mu\nu}} e^{\phi C} e^{\xi^{\mu} \tilde{C}_{\mu}}e^{A^{\mu}B_{\mu}}e^{-{1\over 2}G^{\mu\nu}M_{\mu\nu}}\,.
\end{align}
However, now consider the symmetries generated by $\tilde{C}_{\mu}$. We have 
\begin{align}
    \delta_{(\tilde{C}_{\mu})} \phi = -\tilde{c}_{\mu} x^{\mu}, \qquad  \delta_{(\tilde{C}_\mu)}A_{\mu} = \alpha \tilde{c}_{\nu} \psi^{\nu}{}_{\mu} \,, 
\end{align}
so in contrast to prior to the basis change, the gauge vector now transforms under the generator of the Galileon symmetry. Once we solve the inverse Higgs constraint to eliminate $\psi^{\mu\nu}$, this transformation will be non-linear in the fields and not writable as a standard gauge transformation. In conclusion, it is not possible to set $\alpha = 0$ by a basis change. In the massive gravity context, where the $\alpha=0$ case arises from the vector/scalar sector of the decoupling limit of pseudo-linear massive gravity, while the $\alpha \neq 0$ case comes from dRGT \cite{Bonifacio:2019hrj}, this shows that the decoupling limits of these two cases are distinct in the scalar/vector sector, even though they are identical in the scalar/tensor sector.

\subsubsection{Kinetic Terms and Interactions}
Our previous analysis has been valid in general $D$, but now we restrict to $D=4$. It is well known that the scalar kinetic term cannot arise from the above one-form building blocks: given that $\omega_C=0$ on the inverse Higgs solution, the scalar only appears via the two-derivative object $\Pi_{\mu\nu}$, so every Lagrangian constructed from these building blocks will have at least two derivatives per scalar.  Instead, the kinetic term and leading self-interactions for the scalar arise as Wess-Zumino terms \cite{Goon:2012dy}. We are in a similar situation for the gauge vector kinetic term $F_{\mu\nu}F^{\mu\nu}$: on the inverse Higgs solution we are left with the symmetric part of the pullback of $\omega^{\mu}_{B}$ and the pullback of $\omega^{\mu\nu}_M$ and neither of these can give rise to the desired kinetic term. We must therefore again turn to Wess-Zumino terms.   

To find the kinetic term for the vector, we want to build a five-form that starts at quadratic order in the vector and yields a Lorentz invariant Lagrangian once we pullback to four-dimensional spacetime. We therefore start with
\begin{align} \label{KineticWZ1}
\omega^{(a)}_{5} = \eta_{\alpha \beta} \epsilon_{\mu\nu\rho\sigma} \omega^{\mu}_{P} \wedge \omega^{\nu}_{P} \wedge \omega^{\rho}_{P} \wedge \omega^{\alpha}_{B} \wedge \omega^{\sigma \beta}_{M} \,.
\end{align}
Any other way of distributing the indices that are contracted with the $\epsilon$-tensor is degenerate with this five-form, so this is the most general starting point. For this five-form to yield a Wess-Zumino term, it needs to be closed, and therefore writeable as the exterior derivative of a four-form: $\omega^{(a)}_{5} = d \beta^{(a)}_{4}$. We have 
\begin{align} \label{dKineticWZ1}
    d \omega^{(a)}_{5} ={1\over 4} \alpha \epsilon_{\mu\nu\rho\sigma} dx^{\mu} \wedge dx^{\nu} \wedge dx^{\rho} \wedge d \xi_{\gamma} \wedge dG^{\sigma \beta} \wedge d G^{\gamma}{}_{\beta} \,,
\end{align}
which does not vanish and therefore we need to correct \eqref{KineticWZ1}. Since \eqref{dKineticWZ1} is of the form $\phi AA$, the correction we add to \eqref{KineticWZ1} must also be of this form. Our ansatz for the corrected $\omega^{(a)}_{5}$ is 
\begin{align}
\tilde{\omega}^{(a)}_{5} = \omega^{(a)}_{5} + a_1  \eta_{\alpha \beta}\epsilon_{\mu\nu\rho\sigma} \omega^{\mu}_{P} \wedge \omega^{\nu}_{P} \wedge \omega_{C} \wedge \omega^{\rho \alpha}_{M}\wedge \omega^{\sigma \beta}_{M} \,,
\end{align}
where $a_1$ is a constant to be fixed by demanding $d \tilde{\omega}^{(a)}_{5} = 0$. We have 
\begin{align}
   d \tilde{\omega}^{(a)}_{5} = \, & \frac{\alpha}{4} \epsilon_{\mu\nu\rho\sigma} dx^{\mu} \wedge dx^{\nu} \wedge dx^{\rho} \wedge d \xi_{\gamma} \wedge dG^{\sigma \beta} \wedge d G^{\gamma}{}_{\beta} \nonumber \\
   &- \frac{a_1}{4} \epsilon_{\mu\nu\rho\sigma} dx^{\mu} \wedge dx^{\nu} \wedge dx^{\gamma} \wedge d \xi_{\gamma} \wedge d G^{\rho \beta} \wedge d G^{\sigma}{}_{\beta} \\
   = \, & \frac{\alpha}{4} \epsilon_{\mu\nu\rho\sigma} dx^{\mu} \wedge dx^{\nu} \wedge dx^{\rho} \wedge d \xi_{\gamma} \wedge dG^{\sigma \beta} \wedge d G^{\gamma}{}_{\beta} \nonumber \\
   &+ \frac{a_1}{6} \epsilon_{\mu\nu\rho\sigma} dx^{\mu} \wedge dx^{\nu} \wedge dx^{\rho} \wedge d \xi_{\gamma} \wedge d G^{\sigma \beta} \wedge d G^{\gamma}{}_{\beta}\,,
\end{align}
 where we made use of identities in \cite{Goon:2012dy} to rewrite the $a_1$ term. For closure we therefore fix $a_1 = - \frac{3 \alpha}{2}$. The resulting five-form can now be written in a way that makes its closure manifest:
\begin{align}
    & \tilde{\omega}^{(a)}_5 = -\frac{1}{2}\epsilon_{\mu\nu\rho\sigma} dx^{\mu} \wedge dx^{\nu} \wedge dx^{\rho} \wedge d A^{\beta} \wedge d G^{\sigma}{}_{\beta}   -\frac{3 \alpha}{8} \epsilon_{\mu\nu\rho\sigma} dx^{\mu} \wedge dx^{\nu} \wedge d \phi \wedge d G^{\rho \beta} \wedge d G^{\sigma}{}_{\beta} \nonumber \\
   &  -\frac{1}{8}\epsilon_{\mu\nu\rho\sigma} dx^{\mu} \wedge dx^{\nu} \wedge dx^{\rho} \wedge dx_{\gamma} \wedge d (G^{\beta \gamma} G^{\sigma}{}_{\beta})  + \frac{\alpha}{4} \epsilon_{\mu\nu\rho\sigma} dx^{\mu} \wedge d x^{\nu} \wedge dx^{\rho} \wedge d(G^{\beta \gamma} \xi_{\gamma})\wedge d G^{\sigma}{}_{\beta} \,,
 \end{align}
 from which we can read off the corresponding four-form:
\begin{align}
 &    \tilde{\beta}^{(a)}_4 = \frac{1}{2}\epsilon_{\mu\nu\rho\sigma} A^{\beta} dx^{\mu} \wedge dx^{\nu} \wedge dx^{\rho}  \wedge d G^{\sigma}{}_{\beta}  -\frac{3 \alpha}{8} \epsilon_{\mu\nu\rho\sigma} \phi dx^{\mu} \wedge dx^{\nu} \wedge d G^{\rho \beta} \wedge d G^{\sigma}{}_{\beta} \nonumber \\
   &  -\frac{1}{8}\epsilon_{\mu\nu\rho\sigma} G^{\beta \gamma} G^{\sigma}{}_{\beta} dx^{\mu} \wedge dx^{\nu} \wedge dx^{\rho} \wedge dx_{\gamma}  - \frac{\alpha}{4} \epsilon_{\mu\nu\rho\sigma} G^{\beta \gamma} \xi_{\gamma}  dx^{\mu} \wedge d x^{\nu} \wedge dx^{\rho} \wedge  d G^{\sigma}{}_{\beta} \,.
 \end{align}
 From this four-form, we can construct an action by pulling back to spacetime and integrating. Doing so allows us to read off the Lagrangian, and after a number of integrations by parts and use of the inverse Higgs solution \eqref{GalileonIHC}, we find
\begin{align}
     \mathcal{L}_{2} \propto  \left(-\frac{1}{2}F^{\mu}{}_{\nu} G^{\nu}{}_{\mu} + \frac{1}{4}G^{\mu}{}_{\nu} G^{\nu}{}_{\mu} + \frac{\alpha}{4} G^{\mu}{}_{\rho} G^{\rho}{}_{\nu} \Pi^{\nu}{}_{\mu} \right) \,,
 \end{align}
 with an arbitrary overall normalization which we will fix by canonically normalizing the gauge vector. By contracting the inverse Higgs constraint \eqref{VectorIHC} with $G_{\mu\nu}$ we find 
\begin{align}
G^{\mu}{}_{\nu}G^{\nu}{}_{\mu} = F^{\mu}{}_{\nu}G^{\nu}{}_{\mu} - \alpha G^{\mu}{}_{\rho} G^{\rho}{}_{\nu} \Pi^{\nu}{}_{\mu} \,, 
 \end{align}
 and therefore we can write the Lagrangian as 
\begin{align} \label{L2}
     \mathcal{L}_{2} = \frac{1}{4} F^{\mu}{}_{\nu} G^{\nu}{}_{\mu} = \frac{1}{2}F^{[\mu}{}_{\nu} G^{\nu]}{}_{\mu}  \,,
 \end{align}
 where we fixed the overall constant to canonically normalize the gauge vector. This Lagrangian takes a particularly simple form, even though it came from a somewhat complicated four-form. Given the explicit solution for $G_{\mu\nu}$ in terms of $F_{\mu\nu}$ and $\Pi_{\mu\nu}$, this Lagrangian ties together the kinetic term of the gauge field with an infinite tower of interactions of the schematic form $\sim\sum_{n=1}^\infty F^2 \Pi^n$.
 
 As a consistency check, we can verify that \eqref{L2} is invariant under the symmetries we derived in Section \ref{sec:CosetSymmetries}, up to a total derivative. Invariance under the Galileon symmetries generated by $C$ and $C_\mu$ is manifest, since the gauge field does not transform and $\Pi_{\mu\nu}$, which is the only way the scalar enters \eqref{L2}, is invariant under those symmetries. The constant shift symmetry on the gauge vector generated by $B_{\mu}$ is a symmetry, and in addition the full set of gauge symmetries are preserved since the action depends on the gauge field only through $F_{\mu\nu}$. This was not guaranteed to be the case since we did not include the full set of gauge symmetries in the coset construction, yet they have emerged for this particular Wess-Zumino term. Finally, we have the symmetries generated by $M_{\mu\nu}$ given in \eqref{MSymmetries}, which are not manifest. At linear order in $m_{\mu\nu}$, we have 
\begin{align}
     \delta^{(1)} \mathcal{L}_2 &= \frac{1}{4} \delta^{(1)} F^{\mu}{}_{\nu} G^{\nu}{}_{\mu} +\frac{1}{4}  F^{\mu}{}_{\nu} \delta^{(1)}  G^{\nu}{}_{\mu} \nonumber \\
      &= -\frac{1}{2} m^{\mu}{}_{\nu}G^{\nu}{}_{\mu} - \frac{\alpha}{2} m^{\mu}{}_{\rho}G^{\rho}{}_{\nu}\Pi^{\nu}{}_{\mu} - \frac{1}{2}m^{\mu}{}_{\nu}F^{\nu}{}_{\mu} \nonumber \\
      & = - m^{\mu}{}_{\nu}F^{\nu}{}_{\mu} \,,
 \end{align}
 where in the final step we used the inverse Higgs constraint \eqref{VectorIHC} contracted with $m_{\mu\nu}$, which is a total derivative. At quadratic order in $m_{\mu\nu}$ we have 
\begin{align}
     \delta^{(2)} \mathcal{L}_2 &= \frac{1}{4} \delta^{(1)}  F^{\mu}{}_{\nu} \delta^{(1)}  G^{\nu}{}_{\mu} \nonumber \\
    & =  m^{\mu}{}_{\nu}m^{\nu}{}_{\mu} + \alpha m^{\mu}{}_{\rho}m^{\rho}{}_{\nu}\Pi^{\nu}{}_{\mu} \,,
 \end{align}
 which is also a total derivative. The total varied Lagrangian is therefore a total derivative, as expected for a Wess-Zumino term.

 Given that this leading Wess-Zumino term is of the form $\sum_{n=0} F^2 \Pi^n$, we are motivated to search for other Wess-Zumino terms that give rise to operators of the same form but starting at non-zeroth order in the scalar, i.e. of the form $\sum_{n > 0} F^2 \Pi^n$. We begin by searching for such Wess-Zumino terms that start at linear order in the scalar, of the form $\sum_{n=1} F^2 \Pi^n$. The general five-form that can give rise to such a Wess-Zumino term is 
 \begin{align} \label{FiveFormb}
     \omega_5^{(b)} = \eta_{\alpha \beta} \epsilon_{\mu\nu\rho\sigma} \omega_{P}^{\mu} \wedge \omega_{P}^{\nu} \wedge [b_1& \omega_C \wedge \omega_{M}^{\rho \alpha} \wedge \omega_{M}^{\sigma \beta} + b_2  \omega_{C}^{\rho} \wedge \omega_{B}^{\alpha} \wedge \omega_{M}^{\sigma \beta} \nonumber \\
     &+ b_3 \omega_{C}^{\alpha} \wedge \omega_{B}^{\rho} \wedge \omega_{M}^{\sigma \beta} + b_4 \omega_{C}^{\alpha} \wedge \omega_{B}^{\beta}  \wedge \omega_{M}^{\rho \sigma} ] \,,
 \end{align}
 where the $b$'s are arbitrary constants. Before imposing closure, we first impose gauge invariance. This would be manifest if we included all of the gauge generators in the algebra and coset, but since we have only included the generator of the constant shift symmetry, we need to impose gauge invariance by hand. Most terms in \eqref{FiveFormb} will yield a gauge invariant Lagrangian other than the following terms:
\begin{align} \label{FiveFormbGauge}
     \omega_5^{(b)} \supset -\frac{1}{2}\eta_{\alpha \beta} \epsilon_{\mu\nu\rho\sigma} dx^{\mu} \wedge dx^{\nu} \wedge [b_2 d \xi^{\rho} \wedge dA^{\alpha} \wedge dG^{\sigma \beta} 
     &+ b_3 d \xi^{\alpha} \wedge dA^{\rho} \wedge dG^{\sigma \beta} + b_4 d\xi^{\alpha} \wedge dA^{\beta}  \wedge dG^{\rho \sigma} ] \,,
 \end{align}
 which are closed. By computing the resulting contributions to the Lagrangian by pulling back to spacetime and integrating, we find that gauge invariance fixes $b_3 = 0$ and $b_4 = \frac{1}{2}b_2$.\footnote{Note that we impose gauge invariance for all $\xi^{\mu}$ and $G^{\mu\nu}$ and not just on the inverse Higgs solutions since this would be the case if we had included all generators of the gauge symmetry.} We will therefore work with 
 \begin{align} \label{FiveFormbNew}
     \omega_5^{(b)} = \eta_{\alpha \beta} \epsilon_{\mu\nu\rho\sigma} \omega_{P}^{\mu} \wedge \omega_{P}^{\nu} \wedge \left[ b_1 \omega_C \wedge \omega_{M}^{\rho \alpha} \wedge \omega_{M}^{\sigma \beta} + b_2  \omega_{C}^{\rho} \wedge \omega_{B}^{\alpha} \wedge \omega_{M}^{\sigma \beta} +  \frac{b_2}{2} \omega_{C}^{\alpha} \wedge \omega_{B}^{\beta}  \wedge \omega_{M}^{\rho \sigma} \right] \,.
 \end{align}
 We now turn our attention to closure. This five-form is not closed, rather we have 
\begin{align}
  &   d \omega_5^{(b)} = \frac{1}{8}\epsilon_{\mu\nu\rho\sigma} dx^{\mu} \wedge dx^{\nu} \wedge [ 2 b_1 d \xi_{\alpha} \wedge dx^{\alpha} \wedge dG^{\rho \beta} \wedge d G^{\sigma}{}_{\beta} + 2 b_2 d \xi^{\rho} \wedge dx^{\alpha} \wedge d G_{\alpha}{}^{\beta} \wedge dG^{\sigma}{}_{\beta} \nonumber \\
     & - b_2 d \xi^{\alpha} \wedge dx^{\beta} \wedge dG_{\alpha \beta}\wedge d G^{\rho \sigma} - 2 b_2 \alpha d \xi^{\rho} \wedge d \xi^{\alpha} \wedge dG_{\alpha}{}^{\beta} \wedge d G^{\sigma}{}_{\beta} + b_2 \alpha d \xi^{\alpha} \wedge d \xi^{\beta} \wedge dG_{\alpha \beta} \wedge dG^{\rho \sigma}]\,.
 \end{align}
 This contains two distinct structures with either three copies of $dx$ or two copies. The terms with three copies can be cancelled by setting $b_2 = -b_1$ while the terms with two copies cannot be cancelled while keeping $b_1 \neq 0$. We therefore need to correct the original five-form, as we had to do previously, and our corrections need to start at $A A \phi \phi$. Our new five-form is 
 \begin{align}
     \tilde{\omega}_{5}^{(b)} = \omega_{5}^{(b)} + \eta_{\alpha \beta}\epsilon_{\mu\nu\rho\sigma} \omega_P^{\mu}  \wedge \omega_C \wedge (b_5 \omega_C^{\nu} \wedge \omega_{M}^{\rho \alpha} \wedge \omega_{M}^{\sigma \beta} + b_6 \omega_{C}^{\alpha} \wedge \omega_{M}^{\nu \beta} \wedge \omega_{M}^{\rho \sigma})\,,
 \end{align}
 and closure requires $b_6 = -b_5 = \alpha b_1$. We have therefore found a unique five-form that is closed and can give rise to a Wess-Zumino term of the form $\sum_{n = 1} F^2 \Pi^n$. It is unique because we started with the most general five-form that can give rise to a leading interaction of the form $\phi AA$ with four derivatives. At this point we can find the associated four-form, pull back to spacetime and integrate to find the action. This very quickly becomes cumbersome and won't necessarily provide us with a neat form for the corresponding interactions (to find a neat form we might need to repeatedly use the inverse Higgs solutions and integrate by parts as we did for the previous case). At this stage it therefore makes sense to write down an ansatz and constrain it by imposing invariance under our symmetries, up to total derivatives. Once we find a solution, we know it must match the unique Wess-Zumino term we have found, up to the addition of a total derivative. The benefit of going through the above analysis is to show that only one such Wess-Zumino exists. 

 Given the form of \eqref{L2}, a sensible ansatz is 
\begin{align}
     \mathcal{L}_3 \propto g_3^{(1)} F^{\mu}{}_{\nu}G^{\nu}{}_{\mu} \Pi^{\rho}{}_{\rho} +g_3^{(2)} F^{\mu}{}_{\nu}G^{\nu}{}_{\rho}\Pi^{\rho}{}_{\mu} \,,
 \end{align}
 which contains all possible contractions between the matrices $F, G$ and $\Pi$. Our ansatz is manifestly gauge invariant and invariant under the scalar symmetries. We therefore only need to impose invariance of the corresponding action under the symmetry generated by $M_{\mu\nu}$. If we first consider the variation at linear order in $m_{\mu\nu}$, after repeated use of the inverse Higgs constraint \eqref{VectorIHC} and integration by parts, we find that invariance of the action requires $g_3^{(2)} = -2 g_3^{(1)}$. The resulting Lagrangian can be expressed as 
\begin{align} \label{L3}
    \mathcal{L}_3  \propto F^{[\mu}{}_{\nu}G^{\nu}{}_{\rho}\Pi^{\rho]}{}_{\mu} \,,
 \end{align}
 which is a neat extension of \eqref{L2}. It is then a straightforward task to verify that the action is also invariant at quadratic order in $m_{\mu\nu}$, up to a total derivative. 

 Next we consider Wess-Zumino terms of the form $\sum_{n=2} F^2 \Pi^n$. The relevant five-form is 
 \begin{align}
     \omega_5^{(c)} = \eta_{\alpha \beta} \epsilon_{\mu\nu\rho\sigma} & \omega_{P}^{\mu} \wedge (c_1 \omega_C \wedge \omega_{C}^{\nu} \wedge \omega_{M}^{\rho \alpha} \wedge \omega_{M}^{\sigma \beta} + c_2 \omega_C \wedge \omega_C^{\alpha} \wedge \omega_M^{\nu \beta} \wedge \omega_{M}^{\rho \sigma} \nonumber \\ + &c_3 \omega_{C}^{\nu} \wedge \omega_C^{\rho} \wedge \omega_B^{\alpha} \wedge \omega_M^{\sigma \beta} + c_4 \omega_C^{\nu} \wedge \omega_C^{\alpha} \wedge \omega_{B}^{\beta} \wedge \omega_M^{\rho \sigma} 
    +  c_5 \omega_{C}^{\nu} \wedge \omega_{C}^{\alpha} \wedge \omega_{B}^{\rho} \wedge \omega_M^{\sigma \beta}) \,.
 \end{align}
 The procedure we follow is as before. We first constrain this five-form by ensuring it gives rise to a gauge invariant Lagrangian, which fixes $c_5=0$ and $c_4=c_3$. Closure for the terms with two factors of $dx$ requires $c_3 = -c_2$ and $c_2 = -\frac{1}{2}c_1$ which leaves us with only a single parameter. The terms with only a single copy of $dx$ are not closed but we can correct the original five-form with terms that start at cubic order in the scalar, i.e. start at $AA\phi \phi \phi$ such that it is closed. Our final unique closed five-form is
 \begin{align} \label{ClosedFiveFormC}
     \omega_5^{(c)} = \frac{c_1}{2}\eta_{\alpha \beta} \epsilon_{\mu\nu\rho\sigma}  \omega_{P}^{\mu} \wedge (&2 \omega_C \wedge \omega_{C}^{\nu} \wedge \omega_{M}^{\rho \alpha} \wedge \omega_{M}^{\sigma \beta} - \omega_C \wedge \omega_C^{\alpha} \wedge \omega_M^{\nu \beta} \wedge \omega_{M}^{\rho \sigma}  + \omega_{C}^{\nu} \wedge \omega_C^{\rho} \wedge \omega_B^{\alpha} \wedge \omega_M^{\sigma \beta} \nonumber \\ & + \omega_C^{\nu} \wedge \omega_C^{\alpha} \wedge \omega_{B}^{\beta} \wedge \omega_M^{\rho \sigma} 
    ) + \frac{c_1 \alpha}{4} \eta_{\alpha \beta} \epsilon_{\mu\nu\rho\sigma} \omega_C \wedge \omega_C^{\mu} \wedge \omega_C^{\nu} \wedge \omega_M^{\rho \alpha} \wedge \omega_M^{\sigma \beta}  \,.
 \end{align}
 Again, using this expression to find the associated operators could become very cumbersome. However, now that we know only one Wess-Zumino term of the form $\sum_{n=2} F^2 \Pi^n$ exists, we can write down an ansatz and fix the free parameters using the symmetries. Given the structure of the other Wess-Zumino terms, a sensible ansatz is linear in $G_{\mu\nu}$ and writing down all possible terms leads to:
\begin{align}
     \mathcal{L}_4 \propto g_4^{(1)}F^{\mu}{}_{\nu}G^{\nu}{}_{\mu}  \Pi^{\rho}{}_{\rho}\Pi^{\sigma}{}_{\sigma} + g_4^{(2)}F^{\mu}{}_{\nu}G^{\nu}{}_{\mu} \Pi^{\rho} &{}_{\sigma}\Pi^{\sigma}{}_{\rho} + g_4^{(3)} F^{\mu}{}_{\nu}G^{\nu}{}_{\rho}\Pi^{\rho}{}_{\mu}\Pi^{\sigma}{}_{\sigma}  \nonumber \\
    & + g_4^{(4)}F^{\mu}{}_{\nu}G^{\nu}{}_{\rho}\Pi^{\rho}{}_{\sigma}\Pi^{\sigma}{}_{\mu}+ g_4^{(5)}F^{\mu}{}_{\nu}\Pi^{\nu}{}_{\rho}G^{\rho}{}_{\sigma}\Pi^{\sigma}{}_{\mu} \,.
 \end{align}
 We can now impose invariance under the symmetry generated by $M_{\mu\nu}$, up to a total derivative, at linear order in the symmetry parameter $m_{\mu \nu}$. At linear order we need to use the inverse Higgs constraint \eqref{VectorIHC}. We find
 \begin{align}
g_4^{(5)} = 2 g_4^{(1)}, \quad g_4^{(4)} = - g_4^{(3)}= 4 g_4^{(1)}, \quad g_4^{(2)} = - g_4^{(1)} \,,
 \end{align}
 thereby leaving us with a unique Lagrangian which can be neatly written as 
\begin{align}
    \mathcal{L}_4 \propto F^{[\mu}{}_{\nu}G^{\nu}{}_{\rho}\Pi^{\rho}{}_{\sigma} \Pi^{\sigma]}{}_{\mu} \,,
 \end{align}
 This Lagrangian is guaranteed to be equivalent to the one coming from \eqref{ClosedFiveFormC}, up to a total derivative. Again, this structure is very similar to \eqref{L2} and \eqref{L3}.  
 
 We can now go on and search for Wess-Zumino terms of the form $\sum_{n=3} F^2 \Pi^n$. There are naively a number of structures we can write for the corresponding five-form but there are many degeneracies and we are left with only two independent terms:
 \begin{align}
     \omega_5^{(d)} = d_1 \eta_{\alpha \beta} \epsilon_{\mu\nu\rho\sigma} \omega_C \wedge \omega_C^{\mu} \wedge \omega_C^{\nu} \wedge \omega_M^{\rho \alpha} \wedge \omega_M^{\sigma \beta} + d_2 \eta_{\alpha \beta} \epsilon_{\mu\nu\rho\sigma} \omega_C^{\mu} \wedge \omega_C^{\nu} \wedge \omega_C^{\rho} \wedge \omega_B^{\alpha} \wedge \omega_M^{\sigma \beta} \,.
 \end{align}
 Only the second term is not manifestly gauge invariant and imposing gauge invariance requires us to fix $d_2 = 0$. We are then left with only a single term and the parts with a single copy of $dx$ are not closed and there is no way to add corrections to fix this since we already wrote down all possible terms at that order. We therefore need to also fix $d_1=0$, meaning that there is no closed five-form that starts at $AA \phi \phi \phi$ with the appropriate power counting and therefore no corresponding Wess-Zumino term. For Wess-Zumino terms of the form $\sum_{n \geq 4} F^2 \Pi^n$ we need to write down five-forms that have two copies of the gauge field and at least four copies of the scalar, but this is not possible because each one-form starts at zeroth or linear order in the fields. We have therefore found all possible Wess-Zumino terms of the form $\sum_{n} F^2 \Pi^n$. 
 
 In summary, all Wess-Zumino terms that yield the gauge vector kinetic term and interactions between the gauge vector and the scalar in $D=4$ are
\begin{align}     
\mathcal{L} \supset  F^{[\mu}{}_{\nu}G^{\nu]}{}_{\mu} \, ,  \   F^{[\mu}{}_{\nu}G^{\nu}{}_{\rho}\Pi^{\rho]}{}_{\mu} \, ,  \  F^{[\mu}{}_{\nu}G^{\nu}{}_{\rho}\Pi^{\rho}{}_{\sigma} \Pi^{\sigma]}{}_{\mu} \,.
 \end{align}
These interactions also appeared in \cite{Ondo:2013wka}, and implicitly in \cite{Gabadadze:2013ria}, via a direct calculation of the decoupling limit of dRGT massive gravity. Here we have taken a more bottom-up, symmetry-focused route, using the symmetry identified in \cite{Bonifacio:2019hrj}, and found the same theory. The abelian version of the theory with $\alpha=0$ is given by the replacement $G_{\mu\nu} \rightarrow F_{\mu\nu}$, cf. \eqref{PertGSolution}.  This abelian version appears in the decoupling limit of pseudolinear massive gravity \cite{Hinterbichler:2013eza}.

If we include the Wess-Zumino terms for the pure scalar interactions, which are the standard Galileons \cite{Goon:2012dy}, then the full Lagrangian can be taken as
\bea {\cal L}&=& -{1\over 4}F^{\mu\nu}G_{\mu\nu}  + c_3 \beta\, F^{[\mu_1}_{\ \ \mu_1} G^{\mu_2}_{\ \ \mu_2} \Pi^{\mu_3]}_{\ \ \mu_3}  +c_4 \beta^2 \, F^{[\mu_1}_{\ \ \mu_1} G^{\mu_2}_{\ \ \mu_2} \Pi^{\mu_3}_{\ \ \mu_3}  \Pi^{\mu_4]}_{\ \ \mu_4} \nn\\
&&-{1\over 2} \partial ^{\mu}\phi\partial_{\mu} \phi +d_3\beta {\cal L}_{\rm gal,\phi^3}+d_4\beta^2 {\cal L}_{\rm gal,\phi^4}+d_5\beta^3 {\cal L}_{\rm gal,\phi^5} \,, \label{LagrangianLevelOne} \eea
where 
\be {\cal L}_{\rm gal,\phi^n} \equiv \partial ^{[\mu_1}\phi\partial_{\mu_1} \phi  \Pi^{\mu_2}_{\ \ \mu_2} \Pi^{\mu_3}_{\ \ \mu_3} \cdots \Pi^{\mu_{n-1}]}_{\ \ \mu_{n-1}} \,, \label{SVGLagrangian}
\ee
is the $n$-th order Galileon interaction \cite{Nicolis:2008in}. 
The coefficients $c_3,c_4$ and $d_3,d_4,d_5$ are free dimensionless parameters, and $\beta$ is an overall dimensionful scale carrying mass dimension $[\beta]=-3$. (Note that there is one redundancy in the parameters because one of the dimensionless parameters could be absorbed into $\beta$, but we will leave them all separate for maximum flexibility later on).

This Lagrangian can easily be extrapolated to arbitrary spacetime dimension $D$ by adding more powers of $\Pi$ and anti-symmetrizing:
\bea {\cal L}&= & -{1\over 4}F^{\mu\nu}G_{\mu\nu}+ \sum_{i=3}^{D} c_{i} \beta^{i-2} \, F^{[\mu_1}_{\ \ \mu_1} G^{\mu_2}_{\ \ \mu_2} \Pi^{\mu_3}_{\ \ \mu_3} \cdots \Pi^{\mu_{i}]}_{\ \ \mu_{i}}  \nn\\
&& -{1\over 2} \partial ^{\mu}\phi\partial_{\mu} \phi +\sum_{i=3}^{D+1} d_i\beta^{i-2} {\cal L}_{\rm gal,\phi^i} \,. \label{GeneralDSVG}
\eea
Note that in the abelian case $\alpha=0$, the individual terms in this Lagrangian are special cases of the general abelian $p$-form Galileons of \cite{Deffayet:2010zh}.
 
In a full EFT, there are of course also higher-derivative corrections to this Lagrangian that are not Wess-Zumino and are constructed directly out of the invariant building blocks coming from the coset, but the Wess-Zumino terms written here have fewer derivatives per field and are thus the leading terms in a derivative expansion.

 \section{Level 2: Special Galileon Extensions} \label{sec:LevelTwo}
 We now include the traceless tensor $C_{\mu\nu}$ at level two in the scalar tree. Here we are primarily interested in the existence of a special Galileon extension of the theory we have already found, that is, we ask if there is a special choice of couplings in \eqref{LagrangianLevelOne} such that an enhanced special Galileon symmetry is realized. We therefore take our final level one algebra to be a sub-algebra, and we assume that the scalar sector is a different sub-algebra with the known special Galileon structure for those commutators \cite{Hinterbichler:2015pqa}. We revert back to general dimensions. 

\subsection{Commutators and Jacobi Identities}
The non-zero commutators of our level one sub-algebra are
\begin{align}
[P_{\mu}, C_{\nu}] = \eta_{\mu \nu} C, \qquad [P_{\mu}, M_{\rho \sigma}] = \frac{1}{2}\left(\eta_{\mu \rho} B_{\sigma}  - \eta_{\mu \sigma} B_{\rho} \right), \qquad [C_{\mu}, M_{\rho \sigma}] =- \frac{\alpha}{2} \left( \eta_{\mu \rho} B_{\sigma}  - \eta_{\mu \sigma} B_{\rho} \right) \,.
\end{align}
The commutator between $C_{\mu\nu}$ and Lorentz is given in \eqref{TensorAndLorentz} and its commutator with translations is fixed by the inverse Higgs tree and is given in \eqref{TensorAndTranslations}. We take the remaining commutators of the scalar sub-algebra to be
\begin{align}
[C,C_{\mu\nu}]&=0 \,,
\\
[C_\rho,C_{\mu\nu}]
&=
s_1\left(
\eta_{\rho\mu}C_\nu
+\eta_{\rho\nu}C_\mu
-\frac{2}{D}\eta_{\mu\nu}C_\rho
\right)
+s_2\left(
\eta_{\rho\mu}P_\nu
+\eta_{\rho\nu}P_\mu
-\frac{2}{D}\eta_{\mu\nu}P_\rho
\right) \,,
\\
[C_{\mu\nu},C_{\rho\sigma}]
&=
s_3\left(
\eta_{\rho\mu}J_{\sigma\nu}
+\eta_{\sigma\mu}J_{\rho\nu}
+\eta_{\rho\nu}J_{\sigma\mu}
+\eta_{\sigma\nu}J_{\rho\mu}
\right) \,,
\end{align}
where we allow for different basis choices for the special Galileon algebra which can be related by Galileon duality \cite{deRham:2013hsa,Hinterbichler:2015pqa}.\footnote{Whichever basis we use we have $[C,C_{\mu\nu}] = 0$ for the special Galileon.} The remaining commutators to add are those between $C_{\mu\nu}$ and the leading generators in the vector tree and our ansatz for those commutators is
\begin{align}
[C_{\mu\nu},B_\rho]
&=
s_4\left(
\eta_{\mu\rho}B_\nu
+\eta_{\nu\rho}B_\mu
-\frac{2}{D}\eta_{\mu\nu}B_\rho
\right) \,,
\\
[C_{\mu\nu},M_{\rho\sigma}]
&=
s_5\left(
-\eta_{\mu\rho}M_{\nu\sigma}
-\eta_{\nu\rho}M_{\mu\sigma}
+\eta_{\mu\sigma}M_{\nu\rho}
+\eta_{\nu\sigma}M_{\mu\rho}
+\frac{4}{D}\eta_{\mu\nu}M_{\rho\sigma}
\right) \\
& ~~~~~~~~~~~~ + s_6\left(
-\eta_{\mu\rho}J_{\nu\sigma}
-\eta_{\nu\rho}J_{\mu\sigma}
+\eta_{\mu\sigma}J_{\nu\rho}
+\eta_{\nu\sigma}J_{\mu\rho}
+\frac{4}{D}\eta_{\mu\nu}J_{\rho\sigma}
\right)\,.
\end{align}
We now impose the Jacobi identities and find the following non-zero commutators:
\begin{align}
[P_\mu,C_\nu]
&=\eta_{\mu\nu}C \,,\\
[P_\mu,M_{\rho\sigma}]
&=\frac12\left(
\eta_{\mu\rho}B_\sigma-\eta_{\mu\sigma}B_\rho
\right)\,,\\
[C_\mu,M_{\rho\sigma}]
&=-\frac{\alpha}{2}\left(
\eta_{\mu\rho}B_\sigma-\eta_{\mu\sigma}B_\rho
\right)\,,\\
[P_\rho,C_{\mu\nu}]
&=\eta_{\rho\mu}C_\nu+\eta_{\rho\nu}C_\mu
-\frac{2}{D}\eta_{\mu\nu}C_\rho
+a_3\left(
\eta_{\rho\mu}P_\nu+\eta_{\rho\nu}P_\mu
-\frac{2}{D}\eta_{\mu\nu}P_\rho
\right),\\
[C,C_{\mu\nu}]&=0\,,\\
[C_\rho,C_{\mu\nu}]
&=-a_3\left(
\eta_{\rho\mu}C_\nu+\eta_{\rho\nu}C_\mu
-\frac{2}{D}\eta_{\mu\nu}C_\rho
\right)\notag\\
&\quad+ \alpha \left(\alpha-2 a_3\right)\left(
\eta_{\rho\mu}P_\nu+\eta_{\rho\nu}P_\mu
-\frac{2}{D}\eta_{\mu\nu}P_\rho
\right)\,,\label{SpecialGalVectorTensor} \\ 
[C_{\mu\nu},C_{\rho\sigma}]
&=(a_3-\alpha)^2\left(
\eta_{\rho\mu}J_{\sigma\nu}
+\eta_{\sigma\mu}J_{\rho\nu}
+\eta_{\rho\nu}J_{\sigma\mu}
+\eta_{\sigma\nu}J_{\rho\mu}
\right)\,,\\
[C_{\mu\nu},B_\rho]
&=(a_3-\alpha)\left(
\eta_{\mu\rho}B_\nu+\eta_{\nu\rho}B_\mu
-\frac{2}{D}\eta_{\mu\nu}B_\rho
\right)\,,\\
[C_{\mu\nu},M_{\rho\sigma}]
&=(a_3-\alpha)\left(
\eta_{\mu\rho}M_{\nu\sigma}
+\eta_{\nu\rho}M_{\mu\sigma}
-\eta_{\mu\sigma}M_{\nu\rho}
-\eta_{\nu\sigma}M_{\mu\rho}
-\frac{4}{D}\eta_{\mu\nu}M_{\rho\sigma}
\right)\,.
\end{align}
We again need to check consistency with the remaining gauge symmetries. Since we are taking the gauge symmetries to be an ideal, we need $[C_{\mu\nu}, \text{Gauge}] = \text{Gauge}$ while our sub-algebra has $[C_\mu, \text{Gauge}] = 0$. The Jacobi identity $(B_{\mu\nu}, C_{\alpha}, C_{\rho \sigma})$ then imposes $\alpha (\alpha - 2 a_3) = 0$ so that there is no $P_{\mu}$ on the RHS of \eqref{SpecialGalVectorTensor}. This follows from $[P_{\mu},B_{\rho \sigma}] \neq 0$. We therefore require $a_3 \neq 0$ for the algebra to be non-trivial. The remaining commutators are in principle consistent with the tower of gauge symmetries as long as we have $[C_{\mu\nu}, B_{l}] = B_{l}$. 

To make further progress, we now derive the transformation rules for $\delta_{(C_{\mu\nu})} \phi$ and $\delta_{(C_{\mu\nu})}A_{\mu}$ and see if they commute with a gauge transformation back into a gauge transformation. The scalar transformation is of the schematic form $\delta \phi \sim c ( xx + x \partial \phi + \partial \phi \partial \phi)$ \cite{Hinterbichler:2015pqa} and we can fix the relative coefficients using the above commutators. We find
\begin{align} \label{ScalarCTransformation}
    \delta_{(C_{\mu\nu})} \phi = c_{\mu\nu}(x^{\mu}x^{\nu} -2 a_3 x^{\mu} \partial^{\nu}\phi ) \,,
\end{align}
with the $\partial \phi \partial \phi$ term eliminated since $\alpha (\alpha - 2 a_3) = 0$. This is clearly consistent with the gauge symmetries since the scalar does not transform under those. Guided by the coset, the vector transformation must be of the form\footnote{It is clear that in the passive form of the transformations we have $\delta A \sim c A$ and when we convert to the active form we find the structure in \eqref{VectorCAnsatz}.} 
\begin{align} \label{VectorCAnsatz}
 \delta_{(C_{\mu\nu})} A_\mu =  f_1 c_{\mu\nu} A^{\nu} + f_2 c_{\rho \sigma} x^{\rho} \partial^{\sigma} A_{\mu} \,.
\end{align}
It is then simple to see that if this transformation is to be consistent with the tower of gauge symmetries, we need $f_1 = f_2$ such that if we set $A_{\mu} = \partial_{\mu} \lambda$ we can write the transformation as a total derivative. We can then check consistency of this transformation with the other commutators. The most powerful commutator is $[C_{\mu\nu}, M_{\rho \sigma}]$, and using \eqref{MSymmetries} and \eqref{ScalarCTransformation} we find $f_1 = f_2 = - 2 a_3$ and $\alpha = 0$. If $\alpha \neq 0$ then we don't have the correct structure on the RHS of this commutator. We therefore see that if we want to constrain the Lagrangian \eqref{LagrangianLevelOne} by augmenting the symmetry algebra by the special Galileon, we must work with the abelian version of the algebra we found at level one. If we define $\beta = - 2 a_3$, then our final non-zero commutators for the special Galileon extension are
\begin{align}
[P_\mu,C_\nu]
&=\eta_{\mu\nu}C \,,\\
[P_\mu,M_{\rho\sigma}]
&=\frac12\left(
\eta_{\mu\rho}B_\sigma-\eta_{\mu\sigma}B_\rho
\right)\,,\\
[P_\rho,C_{\mu\nu}]
&=\eta_{\rho\mu}C_\nu+\eta_{\rho\nu}C_\mu
-\frac{2}{D}\eta_{\mu\nu}C_\rho
-\frac{\beta}{2}\left(
\eta_{\rho\mu}P_\nu+\eta_{\rho\nu}P_\mu
-\frac{2}{D}\eta_{\mu\nu}P_\rho
\right),\\
[C_\rho,C_{\mu\nu}]
&=\frac{\beta}{2}\left(
\eta_{\rho\mu}C_\nu+\eta_{\rho\nu}C_\mu
-\frac{2}{D}\eta_{\mu\nu}C_\rho
\right) \,,\notag\\
[C_{\mu\nu},C_{\rho\sigma}]
&=\frac{\beta^2}{4}\left(
\eta_{\rho\mu}J_{\sigma\nu}
+\eta_{\sigma\mu}J_{\rho\nu}
+\eta_{\rho\nu}J_{\sigma\mu}
+\eta_{\sigma\nu}J_{\rho\mu}
\right)\,,\\
[C_{\mu\nu},B_\rho]
&=-\frac{\beta}{2}\left(
\eta_{\mu\rho}B_\nu+\eta_{\nu\rho}B_\mu
-\frac{2}{D}\eta_{\mu\nu}B_\rho
\right)\,,\\
[C_{\mu\nu},M_{\rho\sigma}]
&=-\frac{\beta}{2}\left(
\eta_{\mu\rho}M_{\nu\sigma}
+\eta_{\nu\rho}M_{\mu\sigma}
-\eta_{\mu\sigma}M_{\nu\rho}
-\eta_{\nu\sigma}M_{\mu\rho}
-\frac{4}{D}\eta_{\mu\nu}M_{\rho\sigma}
\right)\,,
\end{align}
and the new transformation rules are
\begin{align} 
    \delta_{(C_{\mu\nu})} \phi &= c_{\mu\nu}(x^{\mu}x^{\nu} + \beta x^{\mu} \partial^{\nu}\phi ) \,, \label{AbSGscalar} \\
     \delta_{(C_{\mu\nu})} A_\mu &=  \beta (c_{\mu\nu} A^{\nu} + c_{\rho \sigma} x^{\rho} \partial^{\sigma} A_{\mu}) \,. \label{AbSGvector} 
\end{align}
This is our final level two algebra, and for $\beta \neq 0$ we see commutators that mix the scalar and vector tree in non-trivial ways. We now turn to deriving the subset of \eqref{LagrangianLevelOne} that realizes this symmetry extension. 

\subsection{Special scalar-vector Galileon interactions}
We now construct the interactions for this $D=4$ theory. We take the Lagrangian \eqref{LagrangianLevelOne} and set $\alpha = 0$, which is equivalent to replacing $G_{\mu\nu}$ with $F_{\mu\nu}$. 

We can then fix the free coefficients by demanding invariance of the action up to a total derivative under the symmetries \eqref{AbSGscalar} and \eqref{AbSGvector}. We find  
\bea {\cal L}&=& -{1\over 4}F^{\mu\nu}F_{\mu\nu}  -{3\over 2} \beta\, F^{[\mu_1}_{\ \ \mu_1} F^{\mu_2}_{\ \ \mu_2} \Pi^{\mu_3]}_{\ \ \mu_3}  -3 \beta^2 \, F^{[\mu_1}_{\ \ \mu_1} F^{\mu_2}_{\ \ \mu_2} \Pi^{\mu_3}_{\ \ \mu_3}  \Pi^{\mu_4]}_{\ \ \mu_4} \nn\\
&&-{1\over 2} \partial ^{\mu}\phi\partial_{\mu} \phi -{1\over 3}\beta {\cal L}_{\rm gal,\phi^3}-{1\over 4}\beta^2 {\cal L}_{\rm gal,\phi^4}-{1\over 5}\beta^3 {\cal L}_{\rm gal,\phi^5} \,, \eea
leaving us with only a single parameter $\beta$ that controls the non-abelian nature of the underlying algebra. These interactions agree with \cite{Kampf:2021bet}. Note that the scalar self-interactions are those of the standard special Galileon, in the field basis in which the symmetry is at most linear in $\phi$, as in \eqref{AbSGscalar} \cite{Hinterbichler:2015pqa}. Since this Lagrangian is fully fixed in terms of a single coupling constant, it is an \textit{exceptional} EFT, the \textit{special scalar-vector Galileon}. In general $D$, the subset of \eqref{GeneralDSVG} invariant under \eqref{AbSGscalar} and \eqref{AbSGvector} is
\bea {\cal L}&=& -{1\over 4}F^{\mu\nu}F_{\mu\nu} -\sum_{i=3}^{D}\frac{i(i-1)}4\,\beta^{i-2}
 F^{[\mu_1}{}_{\mu_1}F^{\mu_2}{}_{\mu_2}
 \Pi^{\mu_3}{}_{\mu_3}\cdots\Pi^{\mu_i]}{}_{\mu_i} \nn\\
&&-{1\over 2} \partial ^{\mu}\phi\partial_{\mu} \phi -\sum_{i=3}^{D+1}\frac{\beta^{i-2}}{i}\,{\cal L}_{\rm gal,\phi^i} \,. 
\eea
In terms of matrices, the vector part of the Lagrangian, the first line, can be written as
\be {\cal L}_{A}={1\over 4}\det\left(\delta+\beta\Pi\right)\Tr\left[(\delta+\beta \Pi)^{-1}F(\delta+\beta \Pi)^{-1}F\right]\, .
\ee

\section{Soft Theorems} \label{sec:SoftTheorems}

In this section we derive the soft theorems associated with all the non-linear symmetries in both the scalar-vector Galileon and the special scalar-vector Galileon theories that we have derived in the previous sections. In this section we work in general dimensions $D$. Our starting point is the Ward-Takahashi identity in the context of the S-matrix, inspired by \cite{He:2014laa, Du:2025glv},
\begin{align}
    \braket{\text{out}|Q S - S Q|\text{in}} = 0 \,,
\end{align}
where $Q$ is the Noether charge for the given symmetry, $S$ is the S-matrix, and $\bra{\text{out}},\ket{\text{in}}$ are respectively the out and in states in the interaction picture. Before diving into the manipulation of S-matrix elements, we first derive the conserved charges $Q$ for each of the non-linear symmetries, up to quadratic order in fields (higher order terms do not contribute to the Ward-Takahashi identity, since the phase oscillates rapidly in the far future/past, away from the collinear limit, and they vanish by the $i\epsilon$ prescription: fix $i\epsilon$ and send time $T \rightarrow \infty$, where $T$ is the time slice at which the charge is evaluated \cite{Hui:2018cag}). We then act with the charges on the external states to obtain soft theorems associated with each symmetry.

\subsection{Set up for the scalar-vector Galileon}

The cubic Lagrangian in our parametrization is given by
\be{\cal L}_3= \left(-{1\over 4}\alpha+{c_3\over 3}\beta\right) F_{\mu\nu}F^{\nu\rho}\Pi_{\rho}^{\ \mu}+{c_3\over 6}\beta F_{\mu\nu}F^{\mu\nu}\Pi^\rho_{\ \rho} +{3\over 4}d_3\beta(\partial\phi)^2\Box\phi \,.\ee
 The non-linear symmetry transformations derived from the coset construction are
\begin{align}
    \delta_{(M_{\mu\nu})} A_\mu = m_{\mu\nu} (x^\nu + \alpha \partial^\nu\phi),\;\;\delta_C \phi  = c + c_\mu x^\mu \,.
\end{align}
This basis for the Lagrangian is not the most ideal for deriving soft theorems, due to the cubic interactions. However, since they vanish on-shell, they can be removed by suitable field redefinitions.\footnote{Cubic amplitudes with helicities $(+-0)$ are not permitted, while those with $(++0)$ have two derivatives \cite{Arkani-Hamed:2017jhn}.} The following field redefinitions remove these cubic vertices
\be \phi\rightarrow \phi -{1\over 4}\left( {1\over 4}\alpha+{c_3\over 3}\beta\right)F_{\mu\nu}F^{\mu\nu}-{3d_3 \over 4}\beta(\partial\phi)^2\,,\ \ \ A_\mu\rightarrow A_\mu+ \left( -{1\over 4}\alpha+{c_3\over 3}\beta\right) \partial^\nu\phi \, F_{\mu\nu}\,,
\ee
and change the symmetry transformations to 
\begin{align}
    \delta_{(M_{\mu\nu})}\phi &= -\frac{\alpha'}{2}m_{\mu\nu} F^{\mu\nu} + \ldots,&
    \delta_{(M_{\mu\nu})} A_{\mu} &= m_{\mu\nu} (x^\nu + \alpha'\partial^\nu\phi) +\ldots \,, \\
    \delta_C \phi &= c + c_\mu x^\mu + \ldots,& 
    \delta_{(C_{\mu})} A_\mu &= \beta' c^\nu \partial_\nu A_\mu + \ldots \,,
\end{align}
where $\ldots$ denotes terms higher than linear order in fields and terms that are translations or gauge transformations of $A_\mu$. The gauge transformations act in the same way before and after these redefinitions. The symmetry parameters in the new basis, $\alpha'$ and $\beta'$, are related to those in the old basis by
\begin{align}
    \alpha' &= \frac{\alpha}{2} + \frac{2 c_3 \beta}{3} \,,\\
    \beta' &= - \frac{\alpha}{4} + \frac{c_3 \beta}{3} - \frac{3 d_3 \beta}{2} \,.
\end{align} 
The divergence of the Noether current, up to quadratic order in fields, is given by Noether's second theorem:
\begin{align}
    \partial_\mu J_{M}^\mu &= \frac{\delta\mathcal{L}}{\delta\phi}\delta_{M}\phi + \frac{\delta\mathcal{L}}{\delta A_\mu} \delta_{M} A_\mu \nonumber\\
    &=  -\biggl(\frac{1}{2}\alpha' m_{\mu\nu}\Box{\phi}F^{\mu\nu} + m_{\mu\rho} x^\rho \partial_\nu F^{\mu\nu} + \alpha' m_{\mu\rho} \partial_\nu F^{\mu\nu}\partial^{\rho}\phi\biggr)   + \ldots \nonumber\\
    &= m_{\mu\nu} \biggl\{- \partial_{\rho}\biggl(x^\nu F^{\mu\rho}\biggr) + F^{\mu\nu} + \alpha'\biggl[\partial_\rho\biggl(\partial^\nu A^\mu \partial^\rho\phi + \partial^\rho A^\mu \partial^\nu\phi \biggr) - \partial^\nu\biggl(\partial_\rho A^\mu \partial^\rho\phi \biggr) \nonumber\\
    &\;\;\;\;-\partial^\mu\biggl(\partial_\rho A^\rho \partial^\nu\phi\biggr) \biggr] \bigg\}+\ldots\\
    \partial_\mu J_{C}^\mu &= \frac{\delta\mathcal{L}}{\delta\phi}\delta_{C} \phi + \frac{\delta\mathcal{L}}{\delta A_\mu} \delta_{C} A_\mu \nonumber\\
    &= - c_\mu \biggl(- x^\mu\Box\phi +\beta' \partial_\nu F^{\rho\nu} \partial^\mu A_\rho \biggr) + \ldots\nonumber\\
&= c_\mu\biggl[
\partial_\nu\biggl(
x^\mu\partial^\nu\phi
-\beta' F^{\rho\nu}\partial^\mu A_\rho
\biggr)
-\partial^\mu\phi
-\frac{\beta'}{4}\partial^\mu
\biggl(F_{\rho\sigma}F^{\rho\sigma}\biggr)
\biggr]+\ldots\,, \label{lasteqfisee}
\end{align}
where $\ldots$ denotes terms higher than quadratic order in fields. The corresponding conserved charges up to quadratic order in fields are therefore
\begin{align}
    Q_{M}
&=
m_{\mu\nu}\int_{\vec{x}}\,
-x^\nu F^{\mu 0}
+2\eta^{0\mu}A^\nu\nonumber\\
&\qquad\qquad + \alpha'\biggl[
\partial^\nu A^\mu\,\partial^0\phi
+\partial^0 A^\mu\,\partial^\nu\phi
-\eta^{0\nu}\biggl(
\partial_\rho A^\mu\,\partial^\rho\phi
\biggr)
-\eta^{0\mu}\biggl(
\partial_\rho A^\rho\,\partial^\nu\phi
\biggr)
\biggr]+\ldots \,,\label{NAMcharge}\\
Q_{C}
&=
c_\mu
\int_{\vec{x}}\,
\biggl[
x^\mu\partial^0\phi
-\eta^{\mu0}\phi
-\beta'
\biggl(
F^{\rho 0}\partial^\mu A_\rho
+\frac14\eta^{\mu0}F_{\rho \sigma}^2
\biggr)
\biggr]+ \ldots \,.\label{NACcharge}
\end{align}
There is no need to consider the constant shift on the scalar since we know that leads to the usual Adler zero condition \cite{Adler:1965ga} stating that all amplitudes vanish when the momentum of an external scalar is taken soft. 
\subsection{Set up for the special scalar-vector Galileon theory}

The special scalar-vector Galileon enjoys the following symmetries
\begin{align}
    &\delta_{(M_{\mu\nu})} A_\mu = m_{\mu\nu} x^\nu,\;\;\delta_C \phi = c + c_\mu x^\mu + c_{\mu\nu}(x^\mu x^\nu + \beta x^\mu\partial^\nu\phi),\;\;\delta_{(C_{\mu\nu})} A_\mu = \beta (c_{\nu\rho} x^\nu \partial^\rho A_\mu + c_{\mu\nu} A^\nu) \,.
\end{align}
Upon removing the cubic vertices using the following field redefinitions
\begin{align}
    \phi \rightarrow \phi + \frac{\beta}{4}\partial_\mu\phi \partial^\mu\phi + \frac{\beta}{8} F^{\mu\nu} F_{\mu\nu} ,\;\;\; A_\mu \rightarrow A_\mu - \frac{\beta}{2} F_{\mu\nu}\partial^\nu \phi \,,
\end{align}
the new symmetry transformations are
\begin{align}
    \delta_{(M_{\mu\nu})} A_\mu &= m_{\mu\nu} (x^\nu - \beta \partial^\nu\phi)+\ldots,&\delta_{(M_{\mu\nu})} \phi &= \frac{\beta}{2} m_{\mu\nu} F^{\mu\nu}+\ldots \\
    \delta_C \phi &= c + c_\mu x^\mu + c_{\mu\nu} x^{\mu} x^{\nu} \ldots,& 
\end{align}
where $\ldots$ denotes terms quadratic and higher order in the fields, and again we have omitted translations and gauge transformations. Again, the form of the gauge symmetries does not change. Notably, the scalar shift symmetry reduces to the ordinary special Galileon symmetry when modding translation out. The divergences of the currents are then
\begin{align}
    \partial_\mu J^\mu_{M} &=  m_{\mu\nu} \biggl\{ -\partial_{\rho}\biggl(x^\nu F^{\mu\rho}\biggr) + F^{\mu\nu} - \beta\biggl[\partial_\rho\biggl(\partial^\nu A^\mu \partial^\rho\phi + \partial^\rho A^\mu \partial^\nu\phi \biggr) - \partial^\nu\biggl(\partial_\rho A^\mu \partial^\rho\phi \biggr) \nonumber\\
    &\;\;\;\;-\partial^\mu\biggl(\partial_\rho A^\rho \partial^\nu\phi\biggr) \biggr] \bigg\}+\ldots \,,\\
    \partial_\mu J_{C}^\mu &= c_\mu \biggl[ \partial_\nu\biggl(x^\mu\partial^\nu\phi\biggr)-\partial^\mu\phi\biggr] 
    +c_{\mu\nu}\partial_\rho \biggl( x^\mu x^\nu \partial^\rho\phi - 2\eta^{\mu\rho} x^\nu \phi\biggr) + \ldots \,,
\end{align}
where we have invoked the traceless condition of $c_{\mu\nu}$ to arrive at the last term. The respective conserved charges are therefore
\begin{align}
    Q_{M}
=&
m_{\mu\nu}\int_{\vec{x}}\,
-x^\nu F^{\mu 0}
+2\eta^{0\mu}A^\nu
\nonumber\\
&-\beta\biggl[
\partial^\nu A^\mu\,\partial^0\phi
+\partial^0 A^\mu\,\partial^\nu\phi
-\eta^{0\nu}\biggl(
\partial_\rho A^\mu\,\partial^\rho\phi
\biggr)
-\eta^{0\mu}\biggl(
\partial_\rho A^\rho\,\partial^\nu\phi
\biggr)
\biggr]
+\ldots \,,\label{AMcharge}\\
Q_{C}
=&
c_\mu\int_{\vec{x}}\,\biggl(
x^\mu\partial^0\phi
-\eta^{\mu0}\phi\biggr)\nonumber
+ c_{\mu\nu}
\int_{\vec{x}}\,
\biggl(
x^\mu x^\nu\partial^0\phi
- 2\eta^{\mu0}x^\nu\phi
\biggr)
+\ldots \,.
\end{align}
Since the conserved charge for the special Galileon shift symmetry lacks terms quadratic in the fields, the resulting soft theorems are enhanced Adler zeros \cite{Cheung:2014dqa, Hinterbichler:2015pqa}.

\subsection{Derivation of the soft theorems}
We now explicitly illustrate the derivation of the soft theorem  associated with the vector two-form symmetry (the derivations for the other symmetries can be obtained in a similar manner). Generically, we can expand the charge in powers of the fields:
\begin{align}
    Q = Q^{(1)} + Q^{(2)}+\ldots,
\end{align}
where $Q^{(n)}$ denotes the term with $n$ powers of the fields. For the two-form symmetry, the terms linear in the field are given by 
\begin{align}
    Q^{(1)}_{M} &=  -\int_{\vec{x}} \left(m_{i \nu}x^{\nu}F^{i0} - 2m_{i0}A^i \right)\nonumber\\
    &= -\int_{\vec{x}} \left[ m_{i0}\biggl(x^0\partial^i A^0 - x^0\partial^0 A^i - 2A^i\biggr)+ m_{i j}x^j(\partial^i A^0 - \partial^0 A^i) \right] \,.
\end{align}
Since $Q$ is conserved, it can be evaluated at any given time slice, and we choose to evaluate it at the far future/past, where the fields are well approximated by the mode expansion of the free theory,
\begin{align}
    A^\mu(t,\vec{x}) = \sum_{h}\int_{\vec{q}} \frac{1}{\sqrt{2E_q}} e^{-i E_q t + i \vec{q}\cdot\vec{x}} \epsilon^{\mu}_{h}(\vec{q}) a_{\vec{q},h} + \text{c.c.}\,.
\end{align}
Here $\epsilon_h^\mu$ is the polarization vector associated with helicity $h$ that labels the vector representation of the little group $ISO(D-2)$. It obeys the following transverse and orthonormal conditions
\begin{align}
    \epsilon_{h}^\mu(p)p_\mu = 0, \quad \eta_{\mu\nu}\epsilon_{h}^{\mu}(p)\epsilon_{h'}^{*\nu}(p) = \delta_{hh'} \,,
\end{align}
which ensures decoupling among different helicity modes in the free theory. The conserved charge after performing the position space integral is then
\begin{align}
    &Q^{(1)}_{M} \nonumber\\
    = &-\lim_{q\rightarrow 0}\biggl\{ m_{i0} \biggl[iT \biggl(\frac{q^i}{2E_q}\epsilon^0(\vec{q}) - \frac{\epsilon^{i}(\vec{q})}{2}\biggr) - \frac{\epsilon^i(\vec{q})}{E_q} \biggr]\sqrt{2E_q} a_{\vec{q}}  -  m_{ij}\partial_{q_j}\biggl[\biggl( \frac{q^i}{2E_q}\epsilon^0(\vec{q}) - \frac{\epsilon^{i}(\vec{q})}{2}   \biggr)\sqrt{2 E_q} a_{\vec{q}}\biggr] + \text{c.c.}\,\biggr\},
\end{align}
where we have kept the helicity sum implicit. Here we can see that the soft limit naturally arises as the spatial integral anchors the momentum at zero value by the Dirac delta function. We expand the S-matrix in terms of the scattering amplitude as
\begin{align}
    \braket{\text{out}|S|\text{in}} = 1 + i (2\pi)^D\mathcal{A}\;\delta^d(\vec{p})\;\delta(E) \,,
\end{align}
where the total energy momentum $(E,\vec{p})$ is defined by $\vec{p}\equiv \vec{p}_{out} - \vec{p}_{in}$ and $E\equiv E_{out} - E_{in}$. Plugging the charge into the conservation matrix element, we have
\begin{align}
   \frac{1}{(2\pi)^D} i\braket{\text{out}|[Q^{(1)}_{M},S] |\text{in}} &=\sum_{h} m_{i0} \biggl[iT \biggl(\frac{q^i}{2E_q}\epsilon^{0}_{h}(\vec{q}) - \frac{\epsilon^{i}_{h}(\vec{q})}{2}\biggr) - \frac{\epsilon^{i}_{h}(\vec{q})}{E_q} \biggr]\mathcal{A}_{E_q,h}(\vec{q})\delta^d(\vec{p}+\vec{q})\delta(E+E_q)\nn\\
    &-m_{i0} \biggl[-iT \biggl(\frac{q^i}{2E_q}\epsilon^{*0}_{-h}(\vec{q}) - \frac{\epsilon^{*i}_{-h}(\vec{q})}{2}\biggr) - \frac{\epsilon^{*i}_{-h}(\vec{q})}{E_q} \biggr]\mathcal{A}_{-E_q, h}(-\vec{q})\delta^d(\vec{p}-\vec{q})\delta(E - E_q)\nn\\
    &-m_{ij}\partial_{q_j}\biggl[\biggl( \frac{q^i}{2E_q}\epsilon^{0}_{h}(\vec{q}) - \frac{\epsilon^{i}_{h}(\vec{q})}{2}   \biggr)\mathcal{A}_{E_q, h}\delta^d(\vec{p}+\vec{q})\delta(E+E_q)\biggr]\nn\\
    &+m_{ij}\partial_{q_j}\biggl[\biggl( \frac{q^i}{2E_q}\epsilon^{*0}_{-h}(\vec{q}) - \frac{\epsilon^{*i}_{-h}(\vec{q})}{2}   \biggr) \mathcal{A}_{-E_q, h}\delta^d(\vec{p} - \vec{q})\delta(E - E_q)\biggr] \,,
\end{align}
where $\mathcal{A}_{E_q, h}(\vec{q})$ denotes the scattering amplitude that corresponds to the process $\braket{\text{out}, (\vec{q}, h)|S|\text{in}}$ while $\mathcal{A}_{-E_q, h}(-\vec{q})$ corresponds to the process $\braket{\text{out}|S|(\vec{q}, -h),\text{in}}$ . To compute the last two lines, we impose the momentum conserving delta functions $\delta^d(\vec{p}\pm\vec{q})$ to eliminate the $\vec{q}$ dependence in the front factor, so that we can commute it with the $q$ derivative. Schematically
\begin{align}
    \lim_{q\rightarrow 0}\partial_{q} [f(q) \delta(p\pm q)] = \lim_{q\rightarrow 0}f(\mp p)\partial_q\delta(p\pm q) =\lim_{q\rightarrow 0}\pm f(\mp p)\partial_p\delta(p\pm q) = \pm f(\mp p)\partial_p\delta(p) \,.
\end{align}
Invoking the delta function chain rule
\begin{align}
    f(x)\delta'(x) = f(0)\delta'(x) - f'(0)\delta(x) \,,
\end{align}
we then have
\begin{align}
   \frac{1}{(2\pi)^D}  i\braket{\text{out}|[Q^{(1)}_{M},S] |\text{in}} &=  \sum_{h}\lim_{p\rightarrow 0}\biggl\{m_{i0}\biggl[iT \biggl(\frac{-p^i}{2E_p}\epsilon^{0}_{h}(-\vec{p}) - \frac{\epsilon^{i}_{h}(-\vec{p})}{2}\biggr) - \red{\frac{\epsilon^{i}_{h}(-\vec{p})}{E_p} \biggr] \mathcal{A}_{E_p,h}}\biggr\}\delta(E)\delta^d(\vec{p})\nn\\
    &-\lim_{p\rightarrow 0} \biggr\{m_{i0}\biggl[-iT \biggl(\frac{p^i}{2E_p}\epsilon^{*0}_{h}(\vec{p}) - \frac{\epsilon^{*i}_{-h}(\vec{p})}{2}\biggr) - \red{\frac{\epsilon^{*i}_{-h}(\vec{p})}{E_p} \biggr]\mathcal{A}_{-E_p,h}}\biggr\}\delta(E)\delta^d(\vec{p})\nn\\
    &-\lim_{p\rightarrow 0}\biggl[m_{ij}\biggl( \frac{-p^i}{2E_p}\epsilon^{0}_{h}(-\vec{p}) - \frac{\epsilon^{i}_{h}(-\vec{p})}{2}   \biggr)\mathcal{A}_{E_p, h}\biggr]\delta(E)\partial_{p_j}\delta^d(\vec{p})\nn\\
    &-\lim_{p\rightarrow 0}\biggl[m_{ij}\biggl( \frac{p^i}{2E_p}\epsilon^{*0}_{-h}(\vec{p}) - \frac{\epsilon^{*i}_{h}(\vec{p})}{2}   \biggr)\mathcal{A}_{-E_p, h}\biggr]\delta(E)\partial_{p_j}\delta^d(\vec{p})\nn\\
    &+\lim_{p\rightarrow 0}\red{\partial_{p_j}\biggl[m_{ij}\biggl( \frac{-p^i}{2E_p}\epsilon^{0}_{h}(-\vec{p}) - \frac{\epsilon^{i}_{h}(-\vec{p})}{2}   \biggr)\mathcal{A}_{E_p, h}\biggr]}\delta(E)\delta^d(\vec{p})  \nn\\ 
    &+\lim_{p\rightarrow 0}\red{\partial_{p_j}\biggl[m_{ij}\biggl( \frac{p^i}{2E_p}\epsilon^{*0}_{-h}(\vec{p}) - \frac{\epsilon^{*i}_{-h}(\vec{p})}{2}   \biggr)\mathcal{A}_{-E_p,h}\biggr]} \delta(E)\delta^d(\vec{p})\label{LHSVector2}\nn\\
    &+\lim_{p\rightarrow 0}\partial_{p_j}\biggl[m_{ij}\biggl( \frac{-p^i}{2E_p}\epsilon^{0}_{h}(-\vec{p}) - \frac{\epsilon^{i}_{h}(-\vec{p})}{2}   \biggr)\mathcal{A}_{E_p, h}E_p\biggr]\delta'(E)\delta^d(\vec{p})\nn\\
    &-\lim_{p\rightarrow 0}\partial_{p_j}\biggl[m_{ij}\biggl( \frac{p^i}{2E_p}\epsilon^{*0}_{-h}(\vec{p}) - \frac{\epsilon^{*i}_{-h}(\vec{p})}{2}   \biggr)\mathcal{A}_{-E_p,h}E_p\biggr] \delta'(E)\delta^d(\vec{p}) +\ldots \,.
\end{align}
We imposed energy conservation stemming from the energy delta function $\delta(E)$ onto the scattering amplitude $\mathcal{A}$. In practice, this corresponds to going to the minimal basis where all on-shell conditions and energy-momentum conservation are imposed. Moreover, we make the spatial momentum dependence implicit, i.e. $\mathcal{A}_{E_p,h}\equiv\mathcal{A}_{E_p,h}(-\vec{p})$, since they are localized to the same value for every amplitude. The last two terms and ellipses come from Taylor expanding the energy delta function $\delta(E\pm E_p) = \delta(E) \pm E_p \delta'(E)+ \ldots$. The highlighted terms will be responsible for the soft theorems, while the un-highlighted terms will be responsible for the Adler zero condition associated with the vector shift symmetry. As we shall see, only the highlighted terms are proportional to the full energy-momentum conserving $\delta^D(p)$, which have counterparts from the quadratic-in-field charge $Q^{(2)}$. 

Moving on to the quadratic-in-field expansion of the conserved charge $Q^{(2)}_{M}$, the position space representation is given by
\begin{align}
    Q^{(2)}_{M} = \alpha'\int_{\vec{x}} \;m_{\mu\nu}
    \biggl[
\partial^\nu A^\mu\,\partial^0\phi
+\partial^0 A^\mu\,\partial^\nu\phi
-\eta^{0\nu}\biggl(
\partial_\rho A^\mu\,\partial^\rho\phi
\biggr)
-\eta^{0\mu}\biggl(
\partial_\rho A^\rho\,\partial^\nu\phi
\biggr)
\biggr] \,,
\end{align}
where the last term can be omitted since it is a pure gauge. Invoking the free mode expansion in the far future/past, the charge is schematically
\begin{align}
    Q^{(2)}_{M} &= \int_{\vec{p},\vec{q}}\;e^{-i (E_p - E_q)T}\int_{\vec{x}} \;e^{i(\vec{p}-\vec{q})\cdot\vec{x}}\biggl[..\biggr] + \text{other ordering}\\
    &= \int_{\vec{p}}\; \biggl[..\biggr] \,, \label{Q2m}
\end{align}
where momentum $\vec{q}$ is localized to $\pm\vec{p}$ by the delta function $\delta^d(\vec{p}-\vec{q})$, and the frequency phase $e^{i(E_p - E_q)T}$ nicely converges to $1$ which makes it possible for $Q^{(2)}$ to be non-trivial, otherwise the integral trivially vanishes due to the oscillation of the phase factor in the far past/future.  

We therefore have the following momentum space representation of the conserved charge
\begin{align}
    Q^{(2)}_{M} = \alpha'm_{\mu\nu}\sum_{h}\int_{\vec{p}} \;\epsilon^{\mu}_{h}(\vec{p}) p^{\nu} \biggl[a^\dag_{\vec{p},0} a_{\vec{p},h}\biggr] + \text{h.c.} + \text{other ordering}\,,
\end{align}
where the first two terms in \eqref{Q2m} merge into one upon mode expansion, while the third term vanishes by the on-shell condition. For the sake of simplicity, we assume that all particles lie in the out state, so only the first ordering contributes to the delta structure $\delta^d(\vec{p})\delta(E)$ that appears in $Q^{(1)}$. When the scalar creation operator hits the out state, it annihilates one of the scalar hard momenta and localizes the momentum of the conserved charge to the annihilated momentum $p_a$, while the annihilation operator creates a vector with the summation of helicity; conversely, when the vector creation operator hits the out state, it annihilates an external vector and localizes the momentum and helicity of the conserved charge to the annihilated momentum and helicity respectively. Effectively, the charge component $Q^{(2)}$ rotates scalars into vectors and vice versa. Therefore the commutator reads
\begin{align}
    - \frac{1}{(2\pi)^D}  i \braket{\text{out}|[Q^{(2)}_{M},S]|\text{in}} = \alpha'm_{\mu\nu}\biggl\{\sum_{h}\sum_{a\in \phi}\biggl[\epsilon^{\mu}_{h}(\vec{p}_a) p_a^{\nu}\biggr] \mathcal{A}_{a,h}^n + \sum_{b\in A_\mu}\biggl[\epsilon^{*\mu}_{h_b}(\vec{p}_b) p_b^{\nu} \biggr] \mathcal{A}_{b}^n \biggr\}\delta(E)\delta^d(\vec{p})\,.
\end{align}

By matching the independent delta function structures between $Q^{(1)}$ and $Q^{(2)}$ while recognising only the highlighted part in $Q^{(1)}$ possesses the full energy-momentum conserving delta function, we arrive at the soft theorems:\footnote{We note that when $\alpha'=0$, all soft theorems reduce to enhanced Adler zeros.}
\begin{keyeqn}
\begin{align}
&\sum_{h} \lim_{\vec{p}\rightarrow 0}\biggl(\frac{\epsilon^0_h(\vec{p})p^{i}}{2E_p}-\frac{\epsilon^{i}_{h}(\vec{p})}{2}\biggr)\mathcal{A}_{E_p, h}^{n+1} = 0,\;\sum_{h} \lim_{\vec{p}\rightarrow 0}\biggl(-\frac{\epsilon^{*0}_{-h}(-\vec{p}) p^{i}}{2E_p} - \frac{\epsilon^{*i}_{-h}(-\vec{p})}{2}\biggr)\mathcal{A}_{-E_p, h}^{n+1} = 0 \,,\\
    &\sum_{h}\lim_{\vec{p}\rightarrow 0}\frac{\epsilon^{i}_{h}(\vec{p})\mathcal{A}_{E_p, h}^{n+1}- \epsilon^{*i}_{-h}(-\vec{p})\mathcal{A}_{-E_p, h}^{n+1}}{E_p} \nonumber\\
    =& -\alpha' \biggl\{\sum_{h}\sum_{a\in \phi}\biggl[ \epsilon^i_{h}(\vec{p}_a) E_a - \epsilon^0_{h}(\vec{p}_a) p^i_a \biggr] \mathcal{A}_{a,h}^n + \sum_{b\in A_\mu}\biggl[ \epsilon^{*i}_{h_b}(\vec{p}_b) E_b - \epsilon^{*0}_{h_b}(\vec{p}_b) p^i_b \biggr] \mathcal{A}_{b}^n \biggr\} \,,\label{VectorEnergySoftTheorem}\\
    &\sum_{h}\lim_{\vec{p}\rightarrow 0}\partial_{p_{[j}}\biggl[\biggl(\frac{\epsilon^0_h(\vec{p})p^{i]}}{2E_p}-\frac{\epsilon^{i]}_{h}(\vec{p})}{2}\biggr)\mathcal{A}_{E_p, h}^{n+1} + \biggl(-\frac{\epsilon^{*0}_{-h}(-\vec{p}) p^{i]}}{2E_p} - \frac{\epsilon^{*i]}_{-h}(-\vec{p})}{2}\biggr)\mathcal{A}_{-E_p, h}^{n+1} \biggr]\nonumber\\
    =& -\alpha'\biggl\{\sum_{h}\sum_{a\in \phi}\biggl[\epsilon^{[i}_{h}(\vec{p}_a) p_a^{j]}\biggr] \mathcal{A}_{a,h}^n + \sum_{b\in A_\mu}\biggl[\epsilon^{*[i}_{h_b}(\vec{p}_b) p_b^{j]} \biggr] \mathcal{A}_{b}^n \biggr\}\label{VectorMomentumSoftTheorem}\,.
\end{align}
\end{keyeqn}
 The first line represents the usual Adler zeros for the soft vector emission, emanating from the constant shift component of the gauge symmetry. The second line represents the spatial-temporal component of the soft theorem, associated with the vector shift symmetry parameter $m_{i0}$. The third line represents the spatial-spatial component of the soft theorem, associated with the vector shift symmetry parameter $m_{ij}$ which is anti-symmetric in $i$ and $j$. At first it might seem strange that our derivation yields the Adler zero associated with the constant shift component of the gauge symmetry, but it is not possible to have invariance under $M_{\mu\nu}$ and not $B_{\mu}$, since the IHCs mandate the commutator $[P, M] = B$. The soft theorems therefore nicely capture the necessity of this commutator by carefully dealing with the momentum-conserving delta functions \cite{Du:2024hol}. The right-hand-side of the soft theorem can be combined into a Lorentz-covariant form
 \begin{align}
...=-\alpha'\biggl\{\sum_{h}\sum_{a\in \phi}\biggl[\epsilon^{[\mu}_{h}(\vec{p}_a) p_a^{\nu]}\biggr] \mathcal{A}_{a,h}^n + \sum_{b\in A_\mu}\biggl[\epsilon^{*[\mu}_{h_b}(\vec{p}_b) p_b^{\nu]} \biggr] \mathcal{A}_{b}^n \biggr\} \,.
 \end{align}
 A notable feature is that the soft theorem requires summing over all the helicities of the inserted vector mode, however in practice, as we shall see in the later section and also noted in \cite{Kampf:2021bet}, different helicity components nicely diagonalise, since $\bra{}$ and $[|$ are linearly independent. It is therefore sufficient to consider soft theorems for each helicity mode. 
 Moreover, the left-hand-side of the spatial soft theorem is gauge invariant, i.e., the factors
 \begin{align}
     \frac{\epsilon^0_h(\vec{p})p^{i}}{2E_p}-\frac{\epsilon^{i}_{h}(\vec{p})}{2}, \quad \;-\frac{\epsilon^{*0}_{-h}(-\vec{p}) p^{i}}{2E_p} - \frac{\epsilon^{*i}_{-h}(-\vec{p})}{2}\,,
 \end{align}
 are invariant under the gauge transformation
 \begin{align}
     \epsilon^0(\vec{p})\rightarrow \epsilon^0(\vec{p}) + \lambda E_p,\;\epsilon^i(\vec{p}) \rightarrow \epsilon^i (\vec{p}) + \lambda p^i\,,
 \end{align}
 for any complex number $\lambda$. The left-hand-side of the spatial-temporal soft theorem is also gauge invariant, thanks to the Adler zero condition mandated by the constant shift symmetry. This completes the derivation of the soft theorems associated with the two-form symmetry for the scalar-vector Galileon theory. 
 
 A parallel construction can be done for the charge $Q_{C}$ and the resulting soft theorems are \footnote{Analogous to the vector case, when $\beta' = 0$, the scalar soft theorems reduce to enhanced Adler zeros.} 
\begin{keyeqn}
    \begin{align}
    \lim_{\vec{p}\rightarrow 0}\frac{\mathcal{A}^{n+1}_{E_p, 0} - \mathcal{A}^{n+1}_{-E_p, 0}}{2E_p} &= -\beta'\sum_{a\in A_\mu} E_a \mathcal{A}^n \,,\label{ScalarEnergySoftTheorem}\\ \lim_{\vec{p}\rightarrow 0}\partial_{p_i}\biggl(\frac{\mathcal{A}^{n+1}_{E_p, 0} + \mathcal{A}^{n+1}_{-E_p, 0}}{2}\biggr) &= \beta'\sum_{a\in A_\mu} p_a^i \mathcal{A}^n \,,\label{ScalarMomentumSoftTheorem}
\end{align}
\end{keyeqn}
where the subscript $0$ in $\mathcal{A}_{E_p, 0}$ denotes the soft scalar (helicity $0$) emission. 

All of these soft theorems are valid in any spacetime dimension $D$. At the level of the solutions, these soft theorems can be shown to be identical to those derived in \cite{Kampf:2021bet} using the spinor helicity formalism in $D=4$. 

The soft theorems for vector emission in special scalar-vector Galileon theory can be directly obtained by setting $\alpha' \rightarrow -\beta$, since the conserved charges for the scalar-vector Galileon \eqref{NAMcharge} and special scalar-vector Galileon \eqref{AMcharge} share identical structures. For the scalar emission, the conserved charge is identical to those derived in \cite{Du:2024hol}, and the resulting soft theorems read
\begin{align}
     \lim_{\vec{p}\rightarrow 0}\frac{\mathcal{A}_{E_p, 0} - \mathcal{A}_{-E_p, 0}}{2E_p} &= 0 \,,\\ \lim_{\vec{p}\rightarrow 0}\partial_{p_i}\biggl(\frac{\mathcal{A}_{E_p, 0} + \mathcal{A}_{-E_p, 0}}{2}\biggr) &= 0 \,,\\
    \lim_{\vec{p}\rightarrow 0}\partial_{p_i}\biggl(\frac{\mathcal{A}_{0,E_p} - \mathcal{A}_{0,-E_p}}{2E_p}\biggr) &= 0 \,,\\
    \lim_{\vec{p}\rightarrow 0}\partial_{p_i}\partial_{p_j} \biggl(\frac{\mathcal{A}_{0,E_p} + \mathcal{A}_{0,-E_p}}{2}\biggr) &=0 \,.
\end{align}
Again, we see a tower structure that is mandated by the IHCs \cite{Du:2024hol}. The solution to these soft theorems is the usual enhanced soft limit of the special Galileon \cite{Cheung:2014dqa, Hinterbichler:2015pqa} 
\begin{align}
    \mathcal{A}\sim\mathcal{O}(p^3) \,.
\end{align}
With the soft theorems at hand, we can now attempt to solve them to find the resulting amplitudes. The primary difference here compared to the derivations in \cite{Kampf:2021bet} is that our soft theorems hold in general dimensions and are explicitly derived from the symmetry via Noether's theorem, rather than fed as an input to the soft bootstrap.

\section{Scattering Amplitudes and Soft Bootstrap} \label{sec:SoftBootstrap}

In this section, we aim to bootstrap the scattering amplitudes of the scalar-vector Galileon theory using the soft theorems derived in Section \ref{sec:SoftTheorems}. Since all cubic vertices can be removed, three-point amplitudes vanish identically, which implies that no poles are present in the four-point or five-point amplitudes. We will work in $D=4$, where we have the benefit of spinor helicity variables:
\begin{align}
    \ket{a}_{\alpha}[a|_{\Bar{\beta}} = p_{\mu}\sigma^{\mu}_{\alpha \Bar{\beta}} = -E_a \sigma^0_{\alpha \Bar{\beta}} +\vec{p}_a\cdot  \vec{\sigma}_{\alpha \Bar{\beta}} \,,
\end{align}
where $\sigma^{\mu}$ are the usual two by two Pauli matrices.\footnote{ The indices are raised (lowered) with respect to the $SL(2,{\mathbb{C}})$ invariant tensor $\epsilon^{\alpha \beta}$ ($\epsilon_{\alpha \beta}$), for example,
\begin{align}
\bra{a}^{\alpha}=\epsilon^{\alpha\beta}\ket{a}_\beta, \quad |a]^{\Bar{\alpha}} = \epsilon^{\Bar{\alpha}\Bar{\beta}}[a|_{\Bar{\beta}} \,.
\end{align}
The inner products are thus
\begin{align}
    \braket{ab}\equiv \bra{a}^\alpha \ket{b}_\alpha, \quad [ab]\equiv [a|^{\Bar{\alpha}}|b]_{\Bar{\alpha}} \quad \langle ab]=[ab\rangle=0 \,.
\end{align}
The little group transformation $SO(2)$ which leaves the momenta unchanged is given by
\begin{align}
    \bra{a}\rightarrow \lambda_a \bra{a},\;|a]\rightarrow \frac 1\lambda_a|a]\,,
\end{align}
 where $\lambda_a$ is a generic complex number for complex momenta. The all-out scattering amplitude transforms covariantly under little group scaling as 
 \begin{align}
     \mathcal{A}(1^{h_1},2^{h_2},\ldots,n^{h_n})\rightarrow \prod_{i=1}^n \lambda_i^{2 h_i} \mathcal{A}(1^{h_1},2^{h_2},\ldots,n^{h_n}) \,,\label{Littlegroupscaling}
 \end{align}
 where $h_i$ denotes the helicity of the particle $i$. A more comprehensive discussion of spinor helicity variables can be found in \cite{Cheung:2017pzi, Elvang:2013cua}.} To write an ansatz for the amplitudes, we need to work with a particular power counting. We build amplitudes with momenta scaling $2n-2$, where $n$ is the number of external legs. We first consider four-point scattering, then move to five-point. We also put all particles in the \textit{out} state.

 \subsection{Four-point amplitudes}
 
 We begin with four-point amplitudes and use the little group scaling \eqref{Littlegroupscaling} to fix the general structures. We assume that the underlying theory is parity invariant. There are six possible independent scattering configurations and they are given by\footnote{We label these particles from $2$ to $5$ rather than $1$ to $4$ since we will be matching the soft limit of a five-point amplitude to a four-point one, and will take particle $1$ soft leaving us with particles $2, \ldots, 5$.}
\begin{align}
    2\,\;3\,\;4\,\;5\;\,&\nonumber\\
    \mathcal{A}(--\phi\phi) &= [23]^2 (c^{--00}_1 s_{23}^2 + c^{--00}_2 s_{24} s_{34}) \,,\\
    \mathcal{A}(-+\phi\phi) &= c^{-+00}_1 [24]^2\braket{34}^2 s_{23} \,,\\
    \mathcal{A}(----)
    &= c^{----}_1 [23]^2[45]^2 s_{23}  + 2\;\text{perms} \,,\\
    \mathcal{A}(--++)
    &= c^{--++}_1 [23]^2\braket{45}^2 s_{23} \,,\\
    \mathcal{A}(+---)
    &= c^{+---}_1 \braket{23}^2[34]^2[35]^2 \,,\\
    \mathcal{A}(\phi\phi\phi\phi)&= c_1^{0000} s_{23}s_{24} s_{34}\,,
\end{align}
where the $c$'s are the Wilson coefficients and the Mandelstam variables are defined by $s_{ab} =  2 \eta_{\mu\nu}p_a^\mu p_b^\nu $. All other amplitudes can be related via parity.

We now impose the soft theorems \eqref{ScalarEnergySoftTheorem} and \eqref{VectorEnergySoftTheorem} (by Lorentz symmetry, the others are automatically satisfied once these are). The right-hand-sides of both equations vanish for $n=3$ due to the vanishing cubic amplitudes, and the soft theorems reduce to Adler zero conditions
\begin{align}
    \mathcal{A}_4\sim \mathcal{O}(p^2)\,,
\end{align}
for both soft scalar and vector external momenta. The quadratic scaling is fixed by the symmetries being linear in the coordinates. Consequently, the only constraint imposed by the four-point soft theorem is given by
\begin{align}
\boxed{
    c_1^{+---} = 0 \,.}\label{4ptSoftTheorems}
\end{align}
All other Wilson coefficients remain unfixed due to the enhanced soft behavior of the amplitudes.
The four-point coefficients derived explicitly from the Lagrangian of scalar-vector Galileon \eqref{LagrangianLevelOne} are given by
\begin{align}
c_1^{--00}
&= \frac{9\alpha^2+54\alpha d_3\beta-16c_3^2\beta^2+72c_3d_3\beta^2}{576},
\qquad
c_2^{--00}
= \frac{-9\alpha^2+16c_3^2\beta^2+12c_4\beta^2}{288}
\,, \nonumber\\
c_1^{-+00}
&= \frac{9\alpha^2-24\alpha c_3\beta+16c_3^2\beta^2+24c_4\beta^2}{576}
\,, \nonumber\\
c_1^{----}
&= \frac{9\alpha^2+24\alpha c_3\beta+16c_3^2\beta^2}{576},
\qquad
c_1^{--++}
= \frac{9\alpha^2+24\alpha c_3\beta+16c_3^2\beta^2}{576},
\qquad
c_1^{+---}=0
\,, \nonumber\\
c_1^{0000}
&= \frac{27d_3^2\beta^2+16d_4\beta^2}{16}
\,.
\label{LagrangianAmp4pt}
\end{align}
with the vanishing of $c_1^{+---}$ agreeing with our soft theorem result. Other relations between these four-point coefficients cannot be fixed by four-point amplitudes alone, but can be fixed at higher points, as we will see.

\subsection{Five-point amplitudes}

We now turn to five-point amplitudes. Since there are no on-shell three-point amplitudes, the five-point amplitudes are also free from poles and are therefore regular in spinor-helicity products.
The on-shell condition for particle 5 imposes the following constraint amongst the Mandelstam variables
\begin{align}
    s_{12} + s_{13} +  s_{14} + s_{23} + s_{24} + s_{34} = 0 \,.
\end{align}
 There are also dimension-dependent identities leading to possible redundancies. In $D=4$, they specifically correspond to Schouten identities 
\begin{align}
\langle ij\rangle|k\rangle+\langle jk\rangle|i\rangle+
\langle ki\rangle|j\rangle=0, \qquad [ij]|k]+[jk]|i]+[ki]|j]=0 \,. 
\end{align}
In general, these constraints are tricky to impose, given that they are non-linear in the spinors. To make progress, we therefore resort to numerics by matching the spinor-helicity amplitudes with a brute-force enumeration of polarization-momenta amplitudes. By plugging in specific external data points, we are able to eliminate the redundancies incurred by the Schouten identities. Moreover, parity invariance imposes further constraints on the $+--+\phi$ and $+-\phi\phi\phi$ configurations. For $+--+\phi$, the scattering amplitude is invariant under parity and relabelling $(1\leftrightarrow 2)$ and $(3\leftrightarrow 4)$, while for $+-\phi\phi\phi$, the amplitude is invariant under parity and relabelling $(1\leftrightarrow 2)$. The most general set of $D=4$ pole-free parity-invariant four-vector-one-scalar amplitudes with 8 derivatives is\footnote{To see that this is the correct power counting for the scalar-vector Galileon we can work with the original basis with cubic vertices and construct double and single exchange diagrams and perform the power counting.}
\begin{align}
    1\;\;2\;\;3\;\;4\;\;5\;\,&\nonumber\\
    \mathcal{A}(----\phi)& =  [12]^2[34]^2\biggl[c_{1}^{----0}(s_{12} + s_{34})^2 + c_{2}^{----0}(s_{14} + s_{23})(s_{13} + s_{24}) \nonumber\\
    &+ c_{3}^{----0}(s_{14} s_{23} + s_{13} s_{24}) \biggr] + \text{2\;perms} \,,
    \,&\\
    \mathcal{A}(+---\phi)& = \langle 12 \rangle^2[23]^2[24]^2  \biggl[ c_{1}^{+---0}(s_{13} + s_{14})+c_{2}^{+---0}(s_{23} + s_{24}) ] \nonumber \\
    & + \langle 13 \rangle \langle 14 \rangle [23][24][34]^2 [ c_{3}^{+---0}(s_{13} + s_{14})+c_{4}^{+---0}(s_{23} + s_{24})\biggr]  + 2\; \text{perms} \,,
    \,&\\
    \mathcal{A}(+--+\phi)& =  \langle 14 \rangle^2[23]^2 \biggl[c_{1}^{+--+0}(s_{14} + s_{23})^2 + c_{2}^{+--+0}(s_{12} + s_{34})(s_{13} + s_{24}) \nonumber \\
   & + c_{3}^{+--+0} (s_{12} s_{34} + s_{13} s_{24})   + c_{4}^{+--+0}s_{14}s_{23}\biggr] \nonumber\\
   &+ c_{5}^{+--+0} \biggl[(\braket{12}^2\braket{34}^2 + \braket{13}^2\braket{24}^2)[23]^4 + ([12]^2[34]^2 + [13]^2[24]^2)\braket{14}^4 \biggr]\nonumber\\
    &+ c_{6}^{+--+0}(\braket{12}\braket{13} \braket{24}\braket{34}[23]^4 - [12][13][24][34]\braket{14}^4)\,,
\end{align}
while the two-vector, three-scalar amplitudes are
\begin{align}
        \mathcal A(--\phi\phi\phi) &={} [12]^2
\biggl[ c_1^{--000}s_{12}^3 + c_2^{--000}s_{12}(s_{34}^2+s_{35}^2+s_{45}^2) + c_3^{--000}s_{34}s_{35}s_{45}
\nonumber\\
&+ \left( c_4^{--000}s_{12}(s_{13}-s_{23})^2 + c_5^{--000}(s_{13}+s_{23})(s_{13}-s_{23})^2 + \text{2 perms}
\right)\biggr]
\nonumber\\
& + c_6^{--000}s_{12} \biggl[ \bigl( [13]^2[24]^2+[14]^2[23]^2 \bigr) \langle34\rangle^2
+\text{2 perms}\biggr],\\
\mathcal A_5(+-\phi\phi\phi)
&= c_1^{+-000} \biggl[ \langle12\rangle^2[23]^2[24]^2 \langle34\rangle^2 + [12]^2\langle13\rangle^2 \langle14\rangle^2[34]^2 + \text{2 perms} \biggr]\nonumber\\
&+ \biggl[ \langle13\rangle^2[23]^2
\biggl( c_2^{+-000}s_{12}^2 + c_3^{+-000}s_{12}(s_{13}+s_{23}) + c_4^{+-000}(s_{13}^2+s_{23}^2) + c_5^{+-000}s_{13}s_{23} \biggr)\nonumber\\
&+ \text{2 perms} \biggr].
\end{align}
Finally, the all-scalar amplitude is given by \cite{Henning:2017fpj}
\begin{align}
    \mathcal{A}_5(\phi\phi\phi\phi\phi) = c_1^{00000}\sum_{i=1}^{5}\prod_{\substack{j=1\\j\ne i}}^{5}s_{ij} + c_2^{00000}\sum_{i<j}s_{ij}^4\,.
\end{align}
In these amplitudes, all dimension-dependent identities, momentum conservation, parity and on-shell conditions have been imposed. We have 13 independent Wilson coefficients for four-vector-one-scalar amplitudes, denoted by $c_{a}^{h_1 h_2 h_3 h_4 0}$, and 11 coefficients for two-vector-three-scalar amplitudes, denoted by $c_a^{h_1 h_2 0 0 0 }$. For the Lagrangian \eqref{LagrangianLevelOne}, the amplitude coefficients are fixed in terms of the Lagrangian parameters as follows:
\paragraph{Four-vector-one-scalar amplitudes}
\begin{align}
c_1^{----0}
&= \frac{1}{13824}\Bigl(
27\alpha^3+36\alpha^2c_3\beta+162\alpha^2d_3\beta
-48\alpha c_3^2\beta^2 \nonumber\\
&\qquad {}+432\alpha c_3d_3\beta^2
-64c_3^3\beta^3+288c_3^2d_3\beta^3\Bigr)
\,, \nonumber\\
c_2^{----0}
&= \frac{-27\alpha^3-36\alpha^2c_3\beta+48\alpha c_3^2\beta^2
+36\alpha c_4\beta^2+64c_3^3\beta^3+48c_3c_4\beta^3}{6912}
\,, \nonumber\\
c_3^{----0}
&= \frac{1}{3456}\Bigl(
-27\alpha^3-36\alpha^2c_3\beta-162\alpha^2d_3\beta
+48\alpha c_3^2\beta^2 \nonumber\\
&\qquad {}-432\alpha c_3d_3\beta^2
+64c_3^3\beta^3-288c_3^2d_3\beta^3\Bigr)
\,, \nonumber\\[4pt]
c_1^{+---0}
&= \frac{27\alpha^3-36\alpha^2c_3\beta-48\alpha c_3^2\beta^2
+72\alpha c_4\beta^2+64c_3^3\beta^3+96c_3c_4\beta^3}{13824}
\,, \nonumber\\
c_2^{+---0}
&= \frac{27\alpha^3+108\alpha^2c_3\beta
+144\alpha c_3^2\beta^2+64c_3^3\beta^3}{13824}
\,, \nonumber\\
c_3^{+---0}
&= \frac{-27\alpha^3-108\alpha^2c_3\beta
-144\alpha c_3^2\beta^2-64c_3^3\beta^3}{6912}
\,, \nonumber\\
c_4^{+---0}
&= \frac{-27\alpha^3-108\alpha^2c_3\beta
-144\alpha c_3^2\beta^2-64c_3^3\beta^3}{6912}
\,, \nonumber\\[4pt]
c_1^{+--+0}
&= \frac{1}{13824}\Bigl(
27\alpha^3+36\alpha^2c_3\beta+162\alpha^2d_3\beta
-48\alpha c_3^2\beta^2 \nonumber\\
&\qquad {}+432\alpha c_3d_3\beta^2
-64c_3^3\beta^3+288c_3^2d_3\beta^3\Bigr)
\,, \nonumber\\
c_2^{+--+0}
&= \frac{-27\alpha^3-36\alpha^2c_3\beta+48\alpha c_3^2\beta^2
+36\alpha c_4\beta^2+64c_3^3\beta^3+48c_3c_4\beta^3}{6912}
\,, \nonumber\\
c_3^{+--+0}
&= \frac{-27\alpha^3-36\alpha^2c_3\beta+48\alpha c_3^2\beta^2
+36\alpha c_4\beta^2+64c_3^3\beta^3+48c_3c_4\beta^3}{3456}
\,, \nonumber\\
c_4^{+--+0}
&= \frac{-27\alpha^2d_3\beta-72\alpha c_3d_3\beta^2
-6\alpha c_4\beta^2-48c_3^2d_3\beta^3-8c_3c_4\beta^3}{576}
\,, \nonumber\\
c_5^{+--+0}
&= 0
\,, \nonumber\\
c_6^{+--+0}
&= 0
\,.
\end{align}
\paragraph{Two-vector-three-scalar amplitudes}
\begin{align}
c_1^{--000}
&= \frac{1}{55296}\Bigl(
27\alpha^3-36\alpha^2c_3\beta+486\alpha^2d_3\beta
-48\alpha c_3^2\beta^2 \nonumber\\
&\qquad {}+36\alpha c_4\beta^2+1944\alpha d_3^2\beta^2
+64c_3^3\beta^3-864c_3^2d_3\beta^3 \nonumber\\
&\qquad {}-48c_3c_4\beta^3+2592c_3d_3^2\beta^3
+216c_4d_3\beta^3\Bigr)
\,, \nonumber\\
c_2^{--000}
&= \frac{1}{55296}\Bigl(
-27\alpha^3+36\alpha^2c_3\beta-486\alpha^2d_3\beta
+48\alpha c_3^2\beta^2 \nonumber\\
&\qquad {}-36\alpha c_4\beta^2-1944\alpha d_3^2\beta^2
-64c_3^3\beta^3+864c_3^2d_3\beta^3 \nonumber\\
&\qquad {}+48c_3c_4\beta^3-2592c_3d_3^2\beta^3
-216c_4d_3\beta^3\Bigr)
\,, \nonumber\\
c_3^{--000}
&= \frac{1}{27648}\Bigl(
-135\alpha^3+180\alpha^2c_3\beta-486\alpha^2d_3\beta
+240\alpha c_3^2\beta^2 \nonumber\\
&\qquad {}-36\alpha c_4\beta^2+5832\alpha d_3^2\beta^2
+3456\alpha d_4\beta^2-320c_3^3\beta^3 \nonumber\\
&\qquad {}+864c_3^2d_3\beta^3-336c_3c_4\beta^3
+7776c_3d_3^2\beta^3+4608c_3d_4\beta^3
+648c_4d_3\beta^3\Bigr)
\,, \nonumber\\
c_4^{--000}
&= \frac{1}{27648}\Bigl(
27\alpha^3-36\alpha^2c_3\beta+162\alpha^2d_3\beta
-48\alpha c_3^2\beta^2 \nonumber\\
&\qquad {}-36\alpha c_4\beta^2+64c_3^3\beta^3
-288c_3^2d_3\beta^3+48c_3c_4\beta^3-216c_4d_3\beta^3\Bigr)
\,, \nonumber\\
c_5^{--000}
&= \frac{1}{27648}\Bigl(
27\alpha^3-36\alpha^2c_3\beta+162\alpha^2d_3\beta
-48\alpha c_3^2\beta^2 \nonumber\\
&\qquad {}-36\alpha c_4\beta^2+64c_3^3\beta^3
-288c_3^2d_3\beta^3+48c_3c_4\beta^3-216c_4d_3\beta^3\Bigr)
\,, \nonumber\\
c_6^{--000}
&= 0
\,, \nonumber\\[4pt]
c_1^{+-000}
&= \frac{27\alpha^3+36\alpha^2c_3\beta-48\alpha c_3^2\beta^2
-36\alpha c_4\beta^2-64c_3^3\beta^3-48c_3c_4\beta^3}{20736}
\,, \nonumber\\
c_2^{+-000}
&= \frac{1}{10368}\Bigl(
54\alpha^3-36\alpha^2c_3\beta+243\alpha^2d_3\beta
+48\alpha c_3^2\beta^2 \nonumber\\
&\qquad {}+36\alpha c_4\beta^2-128c_3^3\beta^3
+432c_3^2d_3\beta^3-96c_3c_4\beta^3+324c_4d_3\beta^3\Bigr)
\,, \nonumber\\
c_3^{+-000}
&= \frac{1}{41472}\Bigl(
351\alpha^3+36\alpha^2c_3\beta+1458\alpha^2d_3\beta
-48\alpha c_3^2\beta^2 \nonumber\\
&\qquad {}+1296\alpha c_3d_3\beta^2+72\alpha c_4\beta^2
-832c_3^3\beta^3+2592c_3^2d_3\beta^3 \nonumber\\
&\qquad {}-480c_3c_4\beta^3+1296c_4d_3\beta^3\Bigr)
\,, \nonumber\\
c_4^{+-000}
&= \frac{1}{13824}\Bigl(
27\alpha^3+36\alpha^2c_3\beta+162\alpha^2d_3\beta
-48\alpha c_3^2\beta^2 \nonumber\\
&\qquad {}+432\alpha c_3d_3\beta^2
-64c_3^3\beta^3+288c_3^2d_3\beta^3\Bigr)
\,, \nonumber\\
c_5^{+-000}
&= \frac{1}{6912}\Bigl(
27\alpha^3+36\alpha^2c_3\beta+162\alpha^2d_3\beta
-48\alpha c_3^2\beta^2 \nonumber\\
&\qquad {}+432\alpha c_3d_3\beta^2
-64c_3^3\beta^3+288c_3^2d_3\beta^3\Bigr)
\,.
\end{align}
\paragraph{All scalar amplitudes}
\begin{align}
c_1^{00000} &= -\frac{\beta^3}{16}\left(27d_3^3+24d_3d_4+5d_5\right)\,, \\
c_2^{00000} &= 0 \,.
\label{LagrangianAmp5pt}
\end{align}
We now want to constrain these amplitudes using the soft theorems derived in Section \ref{sec:SoftTheorems}. For illustrative purposes, we consider the full set of scalar-vector soft theorems for all four-vector-one-scalar configurations. For vector soft theorems, there are two independent hard momenta configurations: $\red{\pm}---\phi$ and $\red{\pm}--+\phi$ up to parity with the soft vector helicity highlighted. For scalar soft theorems, there are three independent hard configurations: $----\red{\phi}$, $---+\red{\phi}$ and $+--+\red{\phi}$ up to parity. 

We start by considering the leading soft behavior of vector emission, namely $\mathcal{O}(p_1)$, since $\mathcal{O}(p_1^2)$ vanishes trivially on the left-hand-side of the two-form soft theorems \eqref{VectorEnergySoftTheorem} and \eqref{VectorMomentumSoftTheorem}. We can then express the above amplitudes as 
\begin{align}
    1\;\;2\;\;3\;\;4\;\;5\;\,&\nonumber\\
    \mathcal{A}(----\phi)&\sim [12]^2[34]^2 \biggl[C^{----0}_1 s_{34}^2 + C^{----0}_2 s_{23} s_{24} \biggr]
    + (2\leftrightarrow 3) + (2\leftrightarrow 4) + \mathcal{O}(p_1^2) \,,\\
    \mathcal{A}(+---\phi)&\sim
    \braket{12}^2[23]^2[24]^2 \biggl[ C^{+---0}_1 s_{34} \biggr]\nonumber \\
    &+ \braket{13}\braket{14}[23][24][34]^2 \biggl[ C^{+---0}_2 s_{34}\biggr]  + (2\leftrightarrow 3) + (2\leftrightarrow 4) +\mathcal{O}(p_1^2) \,,\\
    \mathcal{A}(---+\phi)&\sim
    [12]^2[23]^2 \braket{24}^2\biggl[ C^{---+0}_1 s_{23} + C^{---+0}_2 s_{24} \biggr] + (2\leftrightarrow 3)\nonumber \\
    &+ [12][13]\braket{24}\braket{34}[23]^2 \biggl[ C^{---+0}_3 s_{23} \biggr]   +\mathcal{O}(p_1^2) \,,\\
    \mathcal{A}(+--+\phi)&\sim \braket{14}^2 [23]^2\biggl[C_1^{+--+0} s_{23}^2 + C_2^{+--+0} s_{24} s_{34}  \biggr] + C_3^{+--+0} \biggl( \braket{12}^2\braket{34}^2 + \braket{13}^2\braket{24}^2 \biggr) [23]^4 \nonumber\\
    &+ C_4^{+--+0} \braket{12}\braket{13} \braket{24}\braket{34}[23]^4 
    + \mathcal{O}(p_1^2) \,.
\end{align}
The conversion to the general amplitude coefficients is given by
\begin{align}
\begin{alignedat}{3}
C_1^{----0} &= c_1^{----0},
&\hspace{2em} C_2^{----0} &= c_2^{----0},
&\hspace{2em} C_1^{+---0} &= -c_{2}^{+---0} \,,\\
C_2^{+---0} &= -c_{4}^{+---0},
&\hspace{2em} C_1^{---+0} &= c_{2}^{+---0}-c_{1}^{+---0},
&\hspace{2em} C_2^{---+0} &= -c_{1}^{+---0} \,,\\
C_3^{---+0} &= -c_{3}^{+---0},
&\hspace{2em} C_1^{+--+0} &= c_1^{+--+0},
&\hspace{2em} C_2^{+--+0} &= c_2^{+--+0} \,,\\
C_3^{+--+0} &= c_5^{+--+0},
&\hspace{2em} C_4^{+--+0} &= c_6^{+--+0} \,.
\end{alignedat}
\end{align}
Up to $\mathcal{O}(p_1^2)$, 11 of the 13 coefficients remain in these leading soft amplitudes. To implement the vector soft theorems \eqref{VectorEnergySoftTheorem} and \eqref{VectorMomentumSoftTheorem}, we need the explicit form of the polarization vector, which is given by\footnote{To match the amplitude scaling of \eqref{Littlegroupscaling} it is useful to note that it is the complex conjugate of the polarisation vector that is associated with an out-state particle.}
\begin{align}
    \epsilon^\mu_{-}(\vec{p}_a) &= - \frac{[r|\gamma^\mu\ket{a}}{\sqrt{2}[ra]} \,,\\
    \epsilon^\mu_{+}(\vec{p}_a) &=  \frac{[a|\gamma^\mu\ket{r}}{\sqrt{2}\braket{ra}} \,,
\end{align}
 where $r$ can be any reference spinor that's not parallel to the spinor $a$. The abuse of notation is justified by defining
\begin{align}
[r|\gamma^\mu|a\rangle
&=
\begin{pmatrix} 0 & [r|_{\bar\alpha} \end{pmatrix}
\begin{pmatrix}
0 & (\sigma^\mu)_{\alpha\bar\beta} \\
(\bar\sigma^\mu)^{\bar\alpha\beta} & 0
\end{pmatrix}
\begin{pmatrix} |a\rangle_{\beta} \\ 0 \end{pmatrix}
=
[r|_{\bar\alpha}(\bar\sigma^\mu)^{\bar\alpha\beta}
|a\rangle_{\beta}
\,, \nonumber\\[6pt]
\langle a|\gamma^\mu|r]
&=
\begin{pmatrix} \langle a|^{\alpha} & 0 \end{pmatrix}
\begin{pmatrix}
0 & (\sigma^\mu)_{\alpha\bar\beta} \\
(\bar\sigma^\mu)^{\bar\alpha\beta} & 0
\end{pmatrix}
\begin{pmatrix} 0 \\ |r]^{\bar\beta} \end{pmatrix}
=
\langle a|^{\alpha}(\sigma^\mu)_{\alpha\bar\beta}
|r]^{\bar\beta}
\,,
\end{align}
which requires us to convert two-component spinors to four-component ones as in \cite{Elvang:2013cua}. Since $[r|\gamma^\mu\ket{a}^*=[a|\gamma^\mu\ket{r}$ and $[r a]^*=-\braket{r a}$ for physical momenta, we then have
\begin{align}
    \epsilon_{\pm}^{*\mu}(-\vec{p}_1) &= \epsilon_{\mp}^{\mu}(\vec{p}_1)|_{1\rightarrow 1'} \,,
\end{align}
where $\ket{a'}[a'|\equiv E_a \sigma^0 + \vec{p}_a\cdot\vec{\sigma}$ representing the energy flipped version of spinor-helicity variables. 

\subsubsection{The $\red{\pm} - - - \phi$ configuration}
First, we consider the following pair of soft five-point amplitudes
\begin{align}
    \mathcal{A}(+---\phi), \quad \mathcal{A}(----\phi) \,.
\end{align}
The LHS of \eqref{VectorMomentumSoftTheorem} for such soft vector emission is then
\begin{align}
    \text{LHS}_5 =& -\partial_{p_{1[j}}\biggl(\frac{\epsilon^{i]}_{-}(\vec{p}_1)}{2}\mathcal{A}_{E_1}(----\phi) + \frac{\epsilon^{*i]}_{+}(-\vec{p}_1)}{2}\mathcal{A}_{-E_1}(----\phi)\biggr)\nonumber\\
    &- \partial_{p_{1[j}}\biggl(\frac{\epsilon^{i]}_{+}(\vec{p}_1)}{2}\mathcal{A}_{E_1}(+---\phi) + \frac{\epsilon^{*i]}_{-}(-\vec{p}_1)}{2}\mathcal{A}_{-E_1}(+---\phi)\biggr) \nonumber\\
    =&\frac{1}{2}\partial_{p_{1[j}} \biggl( \frac{[r|\gamma^{i]}\ket{1}}{\sqrt{2}[r1]}[12]^2 + (1\rightarrow 1') \biggr) [34]^2\biggl( C_1^{----0} s_{34}^2 + C_2^{----0}s_{23}s_{24} \biggr)\nonumber\\ 
    &-\frac{1}{2}\partial_{p_{1[j}} \biggl( \frac{\bra{r}\gamma^{i]}|1]}{\sqrt{2}\braket{r1}}\braket{12}^2 + (1\rightarrow 1') \biggr) [23]^2 [24]^2\biggl[ C^{+---0}_1 s_{34}\biggr]\nonumber\\
    &-\frac{1}{2}\partial_{p_{1[j}} \biggl( \frac{\bra{r}\gamma^{i]}|1]}{\sqrt{2}\braket{r1}}\braket{13}\braket{14} + (1\rightarrow 1') \biggr) [23] [24] [34]^2 \biggl[ C^{+---0}_2 s_{34}\biggr] + 2 \;\text{perms}\nonumber\\
    \stackrel{\text{choose}\; r}{=\joinrel=\joinrel=}& -\frac{1}{2\sqrt{2}}\partial_{p_{1[j}} \biggl( [2|\gamma^{i]}\ket{1}[12] + (1\rightarrow 1') \biggr) [34]^2\biggl( C_1^{----0} s_{34}^2 + C_2^{----0}s_{23}s_{24} \biggr)\nonumber\\
    &+\frac{1}{2\sqrt{2}}\partial_{p_{1[j}} \biggl( \bra{2}\gamma^{i]}|1]\braket{12} + (1\rightarrow 1') \biggr) [23]^2 [24]^2\biggl[ C^{+---0}_1 s_{34} \biggr] + 2 \;\text{perms}\nonumber\\
    &+\frac{1}{2\sqrt{2}}\partial_{p_{1[j}} \biggl( \bra{3}\gamma^{i]}|1]\braket{14} + (1\rightarrow 1') \biggr) [23] [24] [34]^2\biggl[ C^{+---0}_2 s_{34} \biggr] + 2 \;\text{perms}\nonumber\\
    \stackrel{\ket{1}[1| = p_\mu \sigma^\mu }{=\joinrel=\joinrel=\joinrel=\joinrel=\joinrel=\joinrel=}& \frac{1}{\sqrt{2}} \biggl\{-[2\gamma^{[i}\gamma^{j]}2][34]^2 \biggl( C_1^{----0} s_{34}^2 + C_2^{----0}s_{23}s_{24} \biggr) \nonumber\\
    &+ \braket{2\gamma^{[i}\gamma^{j]}2} [23]^2 [24]^2\biggl[ C^{+---0}_1 s_{34} \biggr] + \frac{1}{2}\biggl(\braket{3\gamma^{[i}\gamma^{j]}4} + (3\leftrightarrow 4)\biggr)[23] [24] [34]^2\biggl[ C^{+---0}_2 s_{34} \biggr] \nonumber \\
    &+ 2 \;\text{perms}\biggr\} \,. \label{eqfirsjee}
\end{align}
The choice of the reference spinors is such that the spurious pole from the polarization vector is cancelled, leaving a regular expression. Due to gauge invariance of the soft theorem, we can opt to fix the gauge by setting $\epsilon^0_h = 0$.\footnote{Alternatively, notice the $\epsilon^0$ term within the soft theorem has a sign difference between the energy-flipped and the un-flipped soft amplitude, which projects out the spatial momentum component of $p_\mu \sigma^\mu$, leaving only $E_p  \sigma^0$, multiplied by $p^j/E_p$ yields $p^j$, which is the only sector the spatial momentum derivative $\partial_{p_i}$ hits, hence zero by anti-symmetrisation.} For the last equal sign, the amplitude combination with its energy flip counterpart projects out the energy component of the momentum
\begin{align}
    \frac{\ket{1}[1| + \ket{1'}[1'|}{2} = \vec{p}\cdot\vec{\sigma}.
\end{align}
Therefore the soft momentum derivative only hits the spatial component. To compute the RHS of \eqref{VectorMomentumSoftTheorem}, we notice the polarization-momentum factor can be written in terms of
\begin{align}
    \epsilon_+^{[\mu}(\vec{p}_a)p^{\nu]}_a &= \epsilon_-^{*[\mu}(p_a)p_a^{\nu]}=  \frac{1}{4\sqrt{2}}[a\gamma^{[\mu}\gamma^{\nu]}a]\,, \\
    \epsilon_-^{[\mu}(\vec{p}_a)p^{\nu]}_a &= \epsilon_+^{*[\mu}(p_a)p_a^{\nu]}= - \frac{1}{4\sqrt{2}}\braket{a\gamma^{[\mu}\gamma^{\nu]}a}\,,
\end{align}
which are independent of the reference spinor, as they should be by gauge invariance. Therefore the RHS of \eqref{VectorMomentumSoftTheorem} for the four-point hard amplitude is
\begin{align}
    \text{RHS}_4 &= \frac{\alpha'}{4\sqrt{2}}\biggl[ c^{----}_1 \biggl( \braket{5\gamma^{[i}\gamma^{j]}5}[54]^2 [23]^2 s_{23} + 2\;\text{perms} \biggr) - c^{---+}_1 [5\gamma^{[i}\gamma^{j]}5] \braket{52}^2[23]^2[24]^2 \nonumber\\
    &-[2\gamma^{[i}\gamma^{j]}2][34]^2 \biggl(c^{--00}_1 s_{34}^2 + c^{--00}_2 s_{23}s_{24} \biggr) + 2\;\text{perms}\biggr]\,.
\end{align}
The $5$ dependence can be removed by momentum conservation
\begin{align}
    \sum_{a=2}^5 \ket{a}[a| = 0 \,,
\end{align}
which results in
\begin{align}
    \text{RHS}_4 =&\frac{\alpha'}{4\sqrt{2}}\biggl[ c^{----}_1\biggl(\braket{2\gamma^{[i}\gamma^{j]}2}[24]^2+ \braket{2\gamma^{[i}\gamma^{j]}3} [24][34] + (2\leftrightarrow 3) \biggr)[23]^2 s_{23} + (2\leftrightarrow 4) + (3\leftrightarrow 4) \nonumber\\
    -& [2\gamma^{[i}\gamma^{j]}2] [34]^2 \biggl(c^{--00}_1 s_{34}^2 + c^{--00}_2 s_{23}s_{24} \biggr) + (2\leftrightarrow 3) + (2\leftrightarrow 4)\biggr] \,,
\end{align}
where $c_1^{---+}$ vanishes by the four-point soft theorem. Matching both sides, we obtain the following constraints on the Wilson coefficients
\begin{align}
     C_1^{----0} &= \frac{\alpha'}{4} c_1^{--00} ,&
    C_2^{----0} &= \frac{\alpha'}{4} c_2^{--00} ,&
    \frac{1}{2} C_2^{+---0} = -C_1^{+---0} &= \frac{\alpha'}{4} c_1^{----} \,.
\end{align}
One can run the same computation for the spatial-temporal component $m_{i0}$ of the soft theorem \eqref{VectorEnergySoftTheorem} and arrive at the same conclusion by applying the following identity 
\begin{align}
    \braket{a \gamma^{[i}\gamma^{0]} a} = \braket{a \gamma^i \gamma^0 a},
\end{align}
since they ultimately live in the same Lorentz multiplet.
\subsubsection{The $\red{\pm} -- + \phi$ configuration}
The next five-point pair configuration in the soft limit we're considering is
\begin{align}
    \mathcal{A}(---+\phi), \quad
    \mathcal{A}(+--+\phi)  \,.
\end{align}
The LHS of the soft theorem \eqref{VectorMomentumSoftTheorem} for this specific configuration follows by the same procedure
\begin{align}
    \text{LHS}_5 = & -\partial_{p_{1[j}}\biggl(\frac{\epsilon^{i]}_{-}(\vec{p}_1)}{2}\mathcal{A}_{E_1}(---+\phi) + \frac{\epsilon^{*i]}_{+}(-\vec{p}_1)}{2}\mathcal{A}_{-E_1}(---+\phi)\biggr)\nonumber\\
 &-\partial_{p_{1[j}}\biggl(
 \frac{\epsilon_+^{i]}(\vec p_1)}{2}\mathcal{A}_{E_1}(+--+\phi)
 +\frac{\epsilon_-^{*i]}(-\vec p_1)}{2}\mathcal{A}_{-E_1}(--+\phi)\biggr)\nonumber\\
    =& \frac{1}{\sqrt{2}} \biggl\{-[2\gamma^{[i} \gamma^{j]} 2] [23]^2 \braket{24}^2 \biggl[C_1^{---+0} s_{23} + C_2^{---+0}s_{24} \biggr] - (2\leftrightarrow 3)\nonumber\\
    &- \frac{1}{2}\biggl([2\gamma^{[i} \gamma^{j]} 3] + (2\leftrightarrow 3)\biggr)\braket{24}\braket{34}[23]^2 \biggl[ C^{---+0}_3 s_{23} \biggr]\nonumber\\
    &+\braket{4\gamma^{[i}\gamma^{j]}4}[23]^2\biggl[C_1^{+--+0} s_{23}^2 + C_2^{+--+0} s_{24} s_{34}  \biggr] \nonumber \\
    &+ C_3^{+--+0}\biggl(\braket{2\gamma^{[i}\gamma^{j]}2}\braket{34}^2 + \braket{3\gamma^{[i}\gamma^{j]}3}\braket{24}^2\biggr) [23]^4 \nonumber\\
    &+ \frac{1}{2}C_4^{+--+0}\biggl(\braket{2\gamma^{[i}\gamma^{j]}3} + (2\leftrightarrow 3)\biggr)\braket{24}\braket{34}[23]^4 \biggr\}\,.
\end{align}
The RHS of the soft theorem for the same helicity configuration is then
\begin{align}
    \text{RHS}_4 =&\frac{\alpha'}{4\sqrt{2}}\biggl( -c_1^{--++}[5\gamma^{[i}\gamma^{j]}5][23]^2\braket{45}^2 s_{23}+ \braket{4\gamma^{[i}\gamma^{j]}4}[23]^2(c_1^{--00} s_{23}^2 + c_2^{--00}s_{24} s_{34}) \nonumber\\
    &- c_1^{-+00} [3\gamma^{[i}\gamma^{j]}3][23]^2\braket{34}^2 s_{24} - (2\leftrightarrow 3)\biggr)\nonumber\\
=& \frac{\alpha'}{4\sqrt{2}} \biggl\{-[2\gamma^{[i}\gamma^{j]}2][23]^2\braket{24}^2\biggl[\bigl(c_1^{--++}-c_1^{-+00}\bigr)s_{23}-c_1^{-+00}s_{24}\biggr]-(2\leftrightarrow3)\nonumber\\
&-\frac12\biggl([2\gamma^{[i}\gamma^{j]}3]+(2\leftrightarrow3)\biggr)\braket{24}\braket{34}[23]^2\biggl[2c_1^{--++}s_{23}\biggr] \nonumber \\
&+\braket{4\gamma^{[i}\gamma^{j]}4}[23]^2\biggl[c_1^{--00}s_{23}^2+c_2^{--00}s_{24}s_{34}\biggr]\biggr\} \,,
\end{align}
where we used momentum conservation to eliminate the momentum dependence of particle $5$. Matching both sides of the soft theorem results in the constraints
\begin{align}
    C_1^{---+0} &= \frac{\alpha'}{4}(c_1^{--++} - c_1^{-+00}),&
    C_2^{---+0} &= \frac{\alpha'}{4} (-c_1^{-+00}),&
    C_1^{+--+0} &= \frac{\alpha'}{4} c_1^{--00},\nonumber\\
    C_3^{---+0} &= \frac{\alpha'}{2} c_1^{--++}&
    C_2^{+--+0} &= \frac{\alpha'}{4} c_2^{--00}, &
    C_3^{+--+0} &= C_4^{+--+0} =  0 \,.
\end{align}
The soft theorem \eqref{VectorEnergySoftTheorem} is then also satisfied once these constraints have been imposed.
\subsubsection{The $AAAA\red{\phi}$ configuration}

For the scalar soft theorem, since on the RHS we need to sum over all vector momenta, the RHS is zero by momentum conservation if there's only one soft scalar emission from the LHS where all other hard external states are vectors. Therefore, the soft single scalar amplitude enjoys an enhanced soft scaling $\mathcal{O}(p^2)$. We devise a shortcut by choosing a specific set of hard kinematics that allow us to extract soft behavior of the scalar momentum $p_5$. We let
\begin{align}
\ket{1}&=\binom{1}{0},
&\hspace{2em}\ket{2}&=\binom{0}{1},
&\hspace{2em}\ket{3}&=\binom{1}{1},
&\hspace{2em}\ket{4}&=\binom{1}{x} \,,
\nonumber\\
|1]&=\binom{1}{0},
&\hspace{2em}|2]&=\binom{0}{1},
&\hspace{2em}|3]&=\frac{1}{x-1}\binom{-x}{1},
&\hspace{2em}|4]&=\frac{1}{x-1}\binom{1}{-1} \,,
\nonumber\\
\ket{5}&=\binom{u}{v},
&\hspace{2em}|5]&=\binom{\rho}{\sigma},
&\hspace{2em}p_5&=\tau\ket{5}[5| \,,
\nonumber\\
|1](\tau)&=|1]-\tau u|5],
&\hspace{2em}|2](\tau)&=|2]-\tau v|5] \,,
\end{align}
where $x\neq 1,\, u,\, v,\,\rho,\,\sigma $ are generic complex variables and $\tau$ is our soft scaling parameter. The soft momenta shift when turning on $\tau$ are only imposed on $1,2$ while rendering $3,4$ unchanged, analogously to a BCFW shift. The parametrization retains both momentum conservation and on-shell conditions. Now we can explicitly evaluate the soft scaling of the amplitudes with respect to $\tau$ and we find
\begin{align}
\mathcal{A}_{----0}(\tau)
={}&
\frac{x^2-x+1}{(x-1)^6}
\bigl(4c_1^{----0}+c_3^{----0}\bigr)\nonumber\\
&\biggl[
2(x^2-x+1)
-3\tau\bigl(
2\rho u(x^2-x+1)+\rho v(x+1)+\sigma u(x^2+x)+2\sigma v(x^2-x+1)
\bigr)
\biggr]\nonumber\\
&+\mathcal{O}(\tau^2),\\
\mathcal{A}_{+---0}(\tau)
={}&
-\tau\frac{\sigma v x^2}{(x-1)^4}
\bigl(4c_{2}^{+---0}+3c_{3}^{+---0}-c_{4}^{+---0}\bigr) + \mathcal{O}(\tau^2),
\nonumber\\[2mm]
\mathcal{A}_{+--+0}(\tau)
={}&
\frac{x^4}{(x-1)^4}
\biggl[
x^2\bigl(4c_1^{+--+0}+c_3^{+--+0}+c_4^{+--+0}+2c_5^{+--+0}\bigr) 
+ 2 (x-1)\bigl(2c_2^{+--+0}-c_3^{+--+0}\nonumber\\
&-2c_5^{+--+0}\bigr)
\biggr]
-\tau\frac{x^3}{(x-1)^4}
\biggl\{
x^2\bigl(4c_1^{+--+0}+c_3^{+--+0}+c_4^{+--+0}+2c_5^{+--+0}\bigr)
\bigl[\rho ux+3\rho v+\nonumber\\
&
\sigma ux+3\sigma vx\bigr]
 + (x-1)\bigl(2c_2^{+--+0}-c_3^{+--+0}-2c_5^{+--+0}\bigr)
\bigl[2\rho ux+\rho vx+4\rho v+\sigma ux^2+6\sigma vx\bigr]
 \nonumber\\
 &+ (x-1)c_6^{+--+0}
\bigl[2\rho ux+\rho v(x-4)+\sigma ux^2-2\sigma vx\bigr]
\biggr\} + \mathcal{O}(\tau^2).
\end{align}
Demanding $\mathcal{A}\sim\mathcal{O}(\tau^2)$, we have
\begin{align}
c_3^{----0}&=-4c_1^{----0}, \quad c_{4}^{+---0} =4c_{2}^{+---0}+3c_{3}^{+---0} \,, \nonumber \\
c_3^{+--+0}&=2c_2^{+--+0}-2c_5^{+--+0}, \quad 
c_4^{+--+0}=-4c_1^{+--+0}-2c_2^{+--+0}, \quad
 c_6^{+--+0}=0 \,.
\end{align}
\subsubsection{Summary of constraints}
Combining all of the above constraints, we find that all 13 five-point Wilson coefficients are uniquely fixed by the four-point Wilson coefficients. In addition, we also have one four-point coefficient fixed in terms of another. The complete solution to the soft theorems is then given by:
\begin{equation}
\boxed{
\begin{aligned}
c_1^{----}&=c_1^{--++},
&\hspace{2em} c_1^{----0}&=\frac{\alpha'}{4}c_1^{--00},
&\hspace{2em} c_2^{----0}&=\frac{\alpha'}{4}c_2^{--00} \,,
\\
c_3^{----0}&=-\alpha'c_1^{--00},
&\hspace{2em} c_{1}^{+---0}&=\frac{\alpha'}{4}c_1^{-+00},
&\hspace{2em} c_{2}^{+---0}&=\frac{\alpha'}{4}c_1^{--++} \,,
\\
c_{3}^{+---0}&=-\frac{\alpha'}{2}c_1^{--++},
&\hspace{2em} c_{4}^{+---0}&=-\frac{\alpha'}{2}c_1^{--++},
&\hspace{2em} c_1^{+--+0}&=\frac{\alpha'}{4}c_1^{--00} \,,
\\
c_2^{+--+0}&=\frac{\alpha'}{4}c_2^{--00},
&\hspace{2em} c_3^{+--+0}&=\frac{\alpha'}{2}c_2^{--00},
&\hspace{2em} c_4^{+--+0}&=-\frac{\alpha'}{2}
    \bigl(2c_1^{--00}+c_2^{--00}\bigr) \,,
\\
c_5^{+--+0}&=0,
&\hspace{2em} c_6^{+--+0}&=0 \,.
\end{aligned}
}\end{equation}
These constraints precisely match the direct Feynman diagram calculations \eqref{LagrangianAmp4pt} and \eqref{LagrangianAmp5pt}, and nicely illustrate the power of the soft theorems.

\section{Conclusions and Future Directions\label{sec:Conclusions}}

In this paper, we have considered flat-space theories of a scalar field coupled to a gauge field exhibiting extended non-linearly realized symmetries. We set out to classify theories by constructing symmetry algebras that lead to the appropriate number of inverse Higgs constraints required to eliminate all would-be Goldstones other than the scalar and the vector, and that allow for canonical two-derivative kinetic terms for each field. We took the algebras to include the linearly-realized Poincar\'{e} algebra, the $U(1)$ gauge symmetry algebra, and additional generators corresponding to non-linearly realized symmetries. 

We started by taking the scalar to transform non-linearly under a scalar generator $C$ and a vector generator $C_{\mu}$, and the gauge field to transform non-linearly under a two-form generator $M_{\mu\nu}$. Within this class, we showed that the only allowed algebras lead to the scalar-vector Galileon theory which arises in the scalar-vector sector of the decoupling limit of ghost-free massive gravity.  This theory comes in an abelian version that arises in the decoupling limit of pseudo-linear massive gravity, and a non-abelian version that arises in the decoupling limit of dRGT massive gravity. In particular, we showed that the dilaton and DBI scalars cannot couple to a gauge field with a two-form symmetry. We used the coset construction to derive compact expressions for the leading Wess-Zumino terms that correspond to the kinetic terms and the leading interactions which mix the scalar and the vector in non-trivial ways. We then looked for interesting subsets of this theory which have further symmetries, in particular special Galileon symmetries. We found, in the abelian case, a unique algebra where the corresponding interactions are completely fixed in terms of a single coupling constant. We derived the Wess-Zumino interactions for this theory in general dimensions. 

For both the scalar-vector Galileon and its special extension, we then derived the corresponding soft theorems. The presence of linear-in-field terms in the symmetries means that the soft theorems are not necessarily enhanced Adler zeros, rather they can have a non-trivial right-hand side and relate soft $n+1$-point amplitudes to hard $n$-point amplitudes. Our soft theorems hold in $D$ dimensions and the combination of both soft vector insertion and emission entailed by the charge conservation equation is indispensable. To solve the soft theorems, we worked in $D=4$ dimensions, wrote down general amplitudes for scalar-vector scattering at four and five points, and constrained the free amplitude coefficients using the soft theorems. At five-point, we focused our attention on amplitudes with one scalar and four vectors and showed that the soft theorems are powerful enough to fix $\textit{all}$ 13 coefficients that appear in the ansatz in terms of four-point coefficients, as well as an additional constraint between four-point coefficients. The result agrees with a Feynman diagram calculation from the scalar-vector Galileon Lagrangian. 

There are a number of avenues for future work:
\begin{itemize}
    \item It would be interesting to extend our algebraic analysis beyond level one of the vector tree, i.e. going beyond the two-form generator to higher-order mixed symmetry ones. As far as we are aware, there are no known interacting theories of gauge fields with such symmetries. A classification would be a conceptually straightforward extension of what we have done in this work. 
    \item At five points, we used our soft theorems to fix the amplitudes with four vectors and one scalar. The remaining amplitudes are those with two vectors and three scalars, and it would be interesting to fix those amplitudes too and verify that the solutions match the Feynman diagram calculation of the scalar-vector Galileon.  
\item Our analysis has been focused on scalar-vector theories.  A natural extension would be to consider scalar-tensor theories. It is known that coupling scalar EFTs with non-linear symmetries to gravity generically breaks the symmetries \cite{Pirtskhalava:2015nla}. This approach would instead involve looking for symmetry algebras that require the graviton itself to also transform non-trivially, as in \cite{Gabadadze:2012tr}. 
\end{itemize}

\paragraph{Acknowledgements} We would like to thank Sadra Jazayeri and Diederik Roest for useful discussions. The work of DS has been supported by a UKRI Stephen Hawking Fellowship with grant number EP/W005441/1, a Nottingham Research Fellowship from the University of Nottingham, and an STFC Consolidated Grant with grant number ST/C004235/1. ZD is supported by Simons Investigator award SFI-MPS-SIP-00017355. We used Claude to help generate figures, and ChatGPT to help verify the five-point scalar soft theorems. For the purpose of open access, the authors have applied a CC BY public copyright licence to any Author Accepted Manuscript version arising.

\bibliographystyle{JHEP}
\bibliography{ExceptionalSpin}

\end{document}